\documentclass{aa}  

\usepackage{graphicx}
\usepackage{txfonts}
\usepackage{amsfonts}
\usepackage{longtable}
\usepackage{pdflscape}
\usepackage{longtable}
\usepackage{afterpage}
\usepackage{changepage}
\usepackage{geometry}
\usepackage{fancyvrb}
\usepackage{hhline}

\begin{document}

   \title{Determining the true mass of radial-velocity exoplanets with Gaia}
   \subtitle{9 planet candidates in the brown-dwarf/stellar regime and 27 confirmed planets}

   \author{F. Kiefer\inst{1,2} \and
          G. H\'ebrard\inst{1,3} \and
          A. Lecavelier des Etangs\inst{1} \and 
          E. Martioli\inst{1,4} \and
          S. Dalal\inst{1} \and
          A. Vidal-Madjar\inst{1} 
          }

   \institute{
   \inst{1} Institut d’Astrophysique de Paris, Sorbonne Universit\'e, CNRS, UMR 7095, 98 bis bd Arago, 75014 Paris, France \\
   \inst{2} LESIA, Observatoire de Paris, Universit\'e PSL, CNRS, Sorbonne Universit\'e, Universit\'e de Paris, 5 place Jules Janssen, 92195 Meudon, France\thanks{Please send any request to flavien.kiefer@obspm.fr} \\
   \inst{3} Observatoire de Haute-Provence, CNRS, Universite\'e d’Aix-Marseille, 04870 Saint-Michel-l’Observatoire, France \\        
   \inst{4} Laborat\'{o}rio Nacional de Astrof\'{i}sica, Rua Estados Unidos 154, 37504-364, Itajub\'{a} - MG, Brazil
             }

   \date{Submitted on 2020/08/20 ; Accepted for publication on 2020/09/24}

 
  \abstract
  { Mass is one of the most important parameters for determining the true nature of an astronomical object. Yet, many published exoplanets
  lack a measurement of their true mass, in particular those detected thanks to radial 
  velocity (RV) variations of their host star. For those, only the minimum mass, or $m\sin i$, is known, owing to the insensitivity of RVs to the inclination of the 
  detected orbit compared to the plane-of-the-sky. The mass that is given in database is generally that of an assumed edge-on system ($\sim$90$^\circ$), but 
  many other inclinations are possible, even extreme values closer to 0$^\circ$ (face-on). In such case, the mass of the published object could be 
  strongly underestimated by up to two orders of magnitude. In the present study, we use GASTON, a tool recently developed in Kiefer et al. (2019) \& 
  Kiefer (2019) to take advantage of the voluminous Gaia astrometric database, in order to constrain the inclination and true mass of several hundreds of published 
  exoplanet candidates. We find 9 exoplanet candidates in the stellar or brown dwarf (BD) domain, among which 6 were never characterized. We show that 
  30\,Ari\,B\,b,   HD\,141937\,b, HD\,148427\,b, HD\,6718\,b, HIP\,65891\,b, and HD\,16760\,b have masses larger than 
  13.5\,M$_\text{J}$ at  3-$\sigma$. We also confirm the planetary nature of 27 exoplanets among which HD\,10180\,c, d and g. Studying the 
  orbital periods, eccentricities and host-star metallicities in the BD domain, we found distributions  with respect to true masses consistent with other publications. 
  The distribution of orbital periods shows of a void of BD detections below $\sim$100\,days, while eccentricity and metallicity distributions agree
  with a transition between BDs similar to planets and BDs similar to stars about 40--50\,M$_\text{J}$.
  }


   \maketitle
%

\section{Introduction}

A large fraction of exoplanets published in all up-to-date catalogs, such as \verb+www.exoplanet.eu+ (Schneider et al. 2011) or \verb+the NASA exoplanet archive+ 
(Akeson et al. 2013) were detected thanks to radial velocity variations of their host star. If the minimum mass $m\sin i$, with $i$ being a common symbolic notation 
for 'orbital inclination', is located below the planet/brown-dwarf (BD) critical mass of 
13.5\,M$_\text{J}$ such detection has to be considered as a new "candidate" planet. If any observed 
system is inclined according to an isotropic distribution, there is indeed a non-zero probability $1-\cos I_c$, with $I_c$ the inclination of the candidate orbit, that 
the $m\sin i$ underestimates the true mass of the companion by a factor larger than $1/\sin I_c$. With $I_c$=10$^\circ$, this leads already to a factor $\sim$6, 
with a probability of 1.5\%. Such small rate, considering the $\sim$500-1000 planets detected through RVs, implies that only few tens of planets have a 
mass strongly underestimated. However, exoplanets catalogs usually neglect an important number of companions which $m\sin i$ is larger than  
13.5\,M$_\text{J}$. The RV-detected samples of exoplanets in catalogs are partly biased to small $m\sin i$ candidates and thus to small inclinations (see 
e.g. Han 2001 for a related discussion). They likely contains more than few tens of objects which actual mass is larger than 13.5\,M$_\text{J}$.

Recovering the exact mass ratio distribution of binary companions from their mass function, therefore bypassing the issue of unknown inclination by using 
inversion techniques, such as the Lucy-Richardson algorithm (Richardson 1972, Lucy 1974), is a famous well-studied problem (Mazeh \& Goldberg 1992, Heacox 
1995, Sahaf et al. 2017). Such inversion algorithms were applied to RV exoplanets mass distribution (Zucker \& Mazeh 2001b, Jorissen, Mayor \& Udry 2001, 
Tabachnik \& Tremaine 2002). However, it is a statistical problem that cannot determine individual masses. It also lacks strong validation by comparing to exact 
mass distributions in the stellar or planetary regime. Moreover, the distribution of binary companions in the BD/M-dwarf, with orbital periods of 1-10$^4$\,days, 
from which exoplanet candidates could be originating, is not well described. Detections are still lacking in the BD regime -- the so-called BD desert (Marcy et al. 2000) 
-- although this region is constantly being populated (Halbwachs et al. 2000, Sahlmann et al. 2011a, Di\'az et al. 2013, Ranc et al. 2015, Wilson et al. 2016, 
Kiefer et al. 2019). Our sparse knowledge of the low-mass tail of the population of stellar binary companions does not allow disentangling low $\sin i$ 
BD/M-dwarf from real exoplanets. It also motivates to extensively characterize the orbital inclination and true mass of companions in the exoplanet to M-dwarf regime.

The true mass of an individual RV exoplanet candidate can be determined by directly measuring the inclination angle of its orbit compared 
to the plane of the sky. If the companion is on an edge-on orbit ($I_c$$\sim $$90^\circ$), then it is likely transiting and could be detected using photometric monitoring. 
Commonly the transiting exoplanets are detected first with photometry -- with e.g. Kepler, WASP, TESS -- and then characterized in mass with RV. About half 
of the exoplanets observed in RV were detected by transit photometry. The other half are not known to transit and the main options to measure the inclination of 
an exoplanet orbit are mutual interactions in the case of multiple planets systems, and astrometry. 

Astrometry has been used to determine the mass of exoplanet candidates in many studies. Observations with the Hubble Space Telescope 
Fine-Guidance-Sensor (FGS) led to confirm few planets, in particular GJ\,876\,b (Benedict et al. 2002), and $\epsilon$-Eri\,b (Benedict et al. 2006). It led also to 
corrected mass of planet candidates beyond the Deuterium-burning limit, such as HD\,38529\,b with a mass in the BD regime of 17\,M$_\text{J}$ 
(Benedict et al. 2010), and HD\,33636\,b, with an M-dwarf mass of 140\,M$_\text{J}$ (Bean et al. 2007). Hipparcos data were also extensively used to that 
purpose (Perryman et al. 1996, Mazeh et al. 1999, Zucker \& Mazeh 2001a, Sozzetti \& Desidera 2010, Sahlmann et al. 2011a, Reffert \& Quirrenbach 2011, 
Di\'az et al. 2012, Wilson et al. 2016, Kiefer et al. 2019) but only yielded masses in the BD/M-dwarf regime. More recently, Gaia astrometric data were used for the 
first time  to determine the mass of RV exoplanet candidates with different methods: either based on astrometric excess noise for HD\,114762\,b, showing it is 
stellar in nature (Kiefer 2019), either by comparing Gaia proper motion to Hipparcos proper motion on the case of Proxima\,b, confirming its planetary nature 
(Kervella et al. 2020). It is expected that Gaia will provide by the end of its mission the most precise and voluminous astrometry able to characterize exoplanet 
companions and even to detect new exoplanets (Perryman et al. 2014).

In the present study, we aim at assessing the nature of numerous RV-detected exoplanet candidates publicaly available in exoplanets catalogs using the 
astrometric excess noise from the first data release, or DR1, of the Gaia mission (Gaia Collaboration et al. 2016). We use the recently developed method GASTON 
(Kiefer et al. 2019, Kiefer 2019), to constrain, from the astrometric excess noise and RV-derived orbital parameters, the orbital inclination 
and true mass of these companions. 

In Section~\ref{sec:initial_selection}, we define the sample of companions and host-stars selected from this study. In Section~\ref{sec:gaia_input}, we explore the 
Gaia archive for the selected systems and reduce the sample of companions to those with exploitable Gaia DR1 data and astrometric excess noise. We summarize
the GASTON method in Section~\ref{sec:gaston}. The GASTON results are presented in Section~\ref{sec:results}. They are then discussed in 
Section~\ref{sec:discussion}. We conclude in Section~\ref{sec:conclusion}.
 
\section{Initial exoplanet candidates selection}
\label{sec:initial_selection}

In order to measure the inclination and true mass of orbiting exoplanet candidates, complete information on orbital parameters are required. We thus need first to 
select a database in which the largest number of published exoplanets fullfill several criteria. They are the followings:

\begin{itemize}
\item[(1)] A measurement for period $P$, eccentricity $e$, RV semi-amplitude $K$ and star mass $M_\star$ must exist;
\item[(2)] $K$, $P$ and $M_\star$ should be $>$0;
\item[(3)] If $e>0.1$, a measurement for $T_p$ and $\omega$, respectively the time of periastron passage and the angle of periastron, must exist. If 
$e$$<$$0.1$, the orbit is about circular, so the phase is not taken into account in the GASTON method and thus $T_p$ and $\omega$ are spurious parameters;
\item[(4)] A given value for $m\sin i$ (otherwise calculated from other orbital parameters);
\item[(5)] Recently updated.
\end{itemize}

We compared the 3 main exoplanets databases available on-line, which are the \verb+www.exoplanet.eu+ (Schneider et al. 2011), \verb+www.exoplanets.org+ (Han et al. 2014) 
and \verb+NASA exoplanet archive+, applying these above criteria. A complete review on the current state of on-line catalogs has been achieved in Christiansen (2018). 
The result of this comparison is shown in Table~\ref{tab:compare_database}. 

\begin{table*}\centering
\caption{\label{tab:compare_database} Comparing the 3 main databases with respect to the 5 criteria given in the text. The five first lines provide the number of 
exoplanets.} 
\begin{tabular}{lccc}
Criterion & \verb+exoplanet.eu+ & \verb+NASA Exoplanet Archive+ & \verb+exoplanets.org+ \\
\hline
(0) None & 4302 & 4197 & 3262  \\
(1 ) $K$, $P$, $e$, $M_\star$ exist & 752 & 1237 & 967 \\ 
(2 ) $K>0$, $P>0$, $e\geq 0$, $M_\star>0$ & 750 & 1237 & 961 \\ 
(3) + $T_p$ and $\omega$ exist if $e\geq0.1$ &  582 & 909 & 924 \\
(4) + measured $m\sin i$	 &  360	& 580	&  911 \\
(5) Last update considered & 10/08/2020 & 23/07/2020 & June 2018\\	
\hline 
\end{tabular}
\end{table*}

The \verb+www.Exoplanets.org+ although not updated since June 2018 is the most complete database available, with respect to planetary, stellar and orbital parameters, with 
a complete set of orbital data for 911 companions. For comparison, in the NASA exoplanet archive (NEA), there are only 580 exoplanets for which a complete set of parameters 
is given. In the \verb+Exoplanet.eu+ database, a reference in terms of up-to-date data (4302 against 4197 in the NEA on 12th of August 2020), suffers from 
inhomogeneities in the reported data, with e.g. some masses expressed in Earth mass while most are given in Jupiter mass, or radial velocities semi-amplitudes that are only
sparsely reported. We found best to rely on the \verb+www.Exoplanets.org+ database, the most homogeneous, although counting only 3262 objects. It constitutes 
a robust yet not too old reference sample of objects that will remain unchanged in the future, since updates have ceased.

In this database, applying the above criteria, the sample of companions reduces down to 924 companions. A measurement of $m\sin i$ is provided with 
uncertainties for 911 of them, following Wright et al. (2011). There thus remains 13 objects for which the $m\sin i$ was not provided. 
Those planets are all transiting, but for 12 of them no RV signal is detected (Marcy et al. 2014) and $K$ is only an MCMC estimation with large errorbars. We will exclude 
those 12 objects from our analysis. The remaining planet with no $m\sin i$ given in the database is Kepler-76\,b. However, a solid RV-variation detection is reported in 
Faigler et al. (2013), leading to an $m\sin i$ of 2$\pm$0.3\,M$_\text{J}$. We thus keep Kepler-76\,b in our list of targets and insert its $m\sin i$ measurement.

The selected sample also includes 358 exoplanets detected with transit photometry and Doppler velocimetry. These companions with known inclination of their 
orbit -- edge-on in virtually all cases -- will be useful to assess the quality of the inclinations obtained independently with GASTON. The full list of 912 selected companions 
orbiting 782 host-stars are shown in Table~\ref{tab:init_list_targets}.

\begin{table*}\centering\tiny
\caption{\label{tab:init_list_targets} List of selected exoplanet companions (see Section~\ref{sec:initial_selection}). Only the 10 first companions of the sample are 
shown here. The whole sample of 912 companions will be made available online.}
\begin{tabular}{@{}l@{~~}c@{~~}c@{~~}c@{~~}c@{~~}c@{~~}c@{~~}c@{~~}c@{~~}c@{~~}c@{}}
Companion 	&  $m\sin i$ 	& $P$	&  $e$  	&  $\omega$ 	&  $T_p$ 	& $K$		&  $M_\star$		& drift flag	& RV (O$-$C)		& Transit flag \\
			&  (M$_J$)	& (day)	& 		&  ($^\circ$)	&    (JD)	& m\,s$^{-1}$ 	&  ($M_\odot$)		&   (0/1)		&   (m\,s$^{-1}$)	&  (0/1)	 \\
\hline
11 Com b   	&	16.1284$\pm$1.53491	&	326.03$\pm$0.32	&	0.231$\pm$0.005	&	94.8$\pm$1.5		&	2452899.6$\pm$1.6		&	302.8$\pm$2.6		& 	2.04$\pm$0.29	&	0	&	25.5		&	0    \\
11 UMi b    	&	11.0873$\pm$1.10896	&	516.22$\pm$3.25	&	0.08$\pm$0.03	&	117.63$\pm$21.06	&	2452861.04$\pm$2.06	&	189.7$\pm$7.15	&	1.8$\pm$0.25		&	0	&	28		&	0    \\
14 And b   	&	4.68383$\pm$0.22621	&	185.84$\pm$0.23	&	0                          	& 	0				&	2452861.4$\pm$1.5		&	100$\pm$1.3		&	2.15$\pm$0.15		&	0	&	20.3		&	0    \\
14 Her b      	&	5.21486$\pm$0.298409	&	1773.4$\pm$2.5	&	0.369$\pm$0.005	&	22.6$\pm$0.9		&	2451372.7$\pm$3.6		&	90$\pm$0.5		&	1.066$\pm$0.091	&	0	&	5.6		&	0    \\
16 Cyg B b	&	1.63997$\pm$0.0833196	&	798.5$\pm$1      	&	0.681$\pm$0.017	&	85.8$\pm$2		&	2446549.1$\pm$7.4		&	50.5$\pm$1.6		&	0.956$\pm$0.0255	&	0	&	7.3		&	0    \\
18 Del b     	&	10.298$\pm$0.36138	&	993.3$\pm$3.2	&	0.08$\pm$0.01		&	166.1$\pm$6.5		&	2451672$\pm$18		&	119.4$\pm$1.3		&	2.33$\pm$0.05		&	0	&	15.5		&	0    \\
24 Boo b    	&	0.912932$\pm$0.110141	&	30.3506$\pm$0.00775&	0.042$\pm$0.0385	&	210$\pm$115		&	2450008.6$\pm$9		&	59.9$\pm$3.25		&	0.99$\pm$0.16		&	0	&	0.02651	&	0    \\
24 Sex b    	&	1.83564$\pm$0.108126	&	455.2$\pm$3.2		& 	0.184$\pm$0.029	&	227$\pm$20		&	2454758$\pm$30		&	33.2$\pm$1.6		&	1.81$\pm$0.08		&	0	&	4.8		&	0    \\
24 Sex c    	&	1.51716$\pm$0.200171	&	910$\pm$21         	&	0.412$\pm$0.064	&	352$\pm$9		&	2454941$\pm$30		&	23.5$\pm$2.9		&	1.81$\pm$0.08		&	0	&	6.8		&	0    \\
\\
\multicolumn{11}{c}{Continued online...} \\
\hline
\end{tabular}
\end{table*}

\section{Gaia inputs}
\label{sec:gaia_input}

\subsection{Gaia DR1 data for the target list}
\label{sec:gaia_selec}

The GASTON algorithm determine the inclination of RV companion orbits using the Gaia DR1 astrometric excess noise (Gaia Collaboration et al. 2016, Lindegren et 
al. 2016). The most recent Gaia DR2 release cannot be used similarly because it is based on a different definition of the astrometric excess noise 
and moreover cursed by the so-called 'DOF-bug' directly affecting the measurement of residual scatter (Lindegren et al. 2018). For that reason, from Kiefer et al. (2019)
it was decided to rely the GASTON analysis on the more reliable, although preliminary, Gaia DR1 data. 

The list of host stars constituted in Section~\ref{sec:initial_selection} is uploaded in the Gaia archive of the DR1 to retrieve astrometric data around each star, with a
search radius of 5". Among the 782 host stars of our initial sample defined in the previous Section, we found 679 entries in the DR1 catalog. Most stars are reported singles, but among the 679 DR1 
sources, 44 (with 50 reported exoplanets) have a close background star, a visual companion, or a duplicated (but non-identified) source, with a separation to the 
main source smaller than 5".

In particular, 7 stars (with 12 reported exoplanets) have a "visual companion" with a different ID, at less than 5" distance but with an equal magnitude $\pm$0.01. 
This is strongly suspicious, and must be due to duplication in the catalogue. Duplication is only reported in the Gaia DR1 database for one of those sources, YZ Cet. 
We consider safer to exclude these 7 sources from our analysis. However, in general we want to keep those that are marked as duplicate. Duplication 
separates the dataset of a single source into two different IDs. In the worst case scenario, duplication lead to ignore outlying measurements, and thus to 
underestimate the astrometric scatter. This can only be problematic if GASTON leads to characterize a mass in the regime of planets, since underestimating the 
astrometric excess noise implies underestimating the mass. More generally, duplication is not an issue because GASTON characterizes masses in the regime of 
BD or stars, allowing to exclude a planetary nature. 

Finally, we identified three supplementary problematic hosts with a magnitude difference with 
commonly adopted values, as in e.g. SIMBAD, larger than 3. These are Proxima\,Cen, HD\,142 and HD\,28254 (see e.g. Lindegren et al. (2016) for Proxima\,Cen) . 
We also exclude them from our studied sample. We also note the presence of 11 sources with a null parallax, which are also taken off the sample.

The Gaia DR1 sources are divided into two different datasets : the 'primary' and the 'secondary' (Lindegren  et al. 2016). The primary dataset contains two million of targets 
also observed with Tycho/Hipparcos for which there is a robust measurement of parallax and proper motion out of a 24-year baseline astrometry. It is also sometimes referred 
to as the TGAS (for the joint Tycho-Gaia Astrometric Solution) dataset. The secondary dataset  contains 1.141 billion sources that do not have a supplementary 
constraint on position from Tycho/Hipparcos, some of those being also newly discovered objects.  In the secondary dataset, the proper motion and parallax are 
fitted to the Gaia data, leaving from a prior based on magnitude (Michalik et al. 2015b), but they are discarded in the DR1. In Lindegren et al. (2016), it is reported 
that the astrometric residuals scatter is generally larger in the secondary dataset that in the primary dataset (see also Section~\ref{sec:noise_distribution} below). 
We will thus separate those secondary dataset objects from those in the primary dataset in the rest of the study and treat them specifically. 

In total, we constituted a sample of 755 exoplanets with both RV and Gaia DR1 data, orbiting 658 stars of which Table~\ref{tab:list_targets} gives the full list. Among
those, 508 exoplanets orbit 436 stars in the primary dataset, for 247 exoplanets around 222 stars in the secondary dataset. We list among all DR1 parameters
the $G$-band magnitude, the parallax, the belonging to primary or secondary dataset, the source duplication (see e.g. Lindegren et al. 2016), the number of field-of-view
transits $N_\text{FoV}$ (\verb+matched_observations+ in the DR1 catalog), the total number of recorded along-scan angle (AL) measurements $N_\text{tot}$, the 
astrometric excess noise $\varepsilon_\text{DR1}$ and its significance parameter $D_\varepsilon$ (Lindegren et al. 2012).

\begin{table*}\tiny
\caption{\label{tab:list_targets} List of selected stellar hosts from the initial sample and selected from the Gaia DR1 archive (see Section~\ref{sec:gaia_selec}). Only 
the 10 first sources of the sample are shown here. The whole list of 658 stars hosting 755 exoplanet candidates will be made available online.}
\begin{tabular}{l|c@{~~~}c@{~~~}c@{~~~}c@{~~~}c@{~~~}|c@{~~~}c@{~~~}c@{~~~}c@{~~~}c@{~~~}c@{~~~}c@{~~~}c@{~~~}|c}
Source 	&   \multicolumn{5}{c|}{SIMBAD}           			&   \multicolumn{8}{c|}{Gaia DR1}   & Gaia DR2 \\
\hline
		& RA 	&  DEC	&   $V$ 	& $B-V$  &  Sp type	&  G\tablefootmark{a} & $\pi$\tablefootmark{b} & Dataset\tablefootmark{c} & Duplicate\tablefootmark{d} & $N_\text{FoV}$\tablefootmark{e} & $N_\text{tot}$\tablefootmark{f} & $\varepsilon_\text{DR1}$\tablefootmark{g}  & $D_\varepsilon$\tablefootmark{h} & $b-r$\tablefootmark{i} \\
		&		&		& 		&		& 		& 	& (mas)	&	&	&	&	& (mas)& & \\
\hline
11 UMi & 15:17:05.8915 & +71:49:26.0375 & 5.02 &  6.38 & K4III & 4.7 	& 7.47$\pm$0.66	& 1	& false	& 15	& 76	& 2.4 	& 11220 	& 1.51\\
14 And & 23:31:17.4127 & +39:14:10.3105 &   & 6.24 & G8III	& 5.0 	& 13.23	& 2	& false	& 6	& 49	& 5.6 	& 14395 & 1.17 \\
14 Her & 16:10:24.3153 & +43:49:03.4987 &   & 7.57 & K0V & 6.3 	& 55.93$\pm$0.24	& 1	& false	& 18	& 107	& 0.62 	& 258 	& 1.00\\
16 Cyg B & 19:41:51.9732 & +50:31:03.0861 &  6.20  & 6.86 & G3V	& 6.0 	& 47.12$\pm$0.23	& 1	& false	& 15	& 80	& 0.40 	& 173 	& 0.83\\
18 Del & 20:58:25.9337 & +10:50:21.4261 &  5.51  & 6.43 & G6III	& 5.3 	& 13.09	& 2	& false	& 7	& 50	& 3.0 	& 10385 	& 1.08\\
24 Boo & 14:28:37.8131 & +49:50:41.4611 &   & 6.44 & G4III-IVFe-1	& 5.3 	& 10.23$\pm$0.56	& 1	& false	& 31	& 195	& 2.6 	& 5721 	& 1.07\\
24 Sex & 10:23:28.3694 & -00:54:08.0772 &  6.44  & 7.40 & K0IV	& 6.1 	& 13.85	& 2	& false	& 6	& 44	& 0.68 	& 175 	& 1.10 \\
30 Ari B & 02:36:57.7449 & +24:38:53.0027 &  7.09  & 7.59 & F6V 	& 6.9 	& 21.42$\pm$0.60	& 1	& false	& 10	& 71	& 1.8 	& 428 	& 0.68\\
7 CMa & 06:36:41.0376 & -19:15:21.1659 &  3.91  & 5.01 & K1.5III-IVFe1 	& 4.0 	& 50.63	& 2	& false	& 37	& 272	& 6.0 	& 164838 	& 1.24\\
70 Vir & 13:28:25.8082 & +13:46:43.6430 &  4.97  & 5.68 & G4Va 	& 4.9 	& 54.70$\pm$0.88	& 1	& false	& 29	& 213	& 3.2 	& 11554 	& 0.90\\
\\
\multicolumn{15}{c}{Continued online...} \\
\hline
\end{tabular}
\tablefoot{ \\
\tablefoottext{a}{The Gaia recorded flux magnitude in the G-band.} \\
\tablefoottext{b}{The parallax. For the sources from the secondary dataset, the values are given without errorbars since missing from the DR1. They are taken from 
the DR2. For those, we will assume 10\% errorbars in the rest of the study.} \\
\tablefoottext{c}{DR1 primary (1) or secondary (2) dataset.} \\
\tablefoottext{d}{Duplicate source (true) or not (false), as explained in Lindegren et al. (2016).} \\
\tablefoottext{e}{Number of field-of-view transits of the sources  (\texttt{matched\_observations} in the DR1 database).} \\
\tablefoottext{f}{Total number of astrometric AL observations reported (\texttt{astrometric\_n\_good\_obs\_al} in the DR1 database).} \\
\tablefoottext{g}{Astrometric excess noise in mas.} \\
\tablefoottext{h}{Significance of $\varepsilon_\text{DR1}$. Any $D_\varepsilon$$>$$2$ leads to a significant astrometric excess noise with p-value=$e^{-D_\varepsilon/2}$ as 
explained in Lindegren et al. (2012).} \\
\tablefoottext{i}{$G_b$-$G_r$ color index as presented in Lindegren  et al. (2018).} \\
}
\end{table*}

\subsection{Magnitude, color and parallax correlations with astrometric excess noise}
\label{sec:mag_selec}

The astrometric excess noise is the main measured quantity that will be used in this study to derive a constraint on the RV companion masses listed in 
Table~\ref{tab:init_list_targets}. The fundamental hypothesis assumed in GASTON relates the astrometric excess noise to astrometric orbital motion. It is thus crucial to identify 
possible systematic correlations of this quantity with respect to other intrinsic data such as magnitude, color or DR1 dataset that would reveal instrumental or 
modelisation effects. 

\begin{figure}
\includegraphics[width=89.3mm]{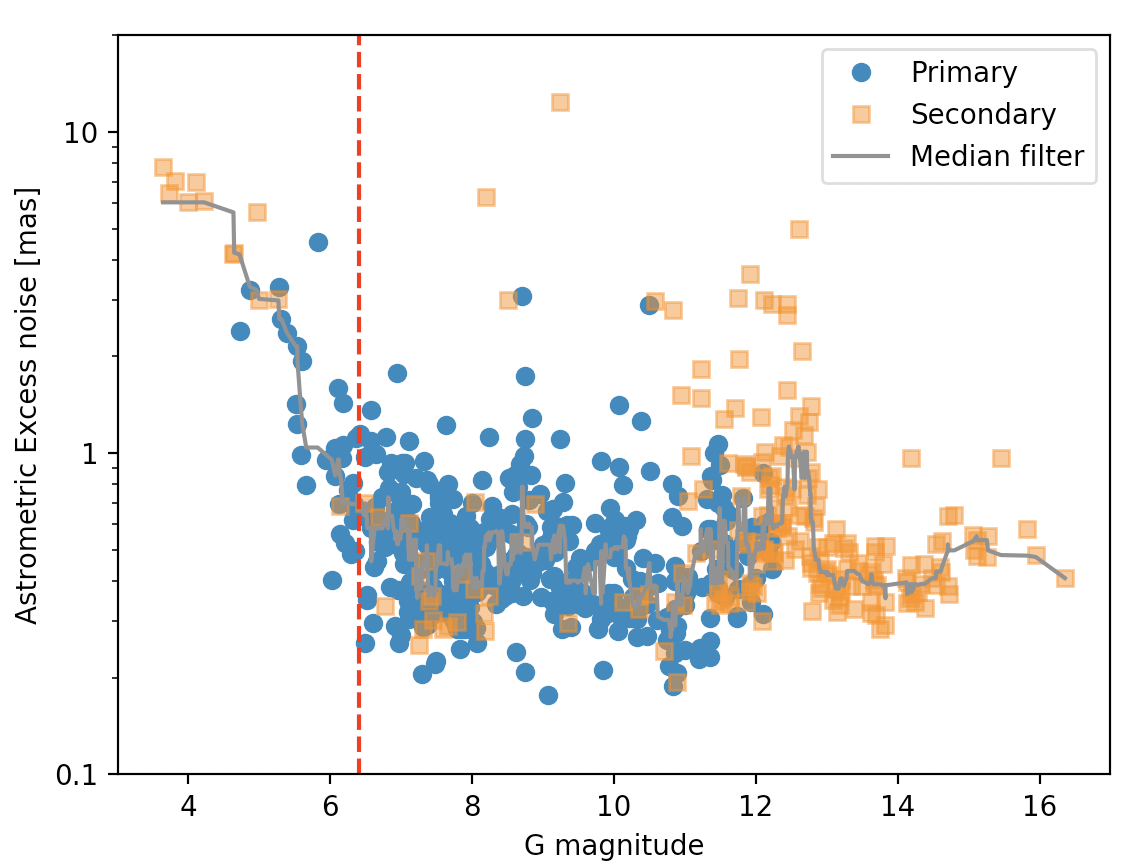}
\caption{\label{fig:mag} Comparing astrometric excess noise to $G$-magnitude for the 658 stars of Table~\ref{tab:list_targets}. The red dashed line shows the 
$G$=6.4 limit discussed in the text. We separate targets from the primary dataset (in blue) and from the secondary dataset (orange). The gray line is a moving 
median filter of the data.}
\end{figure}

\begin{figure*}
\includegraphics[width=89.3mm]{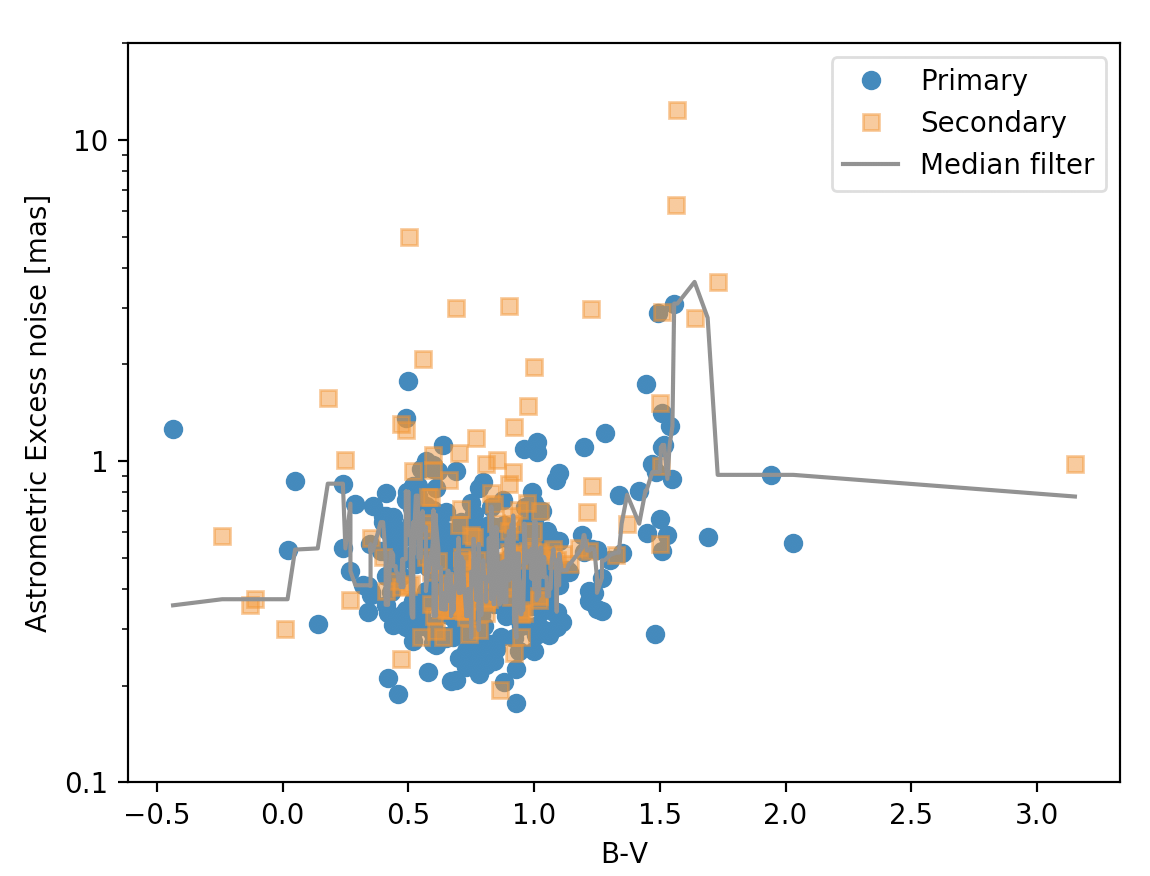}
\includegraphics[width=89.3mm]{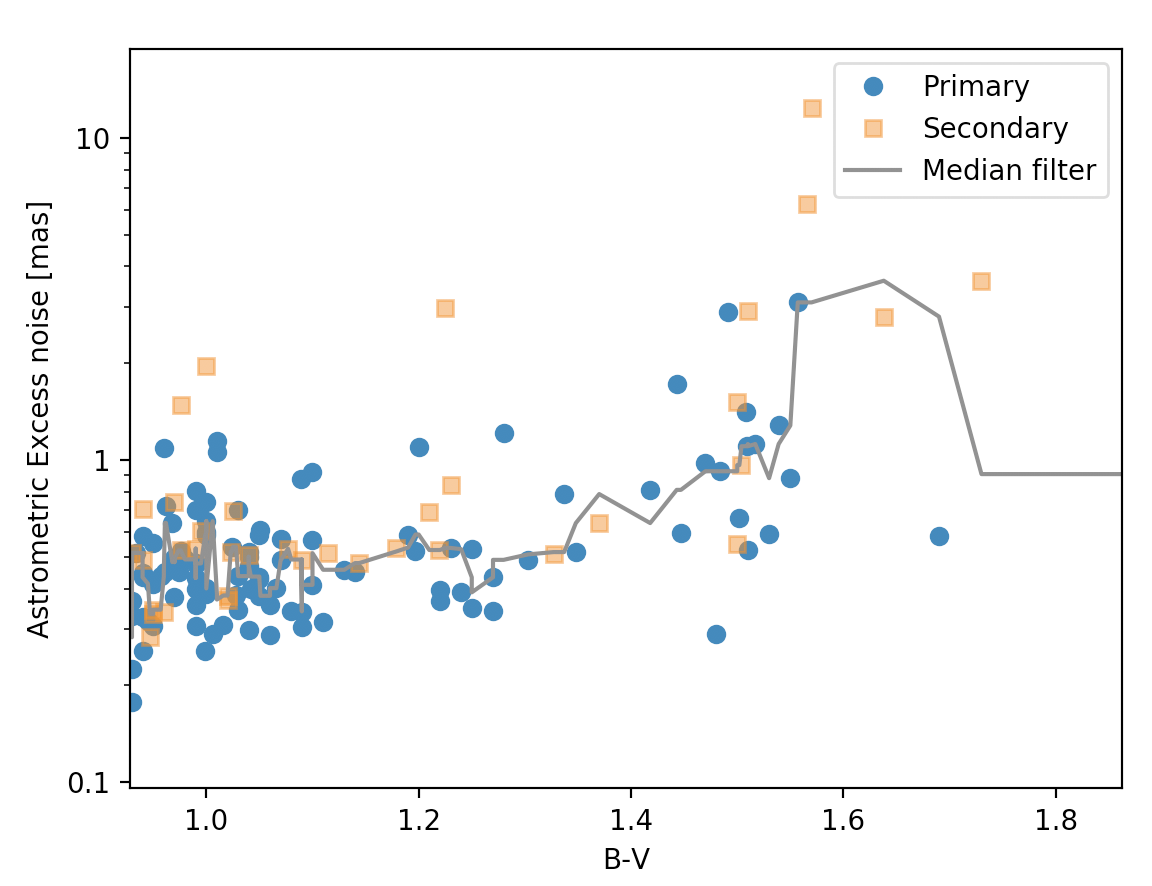}
\caption{\label{fig:BV} Astrometric excess noise with respect to $B-V$ for 498 sources with $B-V$ measurements and $G$$>$$6.4$. The right panel is 
a magnification of the region with positive correlation beyond $B-V$=1. Symbols and color are similar to Fig~\ref{fig:mag}.}
\end{figure*}

\begin{figure*}
\includegraphics[width=89.3mm]{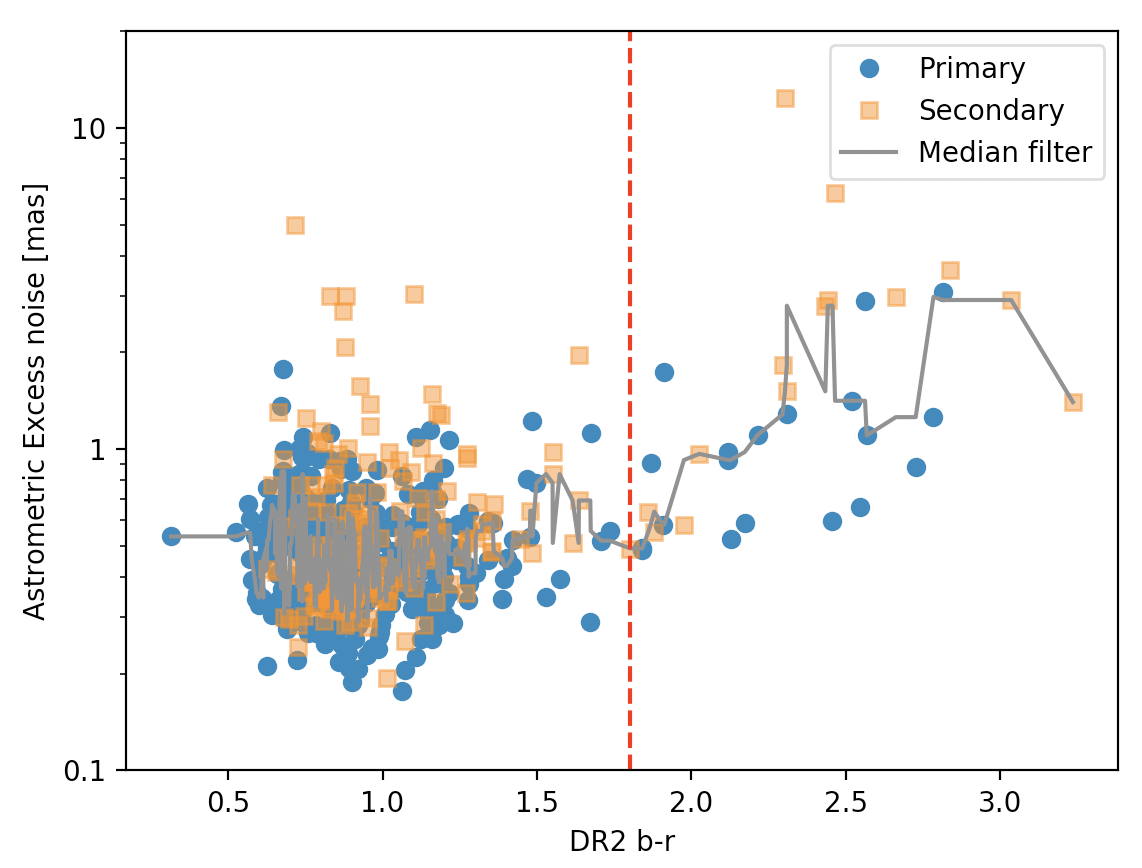}
\includegraphics[width=89.3mm]{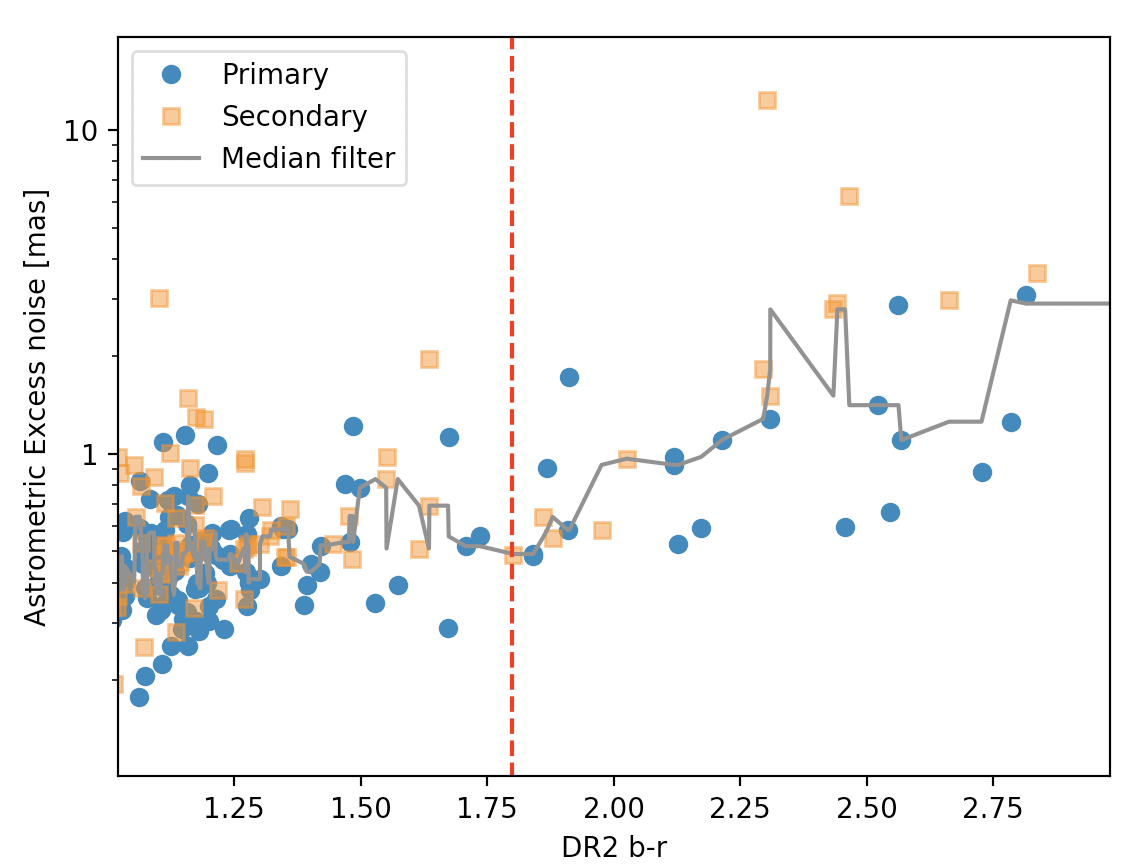}
\caption{\label{fig:DR2br} Astrometric excess noise with respect to $b-r$ color index from Gaia DR2 for the 614 sources with $G$$>$$6.4$. The right panel is 
a magnification of the region with a positive correlation around the limiting color $b-r$=1.8 (red dashed line). Symbols and color are similar to Fig~\ref{fig:mag}.}
\end{figure*}

\begin{figure*}
\includegraphics[width=89.3mm]{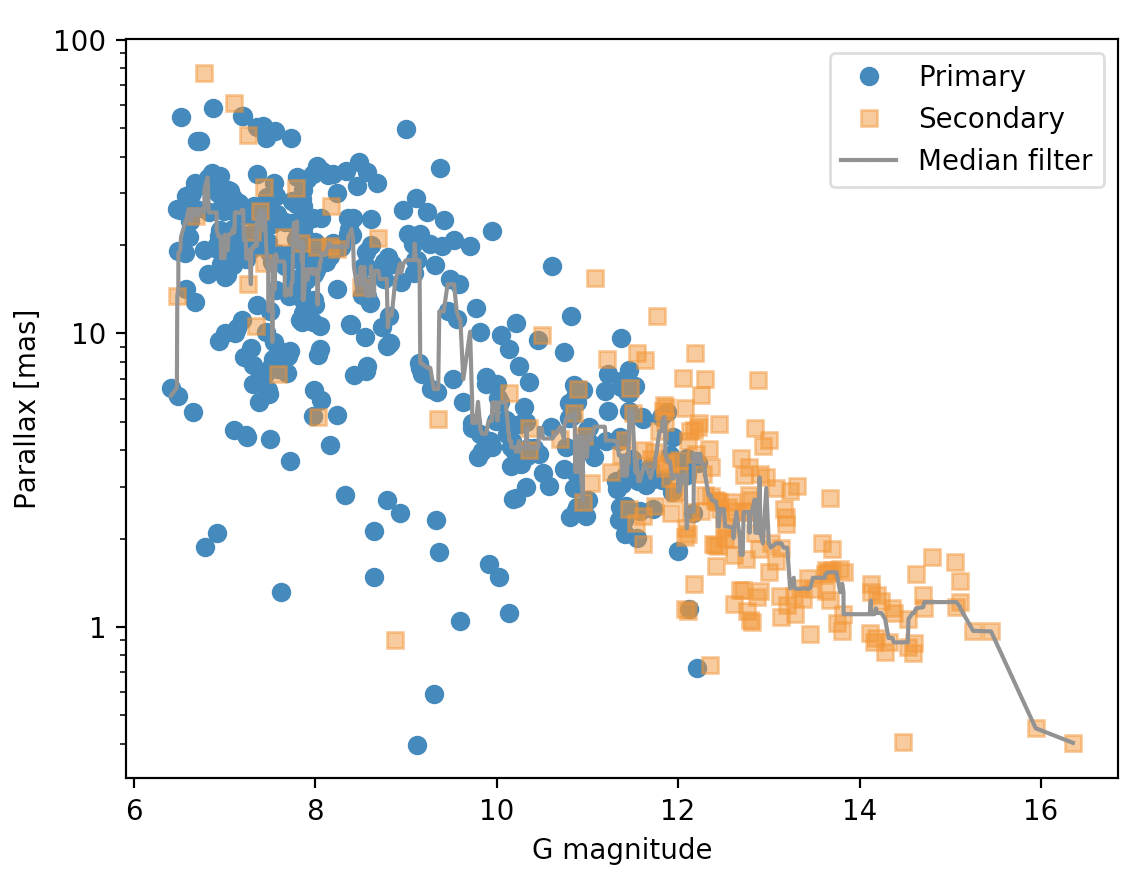}
\includegraphics[width=89.3mm]{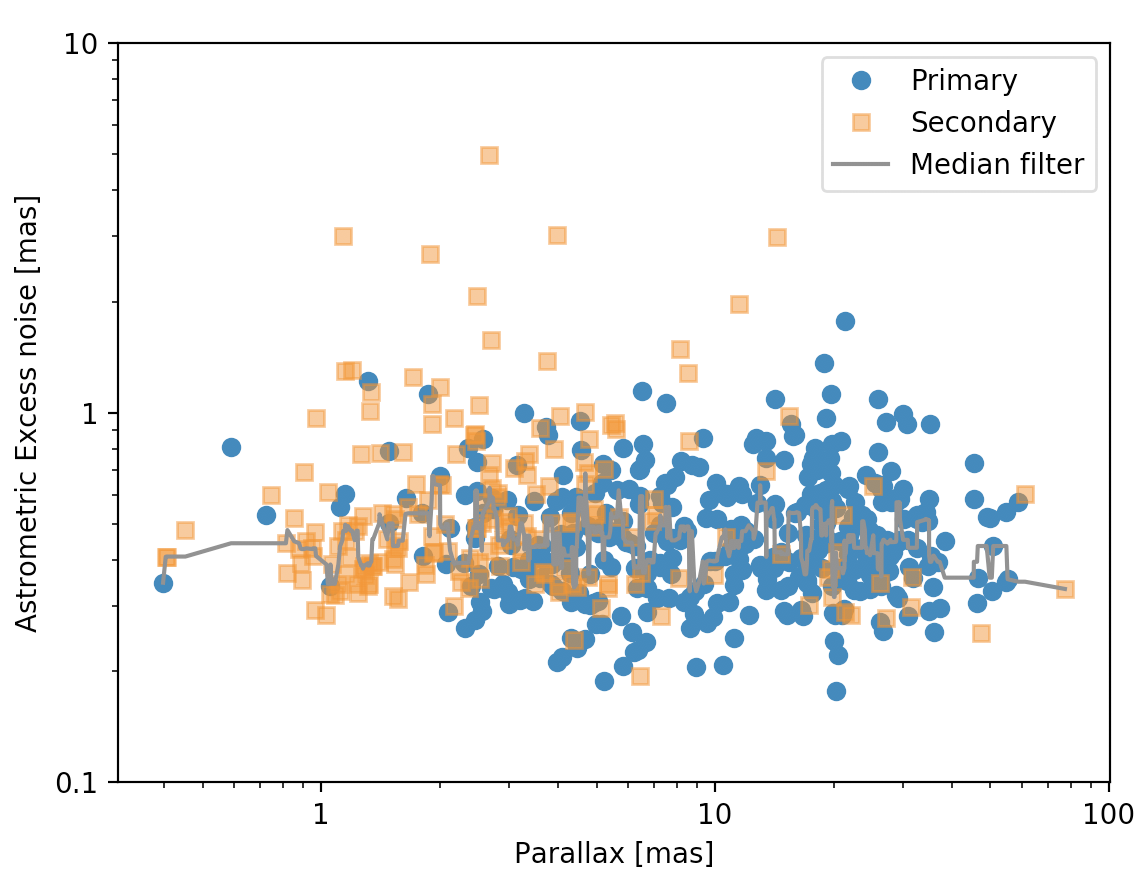}
\caption{\label{fig:parallax} $G$-magnitude with respect to parallax (left-panel) and parallax with respect to astrometric excess noise (right panel) for the 580 
sources with $G$$>$$6.4$ and $b-r$$<$$1.8$. Symbols and color are similar to Fig~\ref{fig:mag}.}
\end{figure*}

As can be seen in Fig~\ref{fig:mag}, stars brigher than magnitude 6.4 show a significant drift of increasing excess noise with decreasing magnitude. This is a sign 
of instrumental systematics  (PSF, jitter, CCD sensibility etc.), that are recognised to occur in Gaia data (Lindegren et al. 2018). With $G$-mag$<$6.4, the 
astrometric excess noise are all larger than 0.4\,mas, with a median value about 0.7\,mas. In the rest of the paper, we will thus exclude any source with a 
$G$-mag$<$6.4, reducing the sample to 614 sources. 

Moreover, the astrometric excess noise also shows a correlation with color indices, i.e. the B-V as found in Simbad for 498 sources with $G$-mag$>$6.4, and the 
Gaia DR2 $b-r$ available in the DR2 database for all the 614 sources with $G$-mag$>$6.4, as plotted in Figs~\ref{fig:BV} and~\ref{fig:DR2br}. A moving median 
filter of the data indeed shows a correlation of $\varepsilon_\text{DR1}$ with B-V beyond 1, and with DR2 $b-r$ beyond 1.4. The B-V index is not available for all
the 614 sources, we thus prefer using the DR2 $b-r$ index as a limiting parameter. As for the magnitude, the astrometric excess noise is larger than 0.4\,mas 
whatever $b-r$ larger than 1.8. A correlation of the along-scan (AL) angle residuals with V-I color was already reported in Section D.2 of Lindegren et al. (2016). These two 
correlations likely have a common optical origin due to the chromaticity of the star centroid location on the CCD. In the rest of the paper, we will thus also 
exclude any source with a $b-r$$>$$1.8$, reducing the sample to 580 sources.

Figure~\ref{fig:parallax} shows that the parallax and magnitude are correlated in both primary and secondary datasets, which is not surprising as we expect 
distant sources to be on average less luminous than sources close-by. Sources of the secondary dataset are located much farther away from the Sun than sources
of the primary dataset, which is expected from the absence of Tycho data for the secondary dataset. The astrometric excess noise is not correlated with 
parallax, but we observe astrometric excess noise measurements on the same order -- and even larger -- than the parallax for sources of the secondary dataset. 
This strongly suggests issues with parallax and proper motion modeling, reminding that those parameters are poorly fitted from rough priors in the 
secondary dataset. We will thus discard from the rest of the study secondary sources for which $\log\pi$$\sim $$\log\varepsilon_\text{DR1}$$\pm $$0.5$. We think wiser to 
postpone their thorough analysis to the future Gaia DR3 release. Moreover, the largest $\varepsilon_\text{DR1}$ in the secondary dataset are generally obtained for small parallax 
($<$10\,mas). This behavior is different from what is observed in the primary dataset where the astrometric excess noise is not correlated with parallax. 

The final sample contains 597 planet candidates orbiting around 524 host stars with $G$$>$$6.4$, $b-r$$<$$1.8$, and for sources in the secondary dataset
with $\log\pi$-$\log\varepsilon_\text{DR1}$>$0.5$. 

\subsection{Distribution of astrometric excess noise}
\label{sec:noise_distribution}

In order to get a sense of how $\varepsilon_\text{DR1}$ is a relevant quantity to characterize a binary or planetary system, it is crucial to understand how the 
astrometric excess noise generally varies with respect to the known or unknown inclination of the gravitational systems observed -- transiting or not -- the 
presence of a long-period outer companion in the system -- presence of RV drift -- and the quality of their observations with Gaia --  primary or secondary 
dataset. We perform here an analysis of the distribution of astrometric excess noise of our selected sample of companions and sources as defined in 
Section~\ref{sec:mag_selec}, with respect to following subsets selection criteria: 
\begin{itemize}
\item Dataset (primary/secondary);
\item All planets around the host star are transiting;
\item At least one planet is not transiting;
\item Detection or hint of an RV drift;
\item No hint of an RV drift.
\end{itemize}

In principle, with orbital inclination fixed to $\sim$90$^\circ$, the semi-major axis of transiting planets host stars should not reach more than a few $\mu$as, 
and remain undetectable in the DR1 astrometric excess noise. The astrometric scatter is dominated by the instrumental and measurement noises on the 
order of $\sim$0.6\,mas (Lindegren et al. 2016). The distribution of $\varepsilon_\text{DR1}$ for transiting planet hosts should be close to the distribution 
of astrometric scatter due to pure instrumental and measurement noises. On the other hand, system with non-transiting planets allow inclinations down to 
0$^\circ$, and host star semi-major axis beyond a few 0.1\,mas. We expect their astrometric excess noise to be generally larger than for systems with 
transiting-only planets. Finally, the detection of a drift in the RV suggest the presence of a hidden outer companion in the system. The astrometric excess 
noise might be systematically larger for those systems, implying that the astrometric signal is not only due to the companion with a well-defined orbit. 
This is however certainly not a rule, as shown e.g. in the case of HD\,114762 (Kiefer 2019) for which the astrometric excess noise is dominated by the 
effect of the short period companion HD\,114762\,Ab.

\begin{figure}
\includegraphics[width=89.3mm]{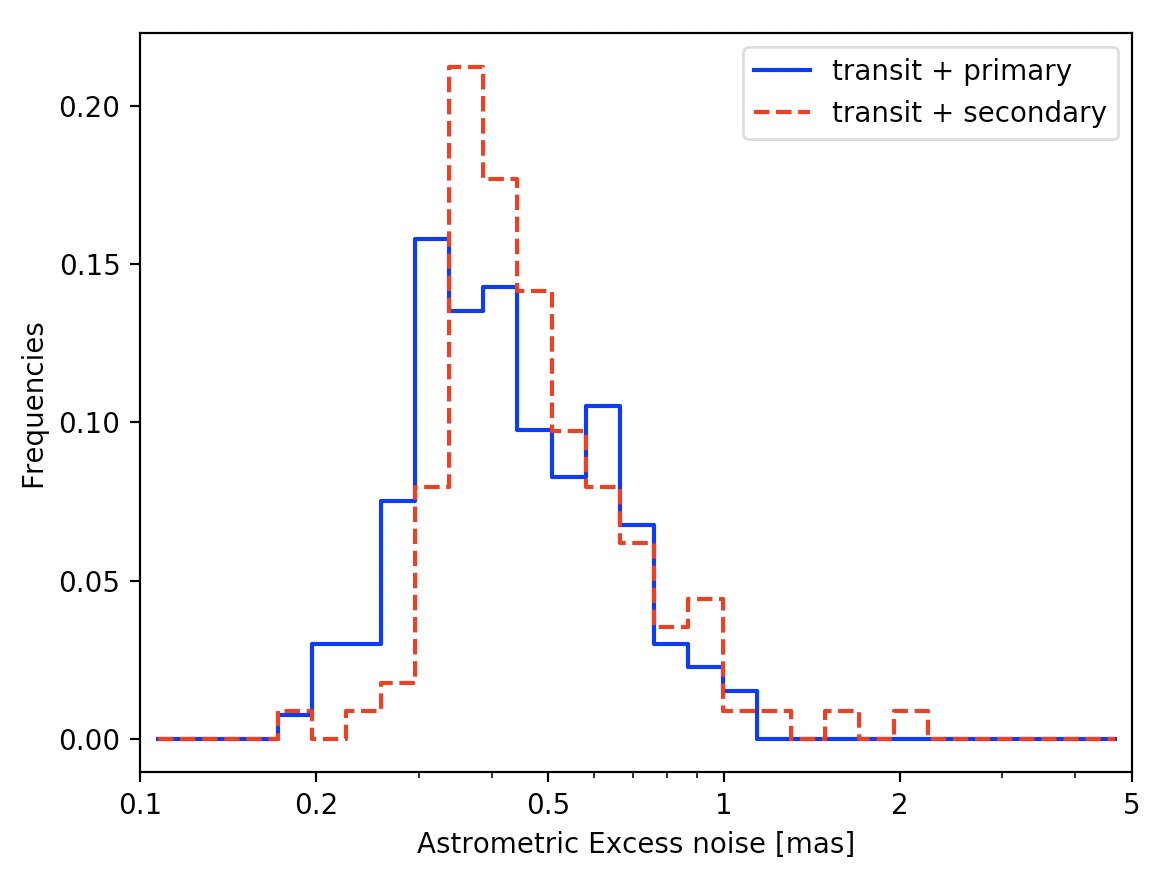}
\includegraphics[width=89.3mm]{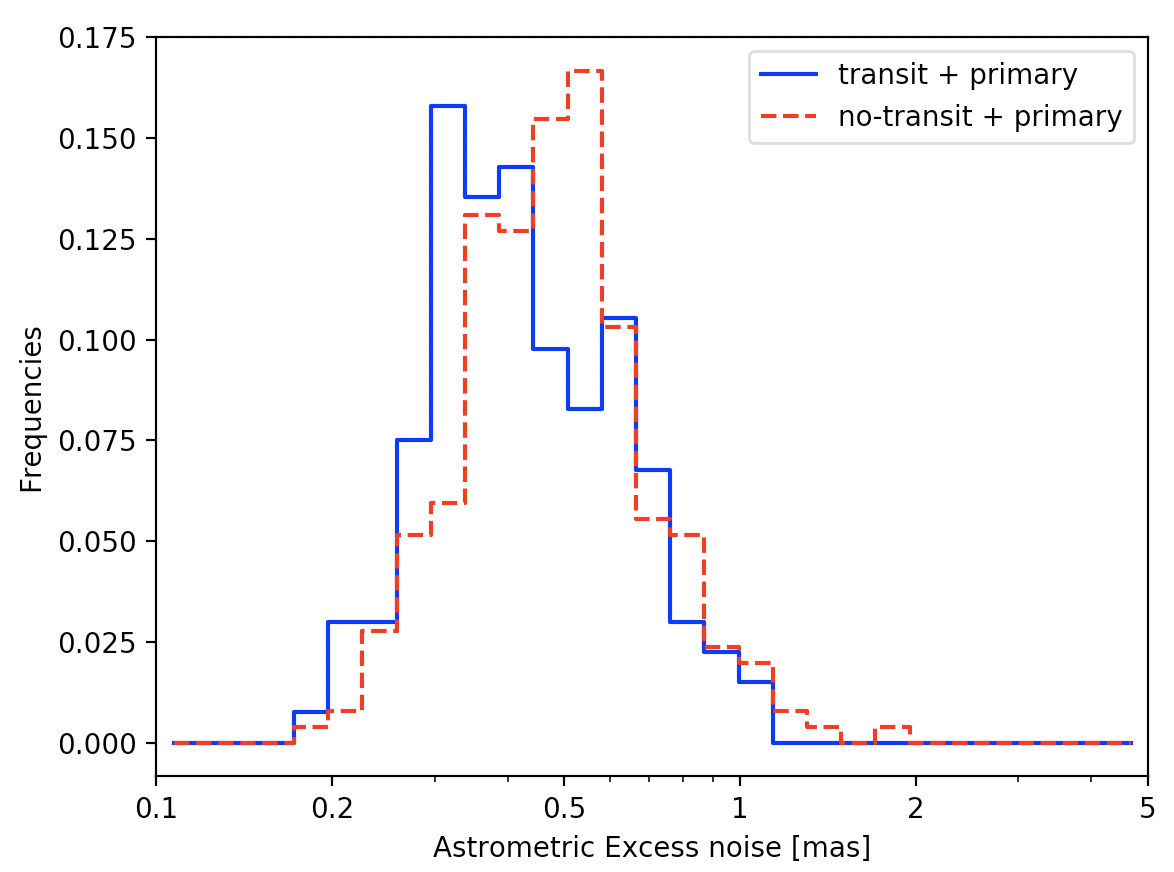}
\caption{\label{fig:compare_epsilon_distribution} Top panel: the astrometric excess noise distribution for sources with only transiting planets, comparing 
primary and secondary dataset. Bottom panel: comparing 'only transiting' and 'non-transiting' subsets within the primary dataset.}
\end{figure}

In Table~\ref{tab:excess_noise_distrib}, we present the 10th, 50th and 90th percentiles of the astrometric excess noise distribution according to the different 
sub-samples defined above. We confront them to the Lindegren et al. (2016) percentiles derived for the whole primary, secondary and Hipparcos 
DR1 datasets (Tables 1 and 2 in Lindegren et al. 2016) based on more than 1 billion sources observed with Gaia. 
In Figure~\ref{fig:compare_epsilon_distribution}, we compare in a first panel the distributions of astrometric excess noise for the sources from the primary and 
secondary datasets with transiting-only planets, and in a second panel, sources from the primary dataset with transiting-only planets to those with at least one 
non-transiting planet. 

The median and 90th percentile value of the $\varepsilon_\text{DR1}$ distribution for all subset in the primary dataset are generally compatible with the Lindegren et al. 
(2016) values. Although Lindegren et al. (2016) study shows that the Hipparcos subset is associated to larger excess noise, Table~\ref{tab:excess_noise_distrib} 
shows that excluding $G$-mag$>$6.4 and $b-r$$>$$1.8$ objects as proposed in Section~\ref{sec:gaia_selec} above leads to decreasing the extent of the astrometric excess noise 
distribution with values agreeing with the primary dataset. The Hipparcos subset excluding bright and late-type sources is thus likely not different from the full 
primary dataset.

We observe a clear distinction in the distributions of $\varepsilon_\text{DR1}$ between the primary and secondary datasets, with significantly higher astrometric excess noise 
in the secondary dataset. This could be well explained by the absence, for the secondary dataset, of the Tycho/Hipparcos supplementary positions 24-years in the 
past that allows deriving robust proper motion and parallax for the sources in the primary dataset. The derivation of proper motion and parallax from Gaia data 
only with Galactic priors based on magnitude (Michalik et al. 2015b, Lindegren et al. 2016) certainly leads to larger scatter in the residuals of the 5-parameters 
solution. 

\begin{table*}
\caption{\label{tab:excess_noise_distrib} The distribution of astrometric excess noise with respect to primary, secondary and Hipparcos datasets in the Gaia DR1 database, 
and varying selection criteria as explained in the text.}
\begin{tabular}{l@{~~}l@{~~}l@{~}cccc}
Reference of data		& DR1 dataset  & \multicolumn{1}{c}{Subset} & $N_\text{star}$ & \multicolumn{3}{c}{Astrometric excess noise}\\ 
					&   			& 		& 			& 10th-percentile 	& Median 				& 90th-percentile  \\ 
					&			&  		&			& (mas) 			& (mas) 				& (mas) \\
\hline
Lindegren et al. (2016) 					& primary 		& 					& 2,057,050		& 0.299	& 0.478 & 0.855 \\
(1,142,719,769 stars)					& secondary 		& 					& 1,140,662,719	& 0.000	& 0.594 & 2.375 \\ 
									& Hipparcos 		&					& 93,635			& 0.347	& 0.572 & 1.185 \\
\hline
This paper sample (Table~\ref{tab:list_targets})  & primary		&	 				& 385	& 0.291	& 0.451 	& 0.751\\ 
(524 stars)							&  			& all transiting planets 	& 133	& 0.271	& 0.399 	& 0.704 \\
									&			& $>$1 non-transiting planet 	& 252	& 0.304	& 0.466 & 0.786 \\ 
									&			& with RV drift 		& 46		& 0.296	& 0.431 	& 0.633 \\
									& 			& no RV drift 			& 339	& 0.291	& 0.453 	& 0.761 \\  \\
									& secondary	&  					& 139	& 0.316	& 0.423 	& 0.776 \\
									&   			& all transiting planets 	& 113	& 0.336 	& 0.438	& 0.791 \\
									&                   	& $>$1  non-transiting planet & 26 & 0.283	& 0.360 	& 0.701 \\ 
									&      		& with RV drift 		& 7		& 0.297	& 0.334 	& 0.511 \\
									&   			& no RV drift 			& 132	& 0.318	& 0.425 	& 0.794 \\  \\
									& Hipparcos 	&					& 246	& 0.307	& 0.466 	& 0.784 \\ 
									&                  	& including G$<$6.4 \& b-r$>$1.8	& 297	& 0.324 	& 0.513 	& 1.048 \\ 
\hline 
\end{tabular}
\end{table*}

For transiting sources of the primary dataset, the 90th percentile of the astrometric excess noise distribution is 0.70\,mas. This is compatible with, and even 
lower than, Lindegren et al. (2016) values of the global DR1 solution. For this subset, the 95-th percentile is 0.81\,mas, still lower than the 90th-percentile of 
Lindegren et al. (2016). This generally small astrometric excess noises of the sources with transiting planets is compatible with statistical noise and the 
non-detection by Gaia of any orbital motion of a star orbited by a planet at short separation ($<$0.1\,a.u. or $P$$<$$50$\,days) and with an edge-on inclination 
of its orbit. 

The systems in the primary dataset with a non-transiting planet have the highest median among all other subsets (0.47\,mas) and the highest 90th percentile 
(0.78\,mas). More importantly, the astrometric excess noise of sources with non-transiting companions is significantly larger than for sources with transiting-only 
planets. This can also be seen in the lower panel of Figure~\ref{fig:compare_epsilon_distribution} with a net shift between the two $\varepsilon_\text{DR1}$ 
distributions. This confirms that $\varepsilon_\text{DR1}$ contains a non-negligible fraction of astrometric motion for systems with a companion which orbital 
inclination is not known.

The $\varepsilon_\text{DR1}$ in the secondary dataset generally reaches larger values than in the primary dataset, with a 90th percentile for the subset of systems
with transiting-only planets $\sim$0.8\,mas. This was expected by the less accurate fit of the proper motion and parallax in the secondary dataset compared
to the primary. However, this is also much smaller than the 2.3\,mas 90th-percentile derived for the whole secondary dataset in Lindegren et  al. (2016). Therefore
once cleaned from problematic systems, in particular those with $\log\pi/\varepsilon_\text{DR1}$$>$$0.5$ (Section~\ref{sec:mag_selec}), the astrometric 
excess noise of remaining objects in the secondary dataset seems robust, with parallax and proper motion most likely well determined (although not published 
in the DR1). 

Interestingly, we find no correlation of the astrometric excess noise distribution with the presence of any drift in the RV data, and even smaller values than in the 
other subsets. This could be due to the smaller number of sources in this category, which if following a inclination probability density function $\sim\sin I_c$ would 
preferentially have inclinations close to 90$^\circ$, and thus smaller astometric motion. It also suggests that the presence of an outer companion does not 
have a strong effect on the astrometric excess noise compared to the enhanced astrometric motion due to a small inclination of a non-transiting companion.

\subsection{Testing the noise model}
\label{sec:noise_scheme}

From the 133 and 113 stars with transiting-only planets from the primary and secondary datasets we can test the model of noise used in the simulations of 
GASTON. In previous studies (Kiefer et al. 2019, Kiefer 2019), we chose to use values based on published estimations of the measurement uncertainty and of the 
typical external noise (including modeling noise and instrument jitter), respectively $\sigma_\text{AL}$=0.4\,mas (Michalik et al. 2015a) and $\sigma_\text{syst}$=0.5\,mas (Lindegren 
et al. 2016). As we showed in the preceding section, the sources with transiting companion must be generally more similar to sources with no astrometric motion. 
Therefore, the astrometric excess noise measured by Gaia for these sources should be close to purely instrumental and photonic stochastic scatter. 

\begin{figure}
\includegraphics[width=88mm]{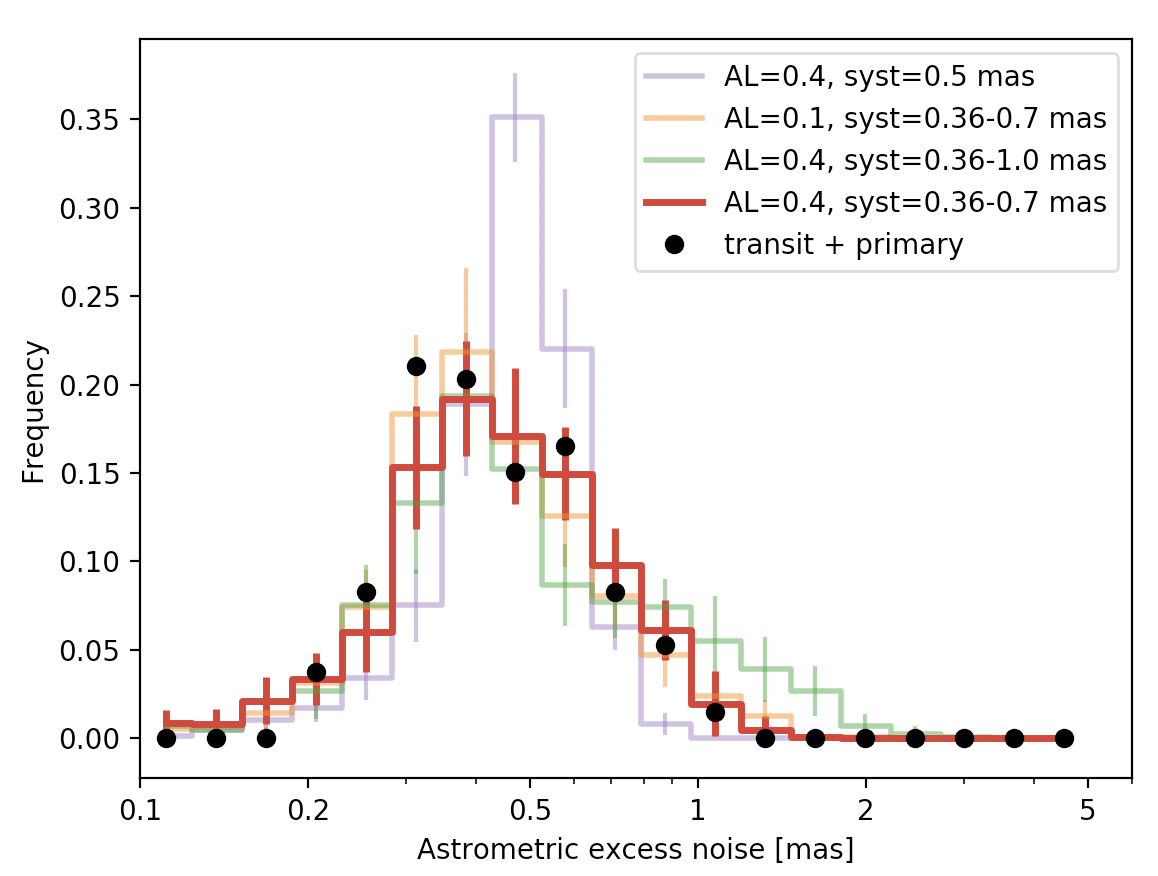}
\includegraphics[width=89.3mm]{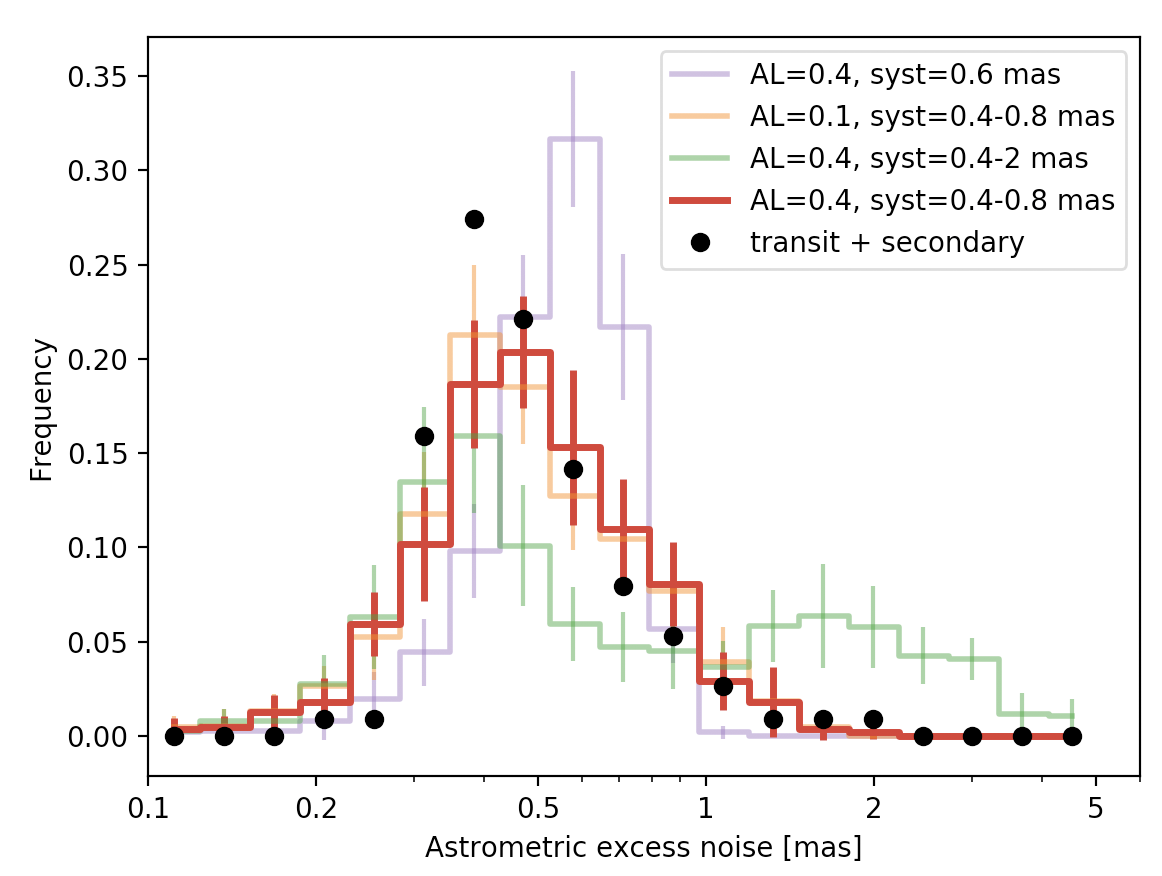}
\caption{\label{fig:noise_model}Comparing measurement and systematical noise models to the distribution of astrometric excess noise measurement for sources with 
transiting companion in the primary dataset (top-panel) and in the secondary dataset (bottom-panel). See text for explanations.}
\end{figure}

The distribution of $\varepsilon_\text{DR1}$ for these 300 sources from primary and secondary datasets is plotted in Fig.~\ref{fig:noise_model}. It is compared to 
simulations of astrometric excess noise of sources with no orbital motion, in the framework of different noise models. We assumed for each simulation random 
numbers of FoV transits ($N_\text{FoV}$) and numbers of measurements per FoV transit ($N_\text{AL}$) in the same ranges as those of the sample presented here, 
e.g. $N_\text{FoV}$=15$\pm$8 and $N_\text{AL}$=7$\pm$2 with Gaussian distribution, and imposing that $48$$>$$N_\text{FoV}$$>$$5$ and 
$9$$>$$N_\text{AL}$$>$$2$. We singled out 5 different noise models of ($\sigma_\text{syst}$,$\sigma_\text{AL}$), either based on literature values, or based 
on the best fit of a bi-uniform distribution of $\sigma_\text{syst}$ with a fixed median to the $\varepsilon_\text{DR1}$ cumulative density function (cdf): 

\begin{itemize}
\item The constant model for the primary dataset, as used in Kiefer et al. 2019 \& Kiefer 2019: $\sigma_\text{AL}$=0.4\,mas from Michalik et al. (2015a), 
and $\sigma_\text{syst}$=0.5\,mas (based on Lindegren et al. 2016),
\item A different constant model for the secondary dataset: $\sigma_\text{AL}$=0.4\,mas as above, and $\sigma_\text{syst}$=0.6\,mas (based on Lindegren et al. 2016),
\item Random $\sigma_\text{syst}$ from a distribution with a median at 0.4\,mas, uniform from 0.36 to 0.40\,mas and from 0.4 to 0.7\,mas for the primary 
dataset, 
\item Random $\sigma_\text{syst}$ from a distribution with a median at 0.45\,mas, uniform from 0.4 to 0.45\,mas and from 0.45 to 0.8\,mas for the secondary 
dataset,
\item Smaller AL angle measurement uncertainty as suggested from Lindegren et al. (2018): $\sigma_\text{AL}$=0.1\,mas.
\end{itemize}

The bi-uniform distributions models were found to lead to the best least square fit of the observed $\varepsilon_\text{DR1}$ cdf. All 5 models are compared to the data in Figure~\ref{fig:noise_model}. A model with a wide range of systematic noise better explain the observed 
distributions for values of astrometric excess noise assumed to be compatible with pure stochastic and systematic noise, e.g. below 0.85\,mas and 2.3\,mas for sources in the primary and secondary 
datasets respectively. The constant noise model tends to overestimate the astrometric excess noise, and the simulated distribution decreases too steeply 
at values closer to 1\,mas. The noise $\sigma_\text{syst}$ taken from a bi-uniform distribution of values within bounds (e.g. 0.36-0.7\,mas in the primary dataset) 
explains better the full distribution of observed astrometric excess noises. The distribution is damped beyond about 0.85-1\,mas for the primary dataset and 
$\sim$1.2-1.4\,mas in the secondary dataset, with less than 1\% of the simulations beyond. 

We can exclude that the $\sigma_\text{syst}$ is much larger than 0.7\,mas in the primary dataset, and 0.8\,mas in the secondary dataset, since that would 
extend the core of the distribution towards larger excess noise, therefore leading to a poorer agreement. Finally, we observe that the AL angle measurement uncertainty 
$\sigma_\text{AL}$ does not have a strong impact on the astrometric excess noise. We tested two values, a small uncertainty 0.1\,mas, as given in Lindegren et al. 
(2018), and a more conservative value as assumed by Michalik et al. (2015) of 0.4\,mas that corresponds well to the typical AL residuals reported in Lindegren et 
al. (2016) of 0.65\,mas ($\sqrt{(\sigma_\text{AL}=0.4)^2+(\sigma_\text{syst}=0.5)^2}$=0.64\,mas). We thus fix $\sigma_\text{AL}$ to 0.4\,mas. 

In conclusion, the systematic noise is typically about 0.5 and 0.6\,mas respectively in the primary and secondary dataset. But it is likely that 
for an individual observed source, $\sigma_\text{syst}$ can be somewhat larger or smaller by a few fraction of mas. We thus adopt a random systematic noise for 
each simulation in GASTON uniformly distributed on both sides of the median at 0.4\,mas down to 0.36\,mas and up to 0.7\,mas for sources in the primary dataset, and 
about the median at 0.45\,mas down to 0.4\,mas and up to 0.8\,mas for sources in the secondary dataset. 

\subsection{Detection threshold and source selection}
\label{sec:threshold}

GASTON cannot be used to characterize the mass of as much as 597 planet candidates. This would require several weeks of calculation, while  
many of them cannot be truly characterized, because the astrometric excess noise is compatible with an edge-on inclination. We thus need to define a 
robust threshold above which $\varepsilon_\text{DR1}$ can be considered significantly non-stochastic -- and thus astrophysical -- and below which it 
could be explained by pure stochastic noise. 

Given that close to 40-50\% of the sources in the Milky Way are part of binary systems (Duquennoy \& Mayor 1991, Raghavan et al. 2010), a consequent fraction 
of Gaia sources -- possibly more than 10\% -- could show a detectable astrometric motion. We thus consider using the 90th-percentiles of Lindegren et al. (2016) 
based on a sample of more than 1 billion stars, as a detection threshold, above which a significant fraction of $\varepsilon_\text{DR1}$ could be imputed to 
astrometric motion. This was assumed in previous work (Kiefer 2019), but the astrometric excess noise distribution of the present sample might differ 
from the sample upon which these percentiles were calculated.

In Sections~\ref{sec:noise_distribution} and~\ref{sec:noise_scheme}, we showed that the 90th-percentile of 0.85\,mas in the primary sample, as derived in Lindegren et al. (2016) is robust, but 
the 2.3\,mas threshold for secondary dataset sources is excessive and could be more reasonably lowered to $\sim$1.2\,mas. This overestimation of the 
90th-percentile in Lindegren et al. (2016) must be due to the inclusion of small magnitude, large $b-r$ color sources and badly modelled parallax, which we did 
clean out from our sample. We will thus use the detections thresholds $\epsilon_{thresh,prim}$=0.85\,mas, and $\epsilon_{thresh,sec}$=1.2\,mas, above which 
the astrometric excess noise would be mainly due to supplementary astrometric motion. This reduces our sample to 28 sources (29 planet candidates) with an 
astrometric detection by Gaia in the DR1. They are the best candidates for orbit inclination and true mass measurement of the companion. They constitutes the 
'detection sample'.

We counted 312 non-transiting planet candidates (around 254 sources) for which the inclination is not constrained from photometry and for which the 
astrometric motion of their host star leads to an astrometric excess noise smaller than the threshold. Even though compatible with pure noise, the astrometric 
excess noise allow constraining the true astrometric extent of the star's orbit. This leads to deriving a minimum inclination and a maximum true mass of the 
exoplanet candidates beyond which they are not compatible with a non-detection. In this sample, we exclude the so-called 'duplicate sources' in the Gaia DR1, because, for such target, possibly large sets of 
astrometric measurements are attributed to another source with another ID, and thus lead to underestimate its AL-angle residuals and its astrometric excess noise. 
This will be of crucial importance if the mass of a companion is found to be smaller than 13.5\,M$_\text{J}$. A larger astrometric excess noise leads to 
a larger mass range. We thus focus on the 227 non-duplicate companions orbiting 187 sources which will constitute the 'non-detection sample'.

The complete list of 29 detected exoplanet candidates is presented in Table~\ref{tab:list_planets}, and the list of 312 non-detected non-transiting planet 
candidates is given in Appendix~\ref{app:non-detection}. Table~\ref{tab:selec_sum} summarizes all source selection steps applied from 
Section~\ref{sec:initial_selection} up to the present section. 

\onecolumn
\thispagestyle{empty}
\newpage
\newgeometry{margin=0.5cm,top=1cm}
\begin{sidewaystable*}\tiny
\caption{\label{tab:list_planets} List of the 29 selected planets which astrometric excess noise overpass the detection threshold (see text). Where uncertainties are 
missing we will assume 10\% errors on the corresponding parameter. The parallax with uncertainties are all taken from the DR1, while those without uncertainties 
are taken from SIMBAD. We expand this table on-line to also include the 227 exoplanet candidates for which the astrometric excess noise is 
smaller than the threshold (see Table~\ref{tab:selec_sum}). } 
\begin{tabular}{@{}l|@{~~}c@{~~}c@{~~}c@{~~}c@{~~}c@{~~}c@{~~}c@{~~}c@{~~}c@{~~}c@{~~}c@{~~}c@{~~}|c@{~~}c@{~~}c@{~~}c@{~~}c@{}}  
\multicolumn{13}{c|}{RV data} & \multicolumn{5}{c}{Gaia DR1 data} \\
\hline
Name & $P$          	& $m\sin i$  		& $K$ 			& $e$ 	& $\omega$ 	& $T_p$ 	& $M_\star$ 	& $a_p$ 	& $\pi$    	& $a_\star\sin i$  & Transit  & Drift  	& $\varepsilon_\text{DR1}$ 	& $N_\text{pts}$ 	& $N_\text{FoV}$ 	& Gaia   & Duplicate \\
	  & (days) 		& (M$_\text{J}$) 	&  (m\,s$^{-1}$) 	& 		& ($^\circ$) 	& -2,450,000	& ($M_\odot$)	&   (au)      & (mas)	&   (mas) 			& flag & flag &  (mas)          	&                      		&				&  dataset & source \\
\hline
30 Ari B b	& 335.1$\pm$2.5 	& 9.88$\pm$0.98 	& 272$\pm$24 	& 0.289$\pm$0.092 	& 307$\pm$18 	& 4538$\pm$20 	& 1.160$\pm$0.040 	& 0.992$\pm$0.012 	& 21.42$\pm$0.60 	& 0.173$\pm$0.019 	& n 	& n 	& 1.8 	& 71 	& 10 	& 1 	& n\\
HAT-P-21 b	& 4.1244810$\pm$0.0000070 	& 4.07$\pm$0.17 	& 548$\pm$14 	& 0.228$\pm$0.016 	& 309.0$\pm$3.0 	& 4995.014$\pm$0.046 	& 0.947$\pm$0.042 	& 0.04943$\pm$0.00073 	& 3.73$\pm$0.51 	& 0.00076$\pm$0.00011 	& n 	& n 	& 0.92 	& 84 	& 10 	& 1 	& n\\
HD 114762 b	& 83.9151$\pm$0.0030 	& 11.64$\pm$0.78 	& 612.5$\pm$3.5 	& 0.3354$\pm$0.0048 	& 201.3$\pm$1.0 	& -110.89$\pm$0.19 	& 0.895$\pm$0.089 	& 0.361$\pm$0.012 	& 25.88$\pm$0.46 	& 0.116$\pm$0.015 	& n 	& n 	& 1.1 	& 130 	& 23 	& 1 	& n\\
HD 132563 B b	& 1544$\pm$34 	& 1.49$\pm$0.14 	& 26.7$\pm$2.2 	& 0.220$\pm$0.090 	& 158$\pm$35 	& 2593$\pm$148 	& 1.010$\pm$0.010 	& 2.623$\pm$0.039 	& 9.30$\pm$0.33 	& 0.0344$\pm$0.0034 	& n 	& n 	& 0.85 	& 130 	& 19 	& 1 	& y\\
HD 141937 b	& 653.2$\pm$1.2 	& 9.48$\pm$0.41 	& 234.5$\pm$6.4 	& 0.410$\pm$0.010 	& 187.72$\pm$0.80 	& 1847.4$\pm$2.0 	& 1.048$\pm$0.037 	& 1.497$\pm$0.018 	& 30.62$\pm$0.35 	& 0.396$\pm$0.023 	& n 	& n 	& 0.93 	& 236 	& 32 	& 1 	& y\\
HD 148427 b	& 331.5$\pm$3.0 	& 1.144$\pm$0.092 	& 27.7$\pm$2.0 	& 0.160$\pm$0.080 	& 277$\pm$68 	& 3991$\pm$15 	& 1.360$\pm$0.060 	& 1.039$\pm$0.017 	& 14.17$\pm$0.42 	& 0.0118$\pm$0.0012 	& n 	& n 	& 1.1 	& 131 	& 17 	& 1 	& n\\
HD 154857 b	& 408.60$\pm$0.50 	& 2.248$\pm$0.092 	& 48.3$\pm$1.0 	& 0.460$\pm$0.020 	& 57.0$\pm$4.0 	& 3572.5$\pm$2.4 	& 1.718$\pm$0.026 	& 1.2907$\pm$0.0066 	& 15.56$\pm$0.39 	& 0.0251$\pm$0.0013 	& n 	& n 	& 0.93 	& 122 	& 19 	& 1 	& n\\
HD 154857 c	& 3452$\pm$105 	& 2.58$\pm$0.15 	& 24.2$\pm$1.1 	& 0.060$\pm$0.050 	& 352$\pm$37 	& 5219$\pm$375 	& 1.718$\pm$0.026 	& 5.35$\pm$0.11 	& 15.56$\pm$0.39 	& 0.1194$\pm$0.0081 	& n 	& n 	& 0.93 	& 122 	& 19 	& 1 	& n\\
HD 164595 b	& 40.00$\pm$0.24 	& 0.0508$\pm$0.0070 	& 3.05$\pm$0.41 	& 0.088$\pm$0.093 	& 145$\pm$135 	& 6280$\pm$12 	& 0.990$\pm$0.030 	& 0.2281$\pm$0.0025 	& 35.11$\pm$0.38 	& 0.000392$\pm$0.000056 	& n 	& n 	& 0.93 	& 62 	& 8 	& 1 	& y\\
HD 16760 b	& 465.1$\pm$2.3 	& 13.29$\pm$0.61 	& 408.0$\pm$7.0 	& 0.067$\pm$0.010 	& 232$\pm$10 	& 4723$\pm$12 	& 0.780$\pm$0.050 	& 1.081$\pm$0.023 	& 14 	& 0.25 	& n 	& n 	& 3.0 	& 62 	& 10 	& 2 	& n\\
HD 177830 b	& 410.1$\pm$2.2 	& 1.320$\pm$0.085 	& 32.64$\pm$0.98 	& 0.096$\pm$0.048 	& 189 	& 254$\pm$49 	& 1.17$\pm$0.10 	& 1.138$\pm$0.033 	& 15.94$\pm$0.37 	& 0.0195$\pm$0.0022 	& n 	& n 	& 0.87 	& 89 	& 19 	& 1 	& n\\
HD 185269 b	& 6.8379$\pm$0.0010 	& 0.954$\pm$0.069 	& 90.7$\pm$4.4 	& 0.296$\pm$0.040 	& 172$\pm$11 	& 3154.089$\pm$0.040 	& 1.28$\pm$0.10 	& 0.0766$\pm$0.0020 	& 19.10$\pm$0.41 	& 0.00104$\pm$0.00012 	& n 	& n 	& 0.97 	& 74 	& 16 	& 1 	& n\\
HD 190228 b	& 1136.1$\pm$9.9 	& 5.94$\pm$0.30 	& 91.4$\pm$3.0 	& 0.531$\pm$0.028 	& 101.2$\pm$2.1 	& 3522$\pm$12 	& 1.821$\pm$0.046 	& 2.602$\pm$0.027 	& 15.77$\pm$0.34 	& 0.1278$\pm$0.0079 	& n 	& n 	& 0.86 	& 116 	& 16 	& 1 	& y\\
HD 197037 b	& 1036$\pm$13 	& 0.807$\pm$0.060 	& 15.5$\pm$1.0 	& 0.220$\pm$0.070 	& 298$\pm$26 	& 1353$\pm$86 	& 1.1 	& 2.1 	& 30.01$\pm$0.32 	& 0.043 	& n 	& n 	& 0.99 	& 107 	& 17 	& 1 	& n\\
HD 4203 b	& 431.88$\pm$0.85 	& 2.08$\pm$0.12 	& 60.3$\pm$2.2 	& 0.519$\pm$0.027 	& 329.1$\pm$3.0 	& 1918.9$\pm$2.7 	& 1.130$\pm$0.064 	& 1.165$\pm$0.022 	& 12.67$\pm$0.44 	& 0.0259$\pm$0.0023 	& n 	& n 	& 0.85 	& 60 	& 11 	& 1 	& n\\
HD 5388 b	& 777.0$\pm$4.0 	& 1.97$\pm$0.10 	& 41.7$\pm$1.6 	& 0.400$\pm$0.020 	& 324.0$\pm$4.0 	& 4570.0$\pm$9.0 	& 1.2 	& 1.8 	& 18.86$\pm$0.32 	& 0.052 	& n 	& n 	& 1.4 	& 327 	& 48 	& 1 	& y\\
HD 6718 b	& 2496$\pm$176 	& 1.56$\pm$0.12 	& 24.1$\pm$1.5 	& 0.100$\pm$0.075 	& 286$\pm$50 	& 4357$\pm$251 	& 0.96 	& 3.6 	& 19.74$\pm$0.41 	& 0.11 	& n 	& n 	& 1.1 	& 199 	& 29 	& 1 	& y\\
HD 7449 b	& 1275$\pm$13 	& 1.31$\pm$0.52 	& 42$\pm$15 	& 0.820$\pm$0.060 	& 339.0$\pm$6.0 	& 5298$\pm$26 	& 1.1 	& 2.3 	& 27.14$\pm$0.41 	& 0.076 	& n 	& n 	& 0.94 	& 61 	& 12 	& 1 	& y\\
HD 95127 b	& 482.0$\pm$5.0 	& 5.04$\pm$0.82 	& 116$\pm$12 	& 0.11$\pm$0.10 	& 40$\pm$38 	& 3200$\pm$50 	& 1.20$\pm$0.22 	& 1.278$\pm$0.079 	& 1.31$\pm$0.58 	& 0.0067$\pm$0.0034 	& n 	& n 	& 1.2 	& 41 	& 8 	& 1 	& n\\
HD 96127 b	& 647$\pm$17 	& 4.01$\pm$0.85 	& 105$\pm$11 	& 0.30$\pm$0.10 	& 162$\pm$18 	& 3969$\pm$31 	& 0.91$\pm$0.25 	& 1.42$\pm$0.13 	& 1.87$\pm$0.83 	& 0.0112$\pm$0.0064 	& n 	& n 	& 1.1 	& 109 	& 16 	& 1 	& n\\
HIP 65891 b	& 1084$\pm$23 	& 6.00$\pm$0.41 	& 64.9$\pm$2.4 	& 0.130$\pm$0.050 	& 356$\pm$16 	& 6015$\pm$49 	& 2.50$\pm$0.21 	& 2.804$\pm$0.088 	& 6.53$\pm$0.37 	& 0.0420$\pm$0.0053 	& n 	& n 	& 1.1 	& 237 	& 35 	& 1 	& n\\
K2-110 b	& 13.86375$\pm$0.00026 	& 0.053$\pm$0.011 	& 5.5$\pm$1.1 	& 0.079$\pm$0.070 	& 90$\pm$122 	& 6863 	& 0.738$\pm$0.018 	& 0.10207$\pm$0.00083 	& 8.6 	& 0.000060 	& n 	& n 	& 1.3 	& 54 	& 9 	& 2 	& y\\
K2-34 b	& 2.9956290$\pm$0.0000060 	& 1.683$\pm$0.061 	& 207.0$\pm$3.0 	& 0.000$\pm$0.027 	& 90 	& 7144.347030$\pm$0.000080 	& 1.226$\pm$0.052 	& 0.04353$\pm$0.00062 	& 3.28$\pm$0.68 	& 0.000187$\pm$0.000040 	& n 	& n 	& 1.00 	& 117 	& 16 	& 1 	& y\\
WASP-11 b	& 3.7224650$\pm$0.0000070 	& 0.540$\pm$0.052 	& 82.1$\pm$7.4 	& 0 	& 90 	& 4473.05588$\pm$0.00020 	& 0.800$\pm$0.025 	& 0.04364$\pm$0.00045 	& 7.49$\pm$0.58 	& 0.000211$\pm$0.000027 	& n 	& n 	& 1.1 	& 60 	& 10 	& 1 	& n\\
WASP-131 b	& 5.3220230$\pm$0.0000050 	& 0.272$\pm$0.018 	& 30.5$\pm$1.7 	& 0 	& 90 	& 6919.82360$\pm$0.00040 	& 1.060$\pm$0.060 	& 0.0608$\pm$0.0011 	& 4.55$\pm$0.56 	& 0.000068$\pm$0.000010 	& n 	& n 	& 0.95 	& 53 	& 7 	& 1 	& n\\
WASP-156 b	& 3.8361690$\pm$0.0000030 	& 0.1305$\pm$0.0087 	& 19.0$\pm$1.0 	& 0.0000$\pm$0.0035 	& 90 	& 4677.7070$\pm$0.0020 	& 0.842$\pm$0.052 	& 0.04529$\pm$0.00093 	& 8.2 	& 0.000055 	& n 	& n 	& 1.5 	& 38 	& 7 	& 2 	& n\\
WASP-157 b	& 3.9516205$\pm$0.0000040 	& 0.559$\pm$0.049 	& 61.6$\pm$3.8 	& 0.000$\pm$0.055 	& 90 	& 7257.803194$\pm$0.000088 	& 1.26$\pm$0.12 	& 0.0528$\pm$0.0017 	& 3.76$\pm$0.66 	& 0.000084$\pm$0.000019 	& n 	& n 	& 0.87 	& 69 	& 9 	& 1 	& n\\
WASP-17 b	& 3.7354330$\pm$0.0000076 	& 0.508$\pm$0.030 	& 59.2$\pm$2.9 	& 0 	& 90 	& 4559.18096$\pm$0.00023 	& 1.190$\pm$0.030 	& 0.04993$\pm$0.00042 	& 2.57$\pm$0.31 	& 0.0000523$\pm$0.0000072 	& n 	& n 	& 0.85 	& 202 	& 25 	& 1 	& n\\
WASP-43 b	& 0.8134750$\pm$0.0000010 	& 1.76$\pm$0.10 	& 550.3$\pm$6.7 	& 0 	& 90 	& 5528.86774$\pm$0.00014 	& 0.580$\pm$0.050 	& 0.01422$\pm$0.00041 	& 11 	& 0.00047 	& n 	& n 	& 2.0 	& 63 	& 7 	& 2 	& n\\
\hline
\end{tabular}
\end{sidewaystable*}
\newgeometry{margin=1.3cm,top=2cm,bottom=2.5cm}
\twocolumn

\begin{table*}\centering
\caption{\label{tab:selec_sum} Number of sources and planets in our sample after the several selection criteria introduced in Sections~\ref{sec:initial_selection}, 
\ref{sec:mag_selec} and \ref{sec:threshold}.}
\begin{tabular}{lccc}
Criterion & Article section & \# stellar hosts & \# planets \\
\hline
\verb+exoplanets.org+ & \ref{sec:initial_selection} & 2466  & 3262 \\
RV planets & \ref{sec:initial_selection} & 782 & 911  \\
sources in the DR1 & \ref{sec:gaia_selec} & 658 & 755 \\
$G$$>$$6.4$ & \ref{sec:mag_selec} & 614 & 705 \\
$b-r$$<$$1.8$ &  \ref{sec:mag_selec} & 580 & 654  \\
\hline
\multicolumn{4}{c}{Separation into primary/secondary Gaia DR1 datasets} \\
\\
primary & \ref{sec:noise_distribution}  & 385 & 442  \\
secondary ; $\log(\pi/\varepsilon_\text{DR1})$$>$$0.5$ & \ref{sec:mag_selec} ; \ref{sec:noise_distribution}   & 139 & 154\\
\hline
\multicolumn{4}{c}{$\varepsilon_\text{DR1}>$ threshold: the detection sample} \\
\\
primary ; $\varepsilon_\text{DR1}$$>$$0.85$\,mas & \ref{sec:threshold} & 24 & 25 \\
secondary ; $\varepsilon_\text{DR1}$$>$$1.2$\,mas & \ref{sec:threshold}  & 4 & 4 \\
\hline
\multicolumn{4}{c}{$\varepsilon_\text{DR1}<$ threshold and non-transiting: the non-detection sample} \\
\\
primary ; $\varepsilon_\text{DR1}$$<$$0.85$\,mas ; no transit ; non duplicate & \ref{sec:threshold}  & 165 & 201 \\
secondary ; $\varepsilon_\text{DR1}$$<$$1.2$\,mas ; no transit ; non duplicate & \ref{sec:threshold} & 18  & 26 \\
\hline
\end{tabular}
\end{table*}

\section{GASTON simulations and new improvements}
\label{sec:gaston}

\subsection{General principle}
In the present study, our goal is to constrain the inclination and true mass of RV planet candidates using the released Gaia astrometric data. To do so, we are applying 
the GASTON method described in Kiefer et al. 2019 \& Kiefer 2019. This algorithm simulates the residuals of Gaia's 5-parameters fit of a source accounting for a supplementary 
astrometric motion due to a perturbing RV-detected companion. It leads to simulated astrometric excess noise  $\varepsilon_\text{simu}$
depending on the actual inclination of the RV-detected orbital motion. It also accounts for measurement noise and modeling errors in the reduction of 
the DR1 through the noise model adopted in Section~\ref{sec:noise_scheme}. These simulations are then compared to the astrometric excess noise actually 
measured by Gaia and reported in the DR1 database (Table~\ref{tab:list_targets}) to derive a matching orbital inclination. 

The GASTON algorithm is embedded into an MCMC process, with \verb+emcee+ (Foreman-Mackey et al. 2013), that allows deriving the posterior distributions of orbital 
inclination and true mass of the RV companion among other parameters. To sum up, the varied physical parameters in the MCMC run are the orbital period $P$, the eccentricity $e$, the longitude of periastron $\omega$, the 
periastron time of passage $T_p$, the inclination $I_c$, the minimum mass $m\sin i$, the star mass $M_\star$, the parallax $\pi$, an hyper-parameter  
$f_\varepsilon$ to scale error bars on $\varepsilon_\text{DR1}^2$, and a jitter term $\sigma_\text{K, jitter}$. Some of these parameters have strong gaussian priors from RV 
($P$, $e$, $T_p$, $\omega$, $m\sin i$), or from other analysis ($M_\star$, $\pi$). The hyper-parameter $f_\varepsilon$ follows a Gaussian prior about 0 with 
a standard deviation of 0.1. The jitter term follows a flat prior between 0 and $\sqrt 3$ assuming that the published uncertainty on $K$ could be 
underestimated by as much as a factor $\sqrt{1+\sigma_\text{K, jitter}^2}$=2. Generally, we adopt a $\text{d}p(I_c)$=$\sin I_c \text{d}I_c$
prior probability distribution for the inclination, assuming the inclination of orbits among RV-candidates is isotropic. If the MCMC converges to an inclination 
strongly different from 90$^\circ$ despite the low prior probability, that implies the data  inputs have a significant weight in the likelihood.

\subsection{Dealing with proper motion and parallax in the simulations}
\label{sec:PM_parallax}

For sources in the primary dataset, we assume that the proper motion fit as performed by Gaia in the DR1 is disentangled from 
the hidden astrometric orbit. We thus assume that the astrometric excess noise is purely composed of noise and orbital motion, and that it is not needed to 
fit out excess parallax and proper motion to the simulated astrometric orbit. This is justified by the addition of past Hipparcos or Tycho-2 positions
in the Gaia's reduction for fitting proper motion of primary dataset sources, thus based on astrometric measurements spanning more than 24 years. 
Given that the orbital periods
of all studied companions are smaller than 14 years, the fit of proper motion to the simulated orbits reduces the amplitude of the simulated residuals -- and 
thus of the astrometric excess noise -- only by a small amount. Numerical simulations show that in the worst case scenario with a 
Tycho-2 position uncertainty of $\sim$100\,mas, the average simulated astrometric excess noise $\varepsilon_\text{simu}$ is lowered by less than 0.2\,mas. 
This offset reduces to less than 0.05\,mas if a 
Hipparcos position ($\sigma_\text{RA,DEC}$$\sim $$1$\,mas) is used instead or if $P$$<$$10$\,days. Hipparcos positions are available for 171 over the 190 
primary sources in our sample, while only 6 sources have a Tycho-2 position with more than 20\,mas of uncertainty and a companion with $P$$>$$10$\,days. 
These 6 sources, HD\,95872, NGC\,2423\,3, BD+20\,2457, HD\,233604, BD+15\,2375, and M67\,SAND\,364, all belongs to the non-detection sample. For those, 
the astrometric excess noise that we simulate with GASTON for a given companion orbit and at a given orbital inclination could be overestimated by up to 
$\sim$0.2 mas. Thus GASTON possibly underestimates the upper-limit on the companion true mass for those stars.

For sources in the secondary dataset, the proper motion given in the DR1 is derived from the Gaia data only, without a supplementary data point from 
Tycho-2 or Hipparcos. An important part of the orbital motion could thus be mistaken for proper motion during the Gaia data reduction of the DR1, especially for 
orbital periods at which the Gaia measurements along the 416-days time baseline of the DR1 campaign could appear almost linear. For sources from the secondary 
dataset, we thus perform a fit of linear motion to the simulated astrometric orbit, from which residuals we derive $\varepsilon_\text{simu}$. 

For sources of both datasets, fitting the parallax to the astrometric orbit does not have a significant effect on $\varepsilon_\text{simu}$ even if $P$$\sim $$365$\,days 
and if the orbital and parallax motions are aligned. Numerical tests of parallax fit to simulated data along an astrometric orbit with 
$P$$\sim $$365$\,days and randomizing along the unknown longitude of ascending node, $\Omega$ from 0 to 2$\pi$, leads to a typical reduction of the 
average $\varepsilon_\text{simu}$ smaller than 0.05\,mas. Fitting parallax to the simulated astrometric orbit is thus unnecessary, leading to negligible deviations
on the simulated astrometric excess noise.

\subsection{Recent improvements}
Since Kiefer (2019), we have done few improvements, with the list below: 

\begin{itemize}
\item The number of walkers was reduced from 200 to 20, as it improved the speed of convergence of the MCMC while leading to equivalent results;
\item The maximum number of iterations is increased to 1,000,000. The MCMC stops whenever the autocorrelation length of every parameters stops progressing 
by more than 1\% and is at least 50 times smaller than the actual number of iterations;
\item The host star and companion magnitude are calculated using a continuous series of model from planetary mass up to stellar mass of 30\,M$_\odot$. We also 
implemented the reflection of stellar light on the surface of the companion. These issues are discussed in Appendix~\ref{app:magnitude}.
\end{itemize}

We highlight here an important effect of the modeling of the light reflected from the companion surface on the motion of the photocenter, and developed 
in more details in the Appendix. For mass ratios $M_c/M_\star$$\sim $$10^{-5}-10^{-3}$ and companion orbit semi-major axis 
$a_c$$<$$0.5$\,a.u, the companion reflected light can become more important than the star's emission in the calculation of the photocenter semi-major axis, 
with $a_\text{ph}$=$L_c a_c + L_\star a_\star$. The astrometric motion of the system observed from Earth can even follow the motion 
of the companion itself rather than the motion of the stellar host. This could lead to wrongly determine the orientation (retrograde/prograde) of the primary star orbit, and 
strongly underestimate the primary star semi-major axis and thus the mass of the companion. This effect cannot be seen in the present study because it is smaller 
than the adopted detection thresholds (Section~\ref{sec:threshold}), but should be taken into account in future analysis of Gaia's time series of systems with 
planets.

Concerning the definition of the parameters explored in the MCMC corresponding to inclination, eccentricity and longitude $\omega$, they are improved compared 
to Kiefer (2019), solving singularity issues at the border of the domain expected for these parameters: 
\begin{itemize}
\item Adopting $\lambda_{I_c}$=$\tan(2 I_c-\pi/2)$ as was used in Kiefer (2019) led the 
StretchMove algorithm of \verb+emcee+ (Foreman-Mackey et al. 2013) to get stuck in low probability regions with large or small inclinations, much wider in terms 
of $\lambda_{I_c}$. We thus considered instead simply varying $I_c$ imposing rigid boundaries at $I_c$=0 and $\pi/2$. 
\item The exploration of the ($e$,$\omega$) space was not optimal, especially around the singularity $e$=0. We thus varied instead 
$\lambda_\text{c}$=$\tan (\frac{\pi}{2} e\cos\omega)$ and $\lambda_\text{s}$=$\tan (\frac{\pi}{2} e\sin\omega)$, with $e$ and $\omega$ being then obtained 
from the simple transformations and combinations of $\lambda_\text{c}$ and $\lambda_\text{s}$.
\end{itemize}

\subsection{Application of GASTON to the defined samples}

We apply GASTON on the 29 candidate exoplanets of the detection sample, orbiting the 28 sources which astrometric excess noise exceed the detection thresholds 
fixed in Section~\ref{sec:threshold} and listed in Table~\ref{tab:list_planets}. For those, with the star's orbit a priori detected in the astrometric data, an inclination 
and true mass could technically be measured. 

For the 227 non-transiting companions of the non-detection sample listed in Table~\ref{tab:list_planets_nondet}, we also used GASTON to derive the lowest inclination 
and largest mass possible for the companion, beyond which the astrometric excess noise would become too large to be compatible with $\varepsilon_\text{DR1}$. 
In order to limit the computation time, and since these calculations only leads to parameter ranges and not strict measurements, we reduced the 
maximum number of MCMC steps in GASTON to 50,000 for these 227 companions. Moreover, conversely to what adopted for the detection sample, for those 227 companions we adopt 
a flat prior for the inclination. The shape of the prior distribution of the inclination tends to dictate the shape of the posterior distributions if the simulated 
astrometric excess noise is compatible with $\varepsilon_\text{DR1}$ for inclinations of 90$^circ$ down to $\sim$0$^\circ$. This prior artificially increases the 
conventional lower limit, such as 3-$\sigma$, for inclination -- and thus decreases the upper-limit on mass. This is typically the case for companions in the 
non-detection sample with $\varepsilon_\text{DR1}$ compatible with noise and $a_\star \sin i$$\ll $$\varepsilon_\text{DR1}$. Adopting a flat prior for the 
inclination favours instead the likelihood -- and thus the data -- to dictate the shape of the posterior distributions down to small inclination. This better reveals 
the variations of the inclination and mass posteriors only due to incompatibilities between the $\varepsilon_\text{simu}$ and $\varepsilon_\text{DR1}$ at 
inclinations close to 0$^\circ$.

In the following, we will only report for the resulting posteriors of the inclination $I_c$ and its deriving parameters: the true mass of the companion, the 
photocenter semi-major axis, and the magnitude difference between the companion and its host star. 

\section{Results}
\label{sec:results}

\subsection{General results}
\label{sec:general}  

Out of the results produced by GASTON, we identified 3 possible situations: 
\begin{itemize}
\item[1. ] Orbits leading to a firm measurement of the RV orbit inclination and the true mass of the companions. This concerns 9 exoplanet candidates out of 29
 in the detection sample. This is summarized in Section~\ref{sec:case1};
\item[2. ] Orbits for which the astrometry cannot constrain the inclination. Because of the noise, producing a measured astrometric excess 
noise compatible with the RV orbital motion is possible for a large range of inclinations. The derived solution follows mainly the $\sin i$ prior distribution of 
inclination, with a median about 60$^\circ$, 1-$\sigma$ confidence interval within 30-80$^\circ$ and a 3-$\sigma$ (99.85\%) percentile larger than 
89.5$^\circ$. Only the upper-limit on the mass and the lower-limit on the inclination is informative. This concerns 18 exoplanet candidates from the detection
sample and the 227 exoplanet candidates from the non-detection sample. This is summarized in Sections~\ref{sec:case3} and ~\ref{sec:non-detection};
\item[3. ]  Companions for which the astrometric excess noise could never be reached in the simulations testing any inclinations from 0.001 to 90$^\circ$. The 
Gaia astrometric excess noise is incompatible with the published RV orbit. Two companions from the detection sample enter this situation, WASP-43\,b and 
WASP-156\,b (see Section~\ref{sec:case5}).
\end{itemize}

For the 29 companions of the detection sample, the results of GASTON according to different situations introduced above are presented in Tables~\ref{tab:results_det} 
\&~\ref{tab:results_5}. Moreover, Table~\ref{tab:results_nondet} summarizes the parameter limits derived for the 227 companions of  the non-detection sample. 
In both tables, we list the resulting corrected mass, astrometric orbit semi-major axis, estimated magnitude difference between the host and the companion, 
MCMC acceptance rate and convergence indicator $N_\text{steps}/\text{max}(\tau_\lambda)$ (see below).

The acceptance rate delivered by \verb+emcee+ allows to quantify the probability of reaching $\varepsilon_\text{DR1}$ 
through all simulations performed during the MCMC process. Typically, if an MCMC performs well, the acceptance rate must reach 0.2-0.4. This is the case for 
all 9 companions entering situation \#1, except one, HD\,96127\,b for which it is 0.06. Low values of the acceptance rate usually imply 
too large steps in the Monte-Carlo process (Foreman-Mackey et al. 2013). We can firmly exclude any "steps issue", since the 
geometry of the parameter space is the same for all systems, and the steps for the different parameters have been tuned such that well behaved cases fullfill the 
0.2-0.4 criterion. Rather we explain this low acceptance rate by the presence of noise in our simulations. A fortuitous pile-up of noise can allow some simulations 
to be compatible with $\varepsilon_\text{DR1}$=1.124\,mas even with an inclination close to 90$^\circ$ and a negligible photocenter orbit. With a $\sin i$-prior 
on inclination favouring the edge-on configuration, this is sufficient to drag the MCMC towards exploring regions where producing such astrometric 
excess noise is not frequent. The low acceptance rate is a reflection of this low frequency. This leads, in the case of HD\,96127\,b, to a 3-$\sigma$ upper-limit 
on the inclination of 89.54$^\circ$. This is the same mechanism that explains the small acceptance rates associated to mass upper-limits for all companions 
entering situation \#2.

The autocorrelation length $\tau_\lambda$ probes the quality of a $\lambda$--parameter exploration by the MCMC during a run. With \verb+emcee+ and its 
Goodman-Weare algorithm (Goodman \& Weare 2010) it can be considered that convergence is reached if at least $N_\text{steps}/\tau_\lambda$$>$$50$ 
(Foreman-Mackey et al. 2013), and at best if $\delta\tau_\lambda/\tau_\lambda$$<$$1$\% for all parameters $\lambda$. The errors on the estimations of the 
posteriors are then reduced by a factor smaller than $1/\sqrt{50}$$\sim $$0.14$. Longer chains obviously produce more accurate results, but are also more time 
consuming. This paper is not aiming perfect accuracy, since only based on a preliminary estimation of one quantity, the astrometric excess noise, by Gaia. We thus 
decided to stop the MCMC whenever $N_\text{steps,max}$ is reached or $N_\text{steps}/\text{max}(\tau_\lambda)$$>$$50$ and 
$\delta\tau_\lambda/\tau_\lambda$$<$$1$\% for all parameters $\lambda$.  With up to 1,000,000 steps and 20 walkers for 10
parameters to explore, the MCMC should have enough time to converge. This allows to identify problematic systems, such as e.g. HD\,96127\,b, for which 
the exploration of the parameter space is inefficient. In Table~\ref{tab:results_det}, we identify 3 companions -- including HD\,96127\,b -- for which GASTON did 
not converge after $N_\text{step}$=1,000,000 iterations, with a maximum autocorrelation length larger than $N_\text{step}/50$ and a small acceptance rate. The 
posteriors for those companions cannot be reliable, and the width of the confidence intervals on their mass is most likely underestimated.

\subsection{Detection sample}
\subsubsection{SItuation \#1: mass measurement for 2 possible massive exoplanets, 2 BDs  and 5 M-dwarfs}
\label{sec:case1}
\begin{figure}[hbt]
\includegraphics[width=87.2mm]{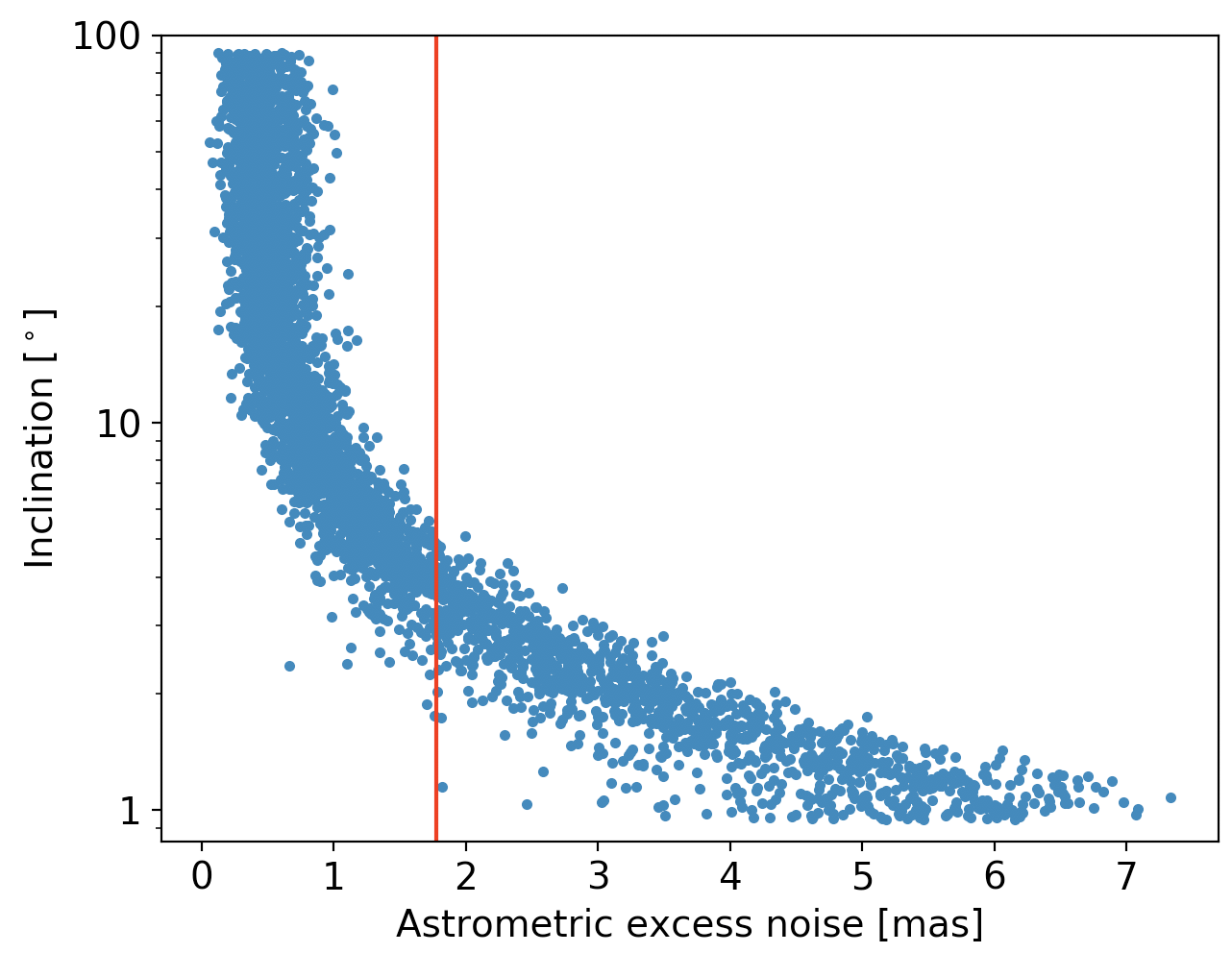}
\includegraphics[width=89.3mm]{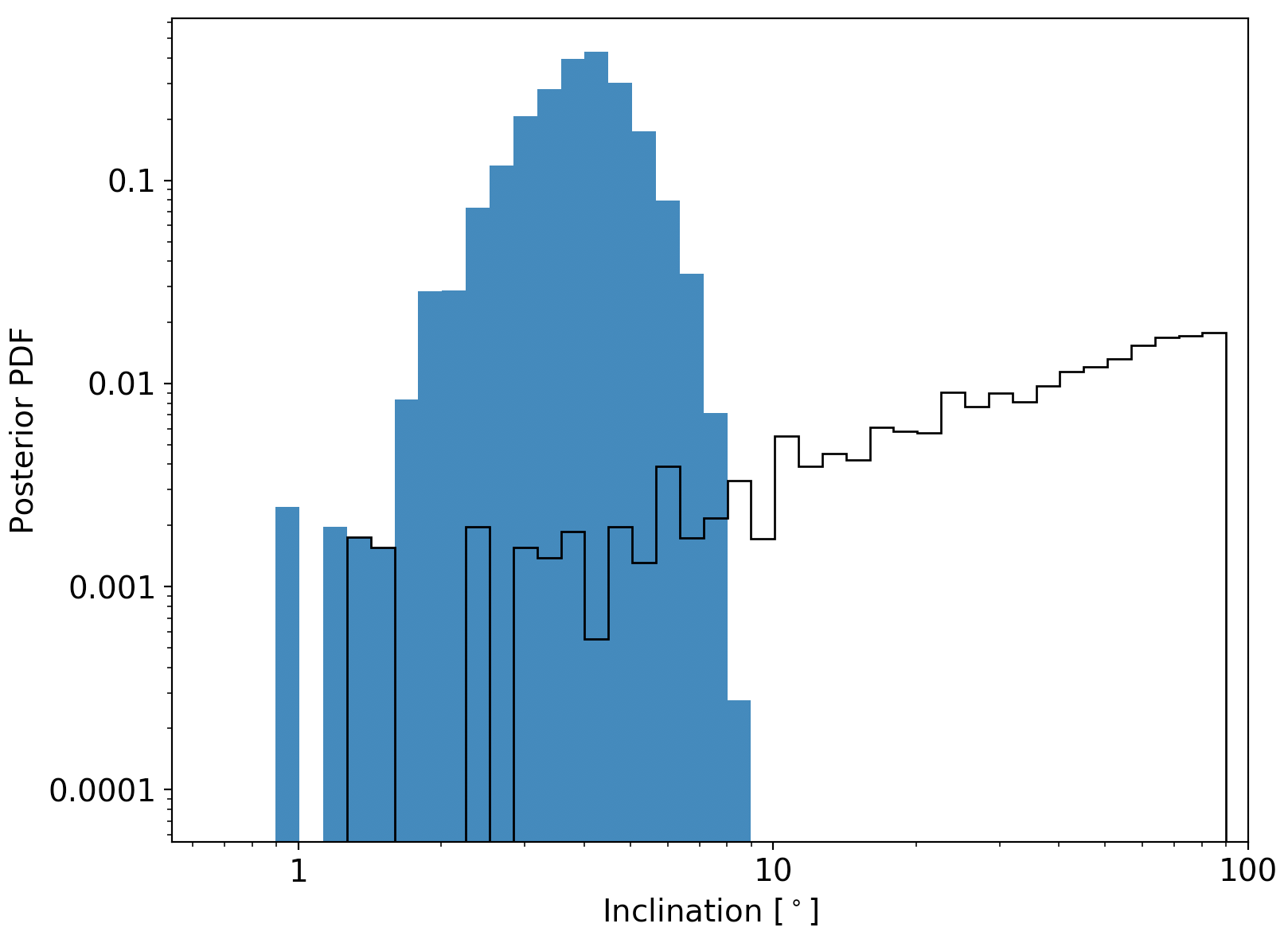}
\caption{\label{fig:case1}Examples of situation \#1 with 30\,Ari\,B\,b, primary dataset, $P$=335\,days, $a_\star\sin i$=0.17\,mas and 
$\varepsilon_\text{DR1}$=1.78\,mas. Top-panel: $I_c$ plotted against the derived $\varepsilon_\text{simu}$ for every simulations (blue points). The red line 
shows $\varepsilon_\text{DR1}$. Bottom-panel: the $I_c$ posterior probability density function (PDF) in blue, compared to the $\sin I_c$ prior PDF (black solid line).}
\end{figure}

\begin{figure*}
\includegraphics[width=179.4mm]{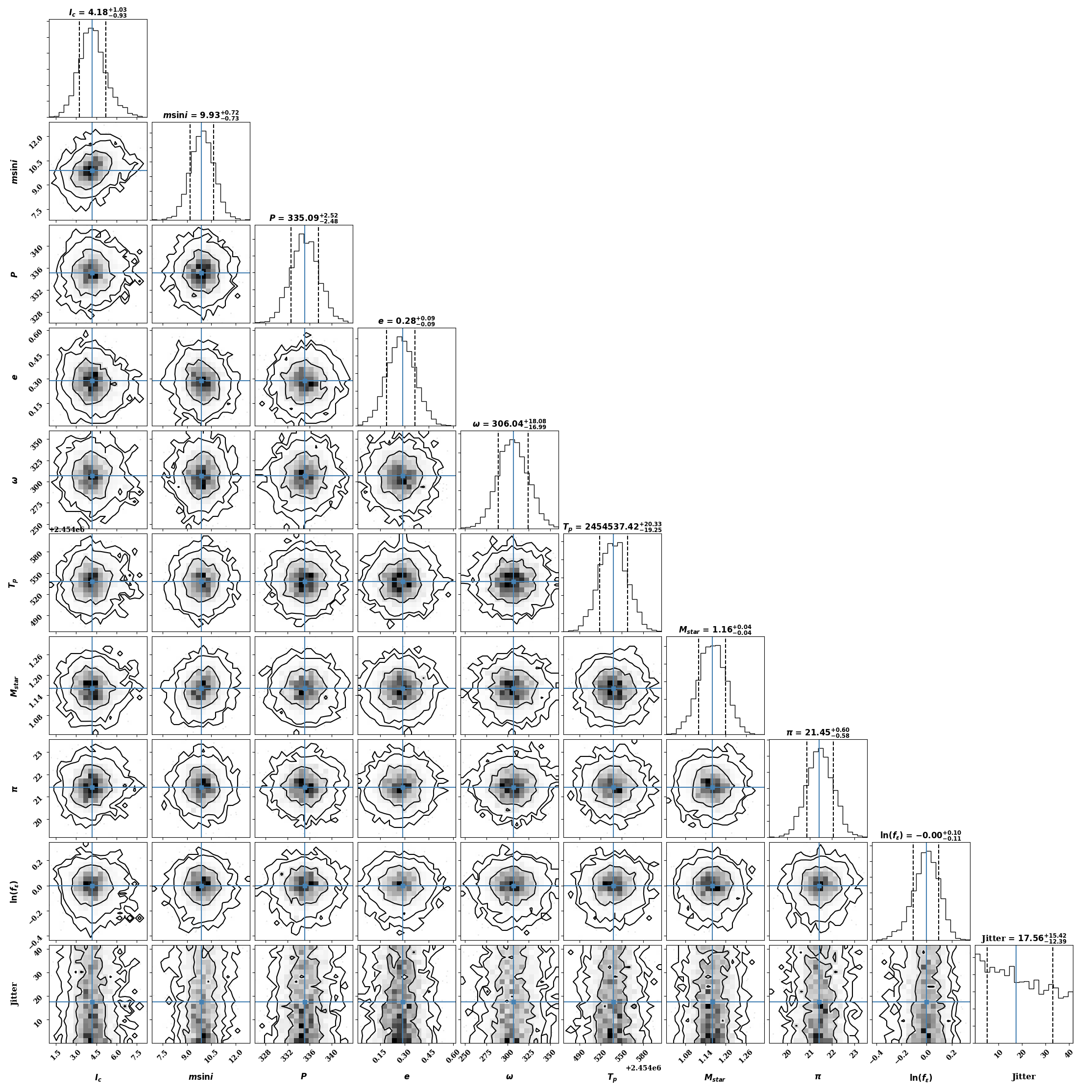}
\caption{\label{fig:mcmc_triangle_1}Corner plot of posterior distributions for all explored parameters with the MCMC for 30\,Ari\,B\,b from situation \#1 
(Section~\ref{sec:case1}). }
\end{figure*}

We illustrates this first case scenario in Figure~\ref{fig:case1} with the example of 30\,Ari\,B\,b for which with a period of 335\,days, the astrometric excess noise 
of 1.78\,mas leads to an inclination of $4.14$$^\circ $$_{-0.90}^{+0.96}$ and a corrected mass of $148_{-27}^{+42}$\,M$_\text{J}$ instead of an 
$m\sin i$=10$\pm$1\,M$_\text{J}$. The top panel of Fig.~\ref{fig:case1} shows the simulated astrometric 
excess noise obtained for 10,000 different values of inclinations from 0.001 to 90$^\circ$. For any inclinations below 1$^\circ$ the true mass of 30\,Ari\,B\,b is 
too large and the magnitude difference with the primary star is smaller than 2.5; these simulations are ignored since they would imply the presence of a 
detectable secondary component in the spectrum of this system, conversely to what observed. The bottom panel of Fig.~\ref{fig:case1} compares the $I_c$ posterior
distribution -- probability density function or PDF -- to the PDF of an ensemble of same size drawn from the assumed prior density function, 
d$p$=$\sin I_c$d$I_c$. This posterior PDF is well distinct from the prior PDF which thus have a minor impact on the posterior distributions output from the MCMC.
The corner plot of all posterior distributions for 30\,Ari\,B\,b is shown in Figure~\ref{fig:mcmc_triangle_1}. 

In this category, all other GASTON runs work similarly as well as 30\,Ari\,B\,b, with the exception of HD\,96127\,b which MCMC run could not converge after 
1,000,000 iterations. In total, the true masses for 9 exoplanet candidates could be determined using GASTON, with 8 orbiting sources from the primary dataset 
and one, HD\,16760\,b, from the secondary dataset. We determined that 7 of the companions are not planets, and two, could be likely brown-dwarf or M-dwarf, 
but the planetary nature cannot be excluded at 3-$\sigma$.

Among the primary sources, we find that HD 5388 b, HD 6718 b, HD\,114762\,b and HD\,148427\,b are constrained within the brown dwarf/M-dwarf domain 
with the 3-$\sigma$ mass-ranges, respectively (57, 150)\,M$_\text{J}$,  (29, 157)\,M$_\text{J}$, (33, 328)\,M$_\text{J}$ and (27, 345)\,M$_\text{J}$. 
We found moreover that 30\,Ari\,B\,b and HIP\,65891\,b are stars in the M-dwarfs mass regime with masses larger than 80\,M$_\text{J}$. 

The two possible planets are HD\,141937\,b and HD\,96127\,b. The true mass of HD\,141937\,b is located just beyond the boundary between massive planets 
and low-mass brown dwarfs with $M$=27.5$^{+6.9}_{-10.8}$\,M$_\text{J}$ at 1-$\sigma$ but a mass possibly as low as $9$\,M$_\text{J}$ at 3-$\sigma$. 

The true mass of HD\,96127\,b is most likely well within the stellar domain with $M$=190$^{+284}_{-184}$\,M$_\text{J}$ and an inclination 
$I_c$=1.364$^{+38.527}_{-0.763}$\,\!$^\circ$ at 1-$\sigma$. Within the 1-$\sigma$ confidence interval, a true mass of HD\,96127\,b as low as 
6\,M$_\text{J}$ could also be compatible with $\varepsilon_\text{DR1}$. However, we already noted that GASTON did not converge for this precise case, 
due to a marginal but possible compatibility of the Gaia DR1 astrometry with an edge-on configuration, as revealed by the low 0.05 acceptance ratio of 
the MCMC run. The $1$-$\sigma$ bounds of HD\,96127\,b's mass are thus questionable and its true nature is still uncertain.

\begin{figure}[hbt]\centering
\includegraphics[width=86.2mm]{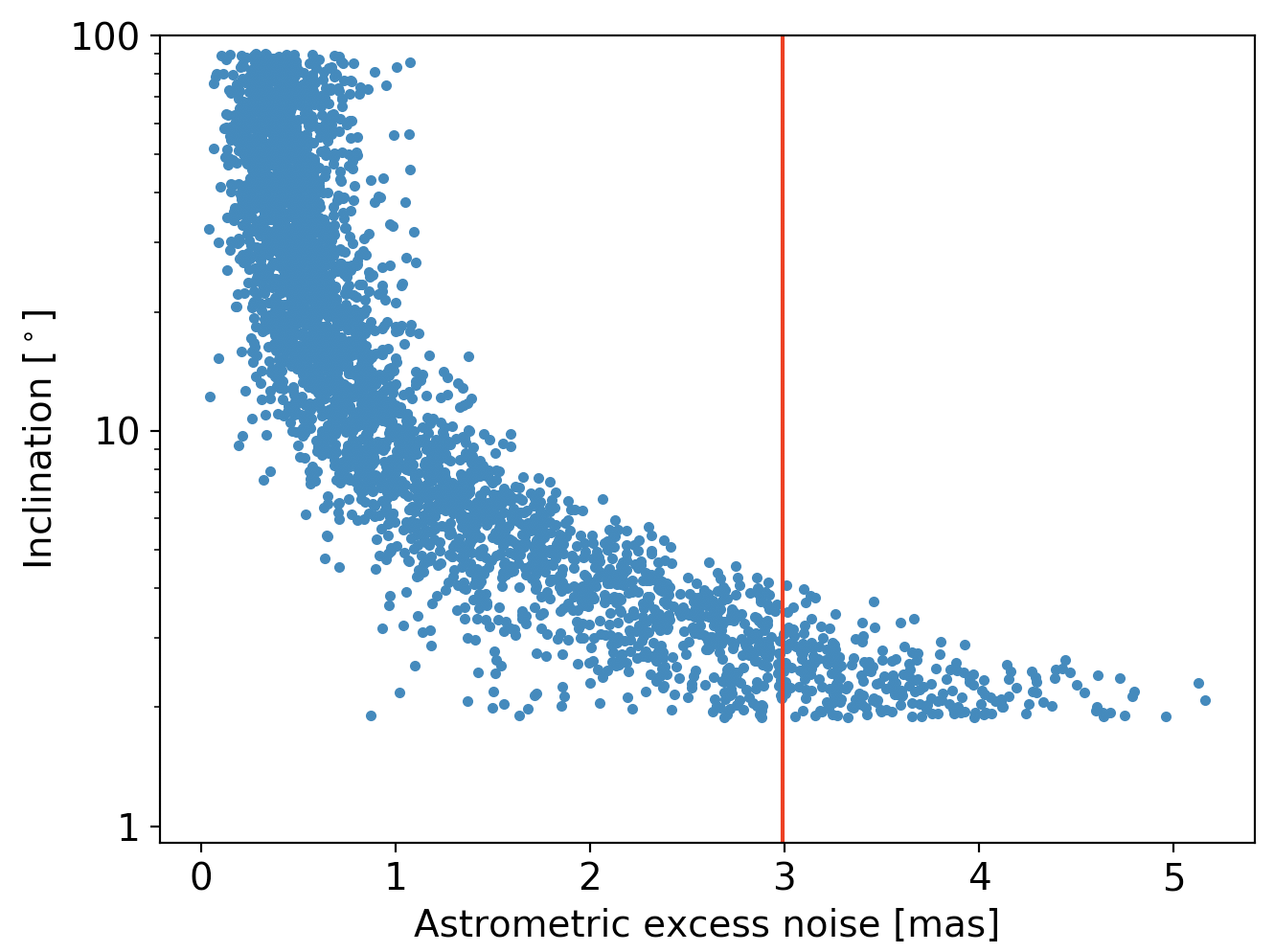}
\includegraphics[width=89.3mm]{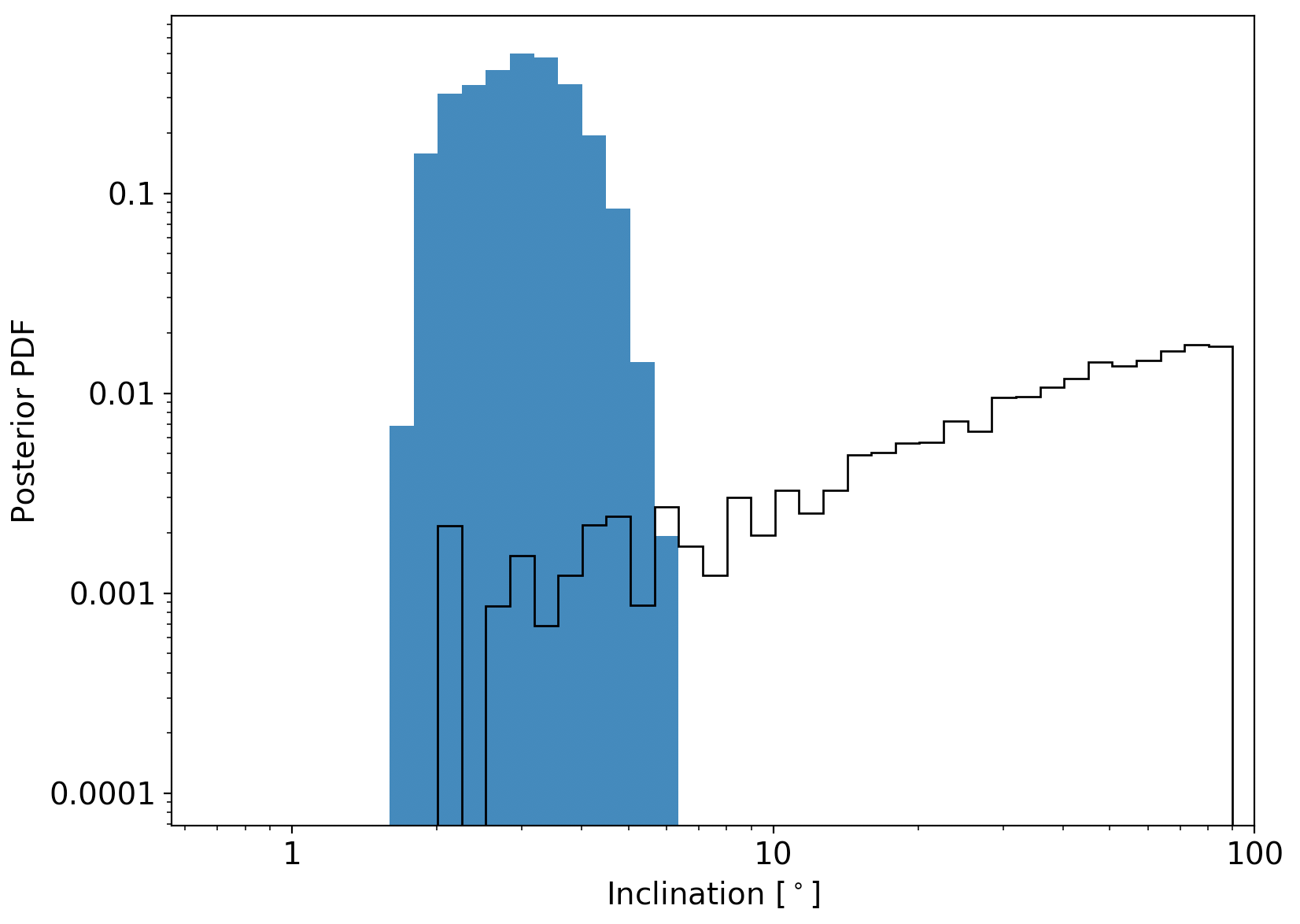}
\caption{\label{fig:HD16760} Same as Fig.~\ref{fig:case1}, illustrating situation \#1, with the secondary dataset source companion HD\,16760\,b ($P$=465\,days, 
$a_\star\sin i$=0.25\,mas and $\varepsilon_\text{DR1}$=3\,mas).}
\end{figure}

The results for the single source from the Gaia DR1's secondary dataset within situation \#1, HD\,16760\,b, are given in table~\ref{tab:results_det}, and illustrated
in Fig.~\ref{fig:HD16760}.
HD\,16760\,b (Bouchy et al. 2009), is the first companion with a possible planetary mass discovered with the SOPHIE spectrograph 
(Perruchot et al.  2013) actually is not a planet. With a parallax of 14\,mas and an astrometric excess noise of 2.99\,mas, we found its astrometry to be rather 
compatible with an M-dwarf which true mass is larger than 13.5\,M$_\text{J}$ at 3-$\sigma$.

\subsubsection{SItuation \#2: upper-limit constraint on companion mass}
\label{sec:case3}

The orbit inclination of 18 companions from 17 different systems in the detection sample cannot be fully determined using GASTON. For those orbits, the 
simulated astrometric excess noise is often compatible with $\varepsilon_\text{DR1}$ from $I_c$=90$^\circ$ down to $\sim$0$^\circ$. Accounting for the 
$\sin I_c$ prior probability distribution on the inclination, the MCMC leads to a posterior distribution for which the 3-$\sigma$ upper-bound on inclination is 
located beyond 89.5$^\circ$. More accurately, the posterior distribution on their orbit inclination and mass are mainly fixed by the $\sin I_c$ prior on inclination. 

As presented in Table~\ref{tab:results_det}, all these candidates are possible planets at the 1-$\sigma$ limit. Excluding the transiting planets which are known
to be bona-fide planets on edge-on orbits, only two of them have a true mass below, but close to, the Deuterium burning limit of 13.5\,M$_\text{J}$. They are 
HD\,164595\,b and HD\,185269\,b with a mass smaller than respectively 12.9 and 12.6\,M$_\text{J}$ at 3-$\sigma$. Those two seem thus likely to be actual 
planets with a mass in the Neptunian (0.06\,M$_\text{J}$ for HD\,164595\,b) and Jupiterian (1.12\,M$_\text{J}$ for HD\,185269\,b) domain.
 
\begin{figure}[ht]\centering
\includegraphics[width=87.3mm]{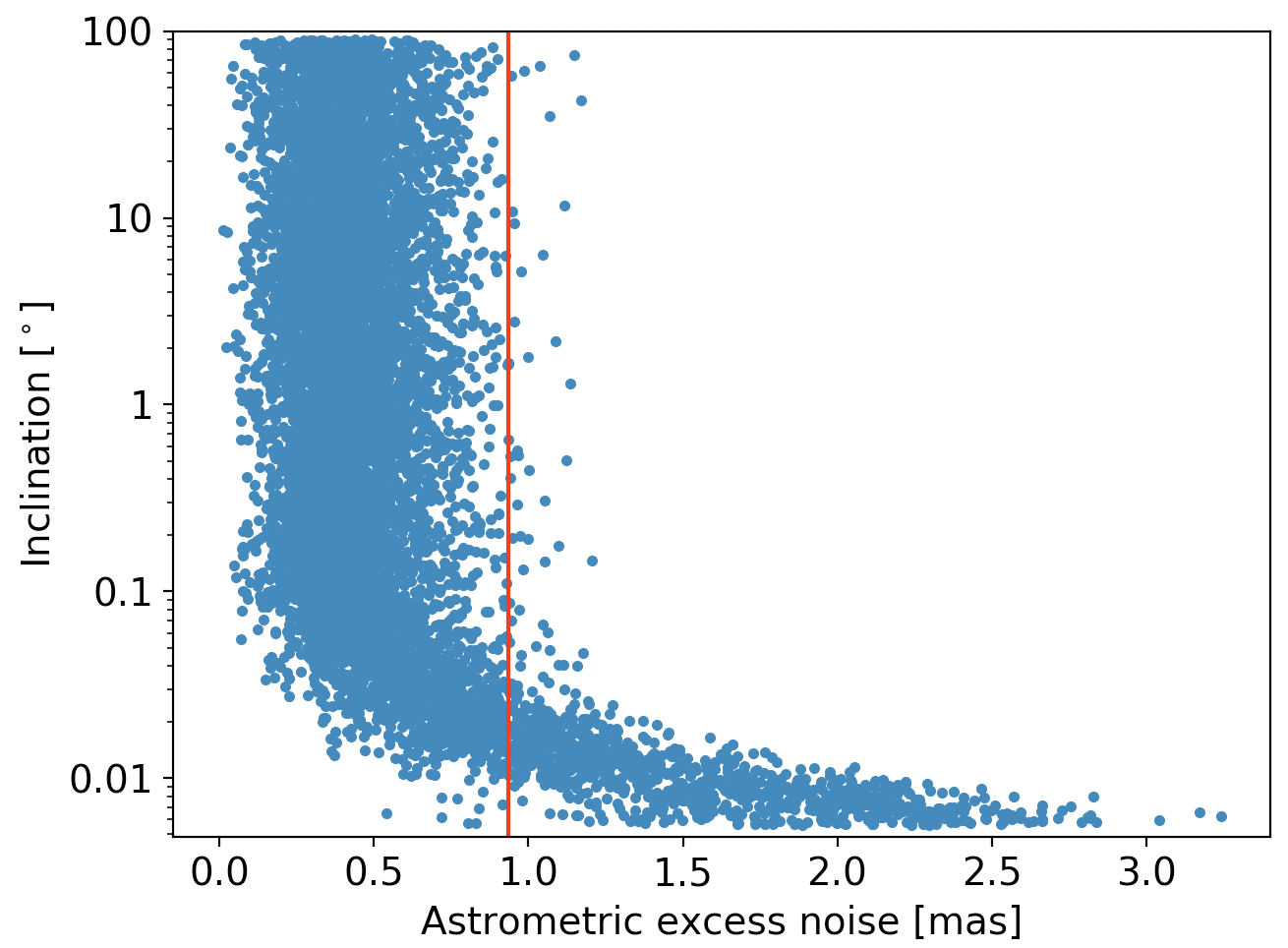}
\includegraphics[width=89.3mm]{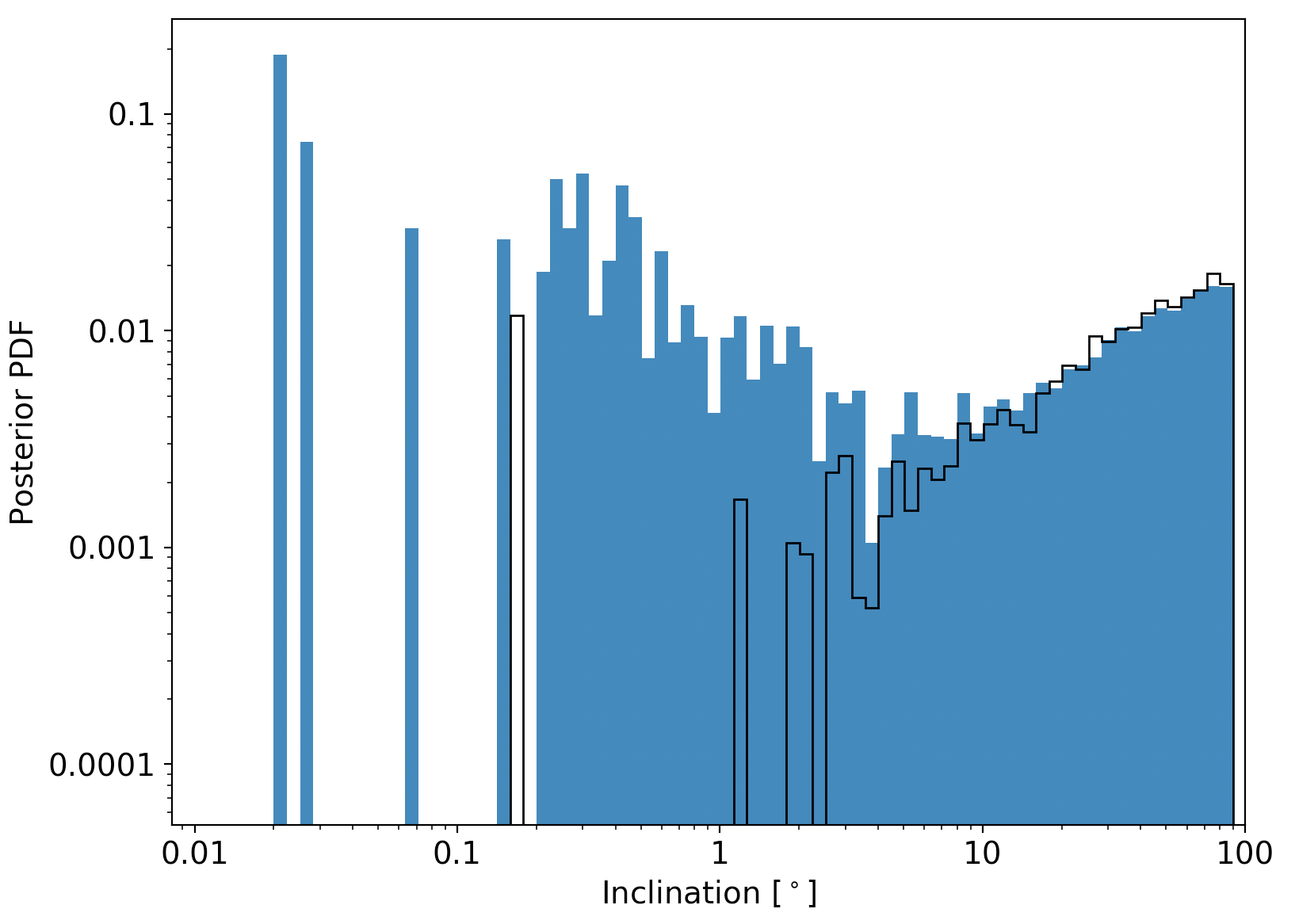}
\caption{\label{fig:HD164595} Illustration of situation \#2 with HD\,164595\,b, $P$=40\,days, $a_\star\sin i$=0.0004\,mas and 
$\varepsilon_\text{DR1}$=0.93\,mas. Many simulations from $I_c$=0.1$^\circ$ up to 90$^\circ$ are compatible with $\varepsilon_\text{DR1}$. This reflects in 
the comparison of the $I_c$ posterior to the $\sin I_c$ prior PDFs shown in black.}
\end{figure} %
\begin{figure}[ht]\centering
\includegraphics[width=86.3mm]{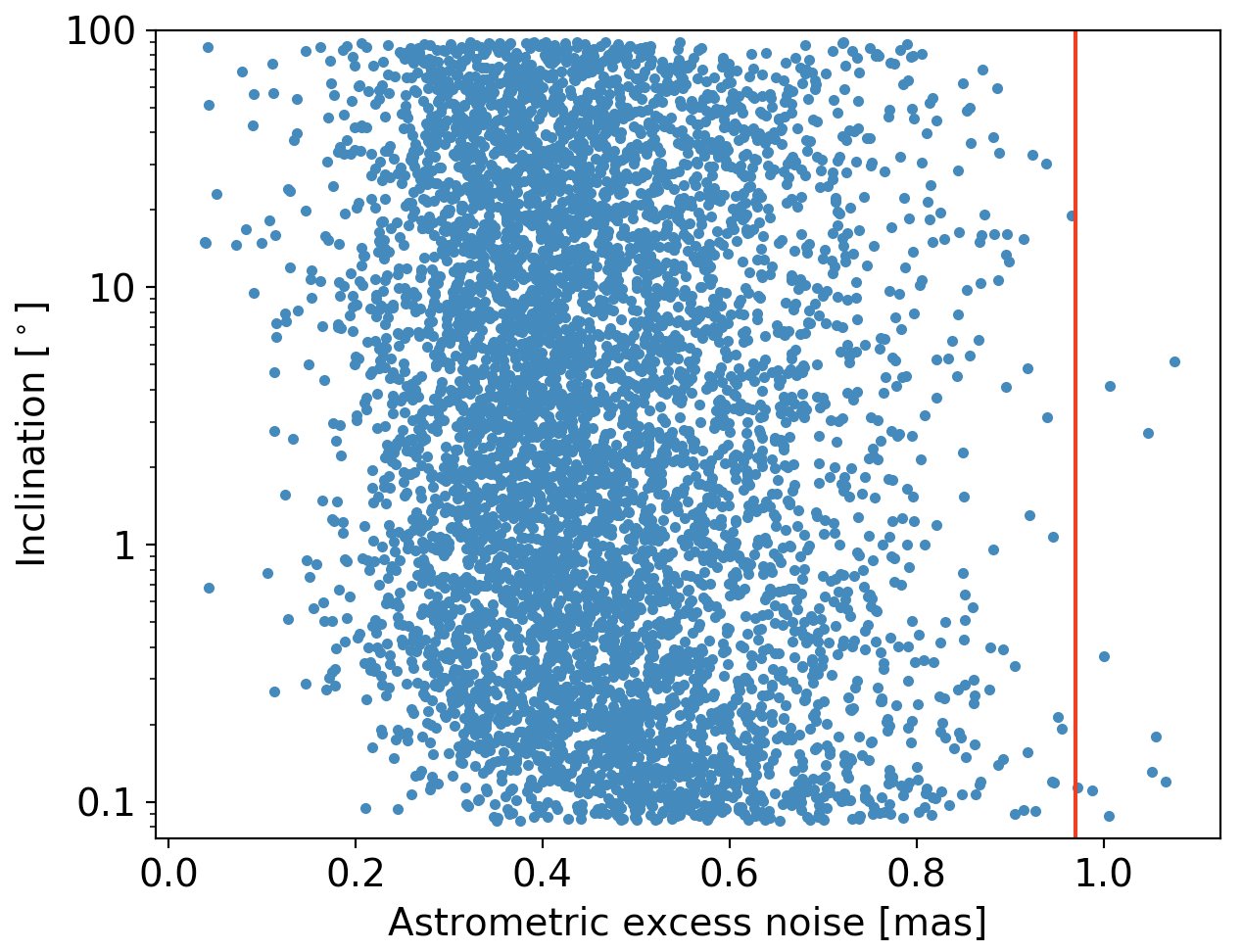}
\includegraphics[width=89.3mm]{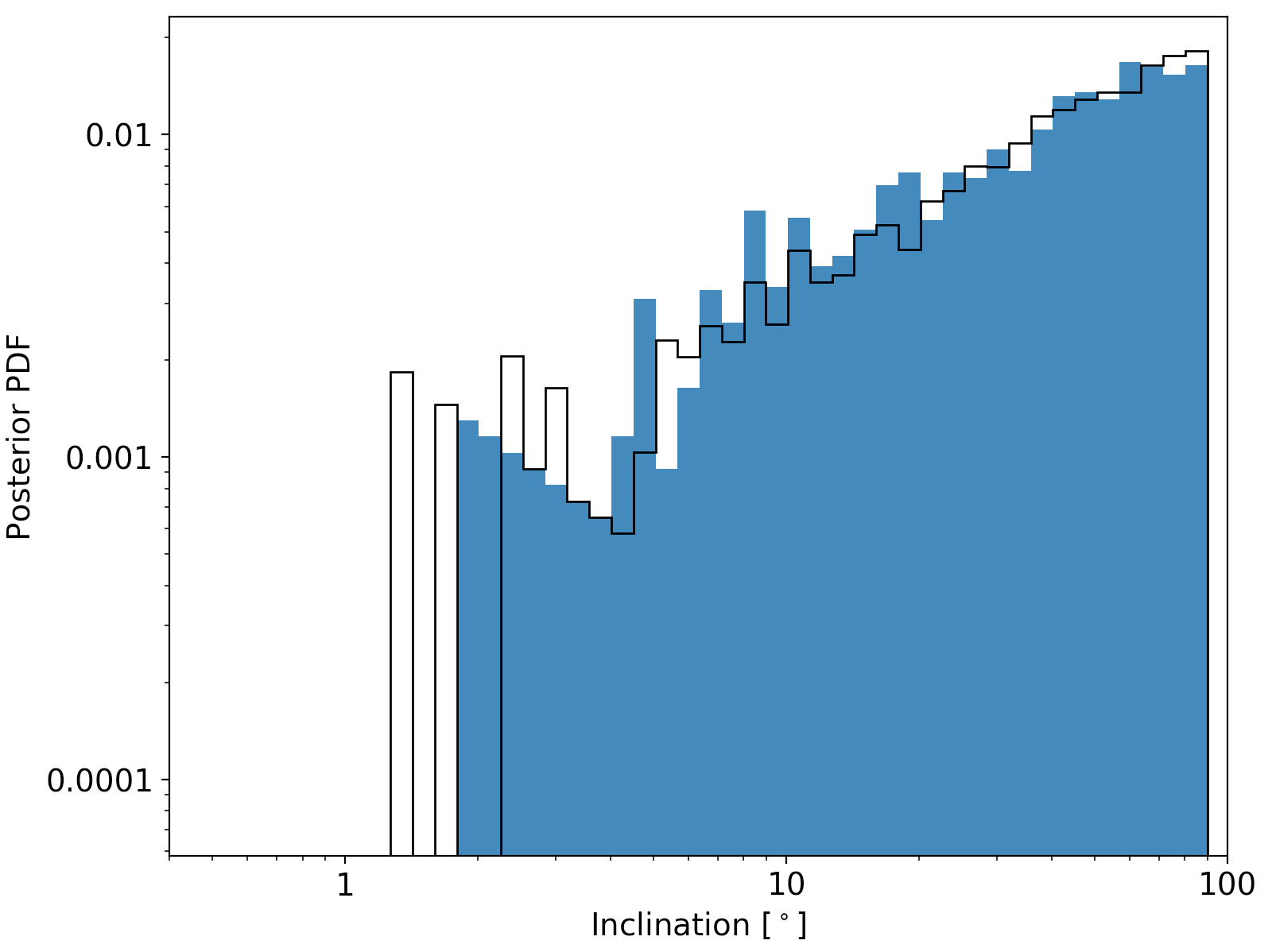}
\caption{\label{fig:HD185269} Same plot as Fig.~\ref{fig:HD164595}, illustrating situation \#2, with HD\,185269\,b ($P$=6.8\,days, $a_\star\sin i$=0.001\,mas 
and $\varepsilon_\text{DR1}$=0.97\,mas). Many simulations from about $I_c$=0.1$^\circ$ up to 90$^\circ$ are compatible with 
$\varepsilon_\text{DR1}$ thanks to noise. Smaller inclinations are rejected due to too large mass and luminosity for the companion. The posterior distribution in 
the bottom panel is fully compatible with the prior $\sin I_c$ PDF.}
\end{figure} 

We note however that HD\,164595 is a duplicate source in the Gaia DR1. Its astrometric excess noise, and thus the mass of HD\,164595\,b, might be 
underestimated (see the discussion on this specific issue in Section~\ref{sec:gaia_selec}). Moreover, for both companions, the simulated 
astrometric excess noise is indeed compatible with $\varepsilon_\text{DR1}$ on a large range of inclinations (Figs.~\ref{fig:HD164595} and~\ref{fig:HD185269}). 
The posterior distributions of $I_c$ and $M_c$ are essentially due to the prior distribution on $I_c$. If the actual prior distribution of $I_c$ is biased 
towards $0^\circ$ (see the related 
discussion in Section~\ref{sec:small_inc}), it cannot be excluded that the masses of HD\,164595\,b and HD\,185269\,b are actually larger than 13.5\,M$_\text{J}$.

Seven companions are transiting planets. They are HAT-P-21\,b, WASP-11\,b, WASP-17\,b, WASP-131\,b, WASP-157\,b and K2-34\,b in the primary dataset, and 
K2-110\,b in the secondary dataset. Gaia DR1 measurements 
all are compatible with the edge-on configurations. The MCMC acceptance rates are smaller than 0.01 with a star semi-major axis smaller than 
1\,$\mu$as. It can be excluded that Gaia will truly detect the reflex motion of these stars due to their transiting exoplanets.

Two exoplanet candidates are part of a common multiple system, HD\,154857\,b and c. The Gaia observations are compatible with an edge-on 
inclination and masses of 2.2 and 2.5\,M$_\text{J}$. At 1-$\sigma$ the posterior distributions, conformally to the $\sin I_c$ prior distribution on $I_c$, allow 
inclinations as low as 20$^\circ$ with masses as large as 6\,M$_\text{J}$, but at 3-$\sigma$ their mass could be as large as 135 and 175\,M$_\text{J}$. Both 
companions are thus possible Jupiter-mass planets with masses within 2-6\,M$_\text{J}$, but their true nature could not be confirmed.

\subsubsection{Situation \#3: incompatible RV orbit and Gaia astrometry}
\label{sec:case5}
The GASTON results for the two companions within this situation are presented in Table~\ref{tab:results_5}. They are WASP-43\,b and WASP-156\,b, both 
transiting planets on compact orbit ($P$=0.8\,days and 3.8\,days). In these two systems, none of the published companions are adequate for explaining 
Gaia's observations. The maximum astrometric excess noise that could be simulated from RV orbital parameters were respectively 1.25 and 1.32\,mas, well 
below the $\varepsilon_\text{DR1}$ of these two sources, respectively 2\,mas and 1.5\,mas. These two sources from the secondary dataset are not mentioned 
as duplicated sources in the DR1 database. 

There are three possible scenarios for explaining this RV-Gaia discrepancy:

\begin{itemize}
\item The value of the astrometric excess noise could depend on the presence of fortuitous outliers. With a number of astrometric measurements 
$\sim$50 per source, outliers of several mas could slightly inflate $\varepsilon_\text{DR1}$ with a discrepancy of a few 0.1\,mas. Outliers larger than 
$4.8$$\times $$\varepsilon_\text{DR1}$$\sim $$10$\,mas (see note 7 in Lindegren et al. 2016) are flagged as "bad" during the AGIS reduction and discarded. Therefore, the 
discrepancy observed in Table~\ref{tab:results_5} for the 2 companions between the highest $\varepsilon_\text{simu}$ and $\varepsilon_\text{DR1}$ of 
0.2-0.7\,mas could be explained by numerous or large outliers. We cannot exclude this possibility without analysing the time series, which will not be available 
until the final Gaia release in a few years.

\item Instrumental and modeling noises larger than those adopted in Section~\ref{sec:noise_scheme} could allow reaching the astrometric excess noise. 
Indeed, for the astrometric data of secondary dataset targets the parallax and proper motion fit is not of good quality, and could individually reach high 
astrometric excess noise, as indicated by the 90th-percentile $\varepsilon_\text{DR1}$=2.3\,mas measured by Lindegren et al. (2016) in the full secondary dataset.
Although plausible, as already discussed in Section~\ref{sec:noise_scheme}, the good match between the distribution of simulated and observed 
$\varepsilon_\text{DR1}$ implies that the instrumental and modeling noise cannot be much larger than the adopted range of 0.4-0.9\,mas in the present sample.

\item A hidden outer companion to the system, unseen in the RV variations, could be responsible for the astrometric signal. This issue is discussed 
in Section~\ref{sec:outer}.
\end{itemize}

Although the presence of outliers cannot be excluded, this RV-Gaia discrepancy motivates the search for supplementary yet hidden companions in these 
systems.

\begin{table}\centering
\caption{\label{tab:results_5} The two exoplanets WASP-43\,b  and WASP-156\,b, which RV orbit is incompatible with Gaia astrometric excess noise. The 
maximum astrometric excess noise that we were able to simulate for these secondary dataset sources is given as $\varepsilon_\text{simu,max}$.}
\begin{tabular}{l@{~~~}c@{~~~}c@{~~~}c@{~~~}}
Parameter & Unit & WASP-43 b & WASP-156 b\\
\hline
Period	& (day) &0.813 & 3.836 \\
$m\sin i$ & (M$_\text{J}$) & 1.761 & 0.131 \\
$a\sin i$ & (mas) &  0.00047 &  0.000055 \\
$\varepsilon_\text{DR1}$ & (mas) &   1.96 &  1.49 \\
$\varepsilon_\text{simu,max}$ & (mas) & 1.25 &  1.32 \\
\hline
\end{tabular}
\end{table}

\subsection{Non-detection sample: 27 confirmed planets}
\label{sec:non-detection}

For a given RV orbit with given $m\sin i$ of the companion, an increasing true mass and thus decreasing orbital inclination imply increasing astrometric motion 
of the star. The non-detection of an astrometric excess noise larger than the defined threshold thus allow deriving an upper-limit on the true mass of the 
companion and a lower-limit on its orbital inclination with GASTON. 

Among the 227 non-transiting companions of the non-detection sample, we constrained true masses lower than 13.5\,M$_\text{J}$ within 3-$\sigma$ confidence 
interval for a total of 27 companions. They are summarized in Table~\ref{tab:results_nondet}. Nine planets have a true mass lower than 5\,M$_\text{J}$, and 19 
have a true mass lower than 10\,M$_\text{J}$. 

\begin{figure}[hbt]\centering
\includegraphics[width=86.2mm]{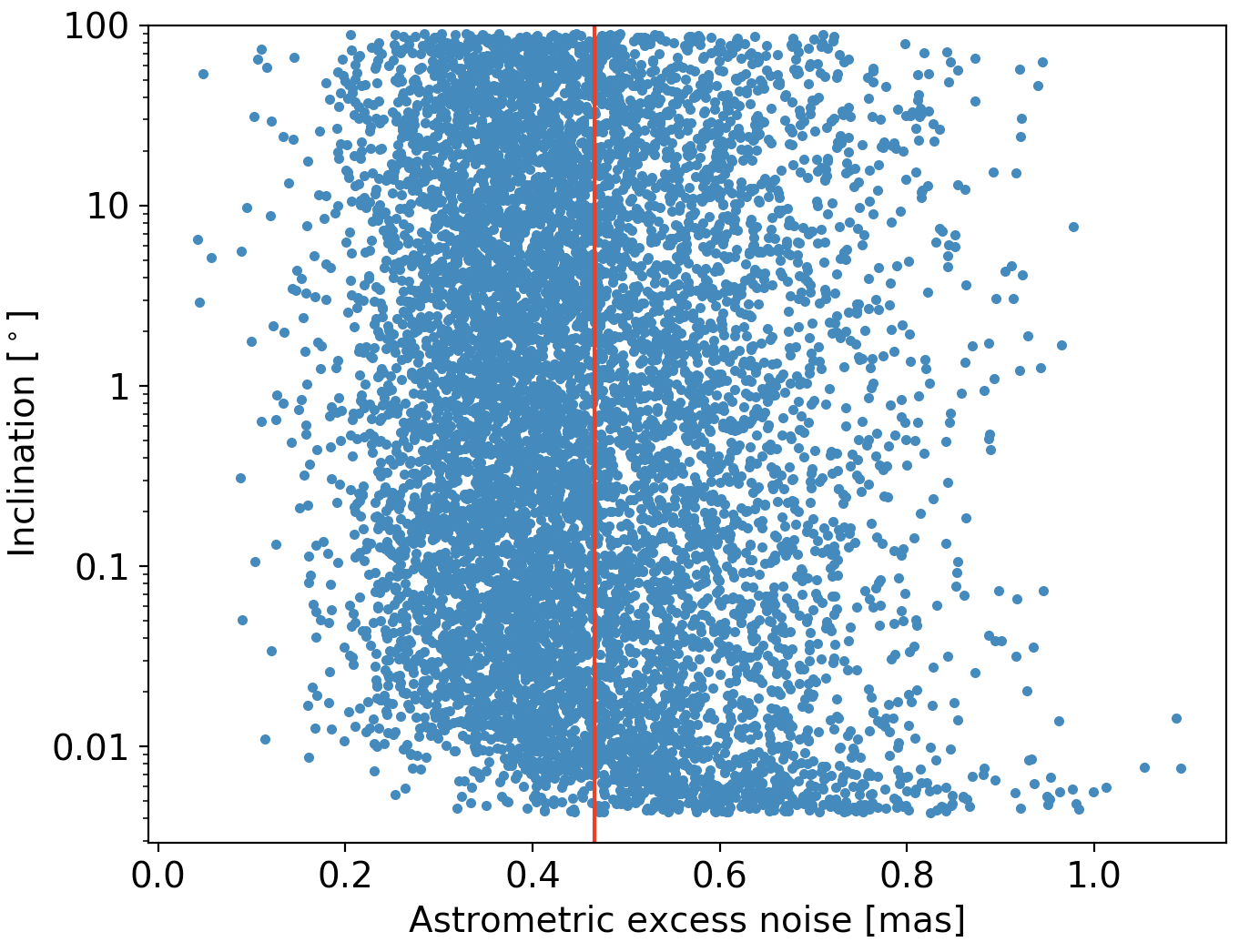}
\includegraphics[width=89.3mm]{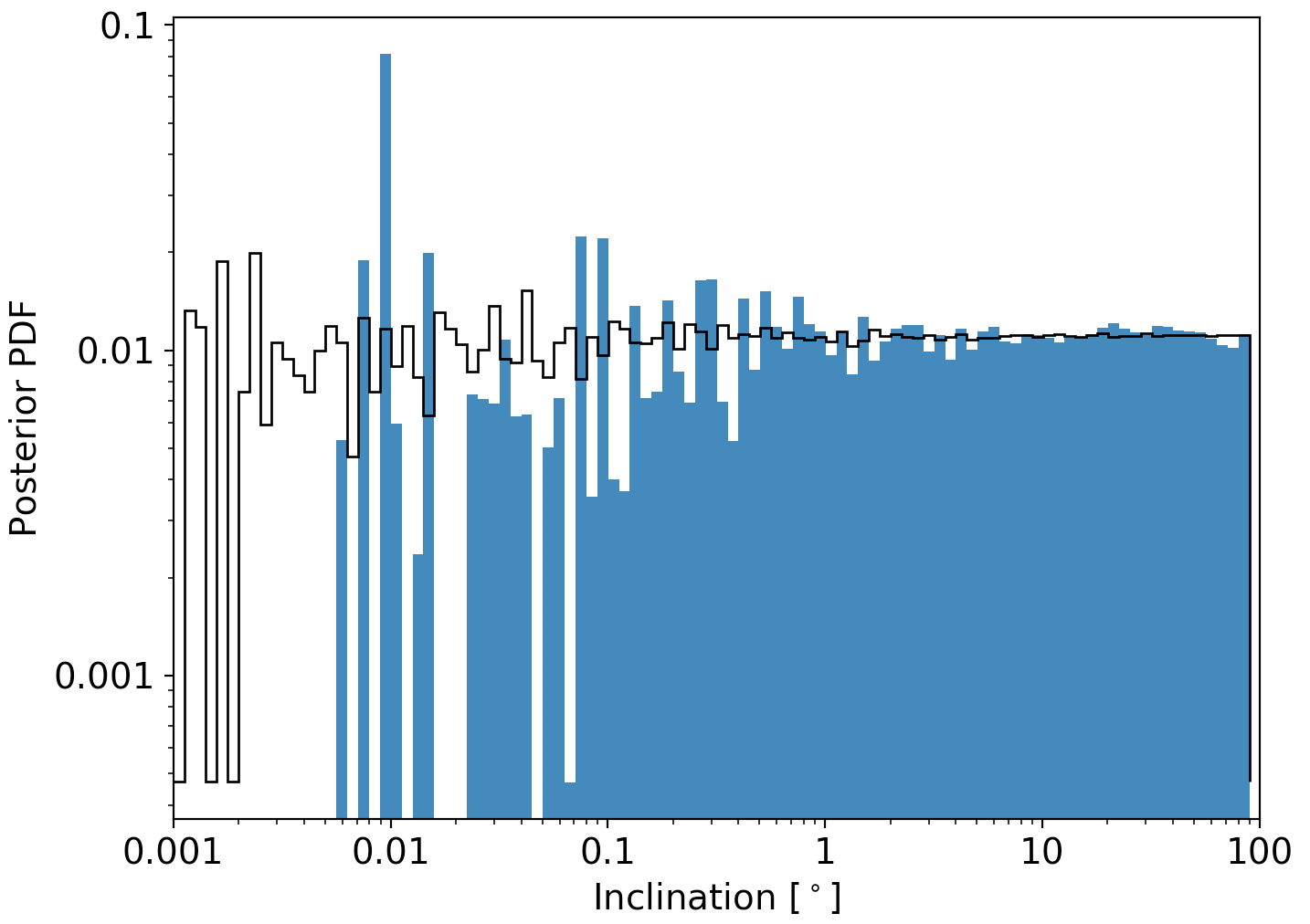}
\caption{\label{fig:HD10180} An example of a confirmed planet from the non-detection sample with HD\,10180\,c ($P$=5.8\,days, $m\sin i$=0.041\,M$_\text{J}$, 
$\varepsilon_\text{DR1}$=0.47\,mas). Top \& bottom panels: same as Fig.~\ref{fig:case1}, only the black line in the bottom panel now represents inclinations drawn from 
a uniform distribution.}
\end{figure}

We confirm that 6 multiple system contains several true planets. They are HD\,10180, HD\,176986, HD\,181433, HD\,215152, HD\,7924, and 
HD\,40307. In the 6-planets system, HD\,10180, we can confirm that the a priori less massive companions c ($P$=5.8\,days, $m\sin i$=0.041\,M$_\text{J}$), d 
($P$=16.4\,days, $m\sin i$=0.037\,M$_\text{J}$), and g ($P$=602\,days, $m\sin i$=0.067\,M$_\text{J}$) are planets with a mass strictly lower than 
12\,M$_\text{J}$ at 3-$\sigma$. Fig.~\ref{fig:HD10180} shows the $\varepsilon_\text{simu}$--$I_c$ relationship and $I_c$ posterior distribution for HD\,10180\,c. A study 
of the effect of mutual inclinations on the stability of this system led to constrain the masses of the planets within a factor of 3, with $I_c$$>$$10^\circ$ for all
planets (Lovis et al. 2010). While our result is not as much as restrictive, it excludes a full face-on inclination with $I_c$$>$$0.2^\circ$ at 3-$\sigma$ and 
confirms planetary mass for planets c, d, and g.

\onecolumn
\thispagestyle{empty}
\newpage
\newgeometry{margin=1.cm,top=2cm}
\begin{sidewaystable*}\tiny\centering
\caption{\label{tab:results_det} GASTON results for the 29 companions in the detection sample, and divided into the two different situations mentioned in the text. }
\begin{tabular}{@{}lcccccccccccc|c@{}}
Planet name & Period & $m\sin i$ & $a\sin i$ & $\epsilon$ & \multicolumn{2}{c}{$a_\text{phot}$} & \multicolumn{2}{c}{$I_c$} & \multicolumn{2}{c}{$M_\text{c,true}$}  & \multicolumn{2}{c|}{$\Delta V$} & MCMC \\
			& (days) & (M$_\text{J}$) & (mas)  & (mas)  	& (mas)  & (mas)  		& \multicolumn{2}{c}{($^\circ$)} &  \multicolumn{2}{c}{(M$_\text{J}$)}  & &   & Acceptance \\
	 		&		&			&		&		& 1-$\sigma$ & 3-$\sigma$ &		1-$\sigma$ & 3-$\sigma$		&	1-$\sigma$ & 3-$\sigma$  &  1-$\sigma$ & 3-$\sigma$& rate \\
\hline
\multicolumn{14}{c}{First situation: strong constraint on inclination and mass} \\
\multicolumn{13}{l|}{Primary dataset:} \\
30 Ari B b & 335.1 & 9.878 & 0.1728 & 1.777 & $2.391_{-0.444}^{+0.592}$ & $(1.347,5.948)$ & $4.181_{-0.931}^{+1.031}$ & $(1.643,7.584)$ &$147.4_{-29.5}^{+41.3}$ & $(78.92,412.5)$ & $11.14_{-1.47}^{+1.05}$ & $(6.888,16.46)$ & 0.2321  \\
HD 114762 b & 83.92 & 11.64 & 0.1161 & 1.088 & $1.339_{-0.353}^{+0.302}$ & $(0.4206,2.597)$ & $4.940_{-0.942}^{+1.773}$ & $(2.517,15.76)$ &$147.0_{-42.0}^{+39.3}$ & $(41.93,323.9)$ & $9.350_{-1.413}^{+1.614}$ & $(5.704,22.37)$ & 0.1758  \\
HD 141937 b & 653.2 & 9.475 & 0.3955 & 0.9337 & $1.126_{-0.401}^{+0.272}$ & $(0.3798,2.047)$ & $20.52_{-4.16}^{+12.47}$ & $(11.01,88.85)$ &$27.42_{-9.86}^{+6.78}$ & $(9.061,50.94)$ & $24.38_{-0.61}^{+1.29}$ & $(22.82,29.59)$ & 0.1340  \\
HD 148427 b & 331.5 & 1.144 & 0.01182 & 1.092 & $1.329_{-0.310}^{+0.332}$ & $(0.4964,2.936)$ & $0.5120_{-0.1082}^{+0.1635}$ & $(0.2314,1.413)$ &$136.5_{-33.7}^{+37.2}$ & $(47.74,336.0)$ & $12.34_{-1.32}^{+1.26}$ & $(8.368,24.50)$ & 0.1866  \\
HD 5388 b & 777.0 & 1.965 & 0.05154 & 1.365 & $2.182_{-0.255}^{+0.330}$ & $(1.519,3.923)$ & $1.356_{-0.191}^{+0.195}$ & $(0.7610,2.050)$ &$87.02_{-10.80}^{+13.99}$ & $(60.02,163.3)$ & $14.95_{-1.82}^{+4.58}$ & $(10.81,23.13)$ & 0.2394  \\
HD 6718 b & 2496 & 1.559 & 0.1087 & 1.121 & $4.201_{-0.881}^{+1.092}$ & $(1.982,9.951)$ & $1.488_{-0.310}^{+0.410}$ & $(0.6181,3.688)$ &$62.79_{-13.80}^{+16.98}$ & $(27.74,159.7)$ & $21.22_{-5.60}^{+1.20}$ & $(9.348,24.13)$ & 0.2073  \\
\tablefootmark{$\dagger$}HD 96127 b & 647.3 & 4.007 & 0.01116 & 1.124 & $0.6640_{-0.6457}^{+0.7888}$ & $(0.0009983,2.231)$ & $1.364_{-0.763}^{+38.527}$ & $(0.4084,89.54)$ &$190.2_{-184.0}^{+284.1}$ & $(3.359,679.2)$ & $8.008_{-3.522}^{+16.829}$ & $(2.526,26.81)$ & 0.05775  \\
HIP 65891 b & 1084 & 6.001 & 0.04197 & 1.146 & $2.040_{-0.341}^{+0.409}$ & $(1.103,4.026)$ & $1.184_{-0.207}^{+0.256}$ & $(0.5929,2.161)$ &$312.3_{-57.4}^{+74.2}$ & $(168.4,713.7)$ & $10.85_{-0.63}^{+0.65}$ & $(7.196,13.52)$ & 0.2000 \\
\multicolumn{13}{l|}{} \\
\multicolumn{13}{l|}{Secondary dataset:} \\
HD 16760 b & 465.1 & 13.29 & 0.2531 & 2.990 & $4.467_{-0.758}^{+1.053}$ & $(2.758,7.364)$ & $3.164_{-0.762}^{+0.810}$ & $(1.796,5.549)$ &$291.9_{-69.4}^{+120.7}$ & $(151.8,580.9)$ & $5.154_{-1.124}^{+0.926}$ & $(2.517,8.215)$ & 0.2521 \\
\hline
\multicolumn{14}{c}{Second situation: lower and upper limits on inclination and mass} \\
\multicolumn{13}{l|}{Primary dataset:} \\
HAT-P-21 b & 4.124 & 4.073 & 0.0007560 & 0.9171 & $<$0.0014260  & $<$0.01209 & $>$32.88 & $>$4.067 & $<$7.542 & $<$57.72 & $>$14.79 & $>$13.58 & 0.009911  \\
HD 132563 B b & 1544 & 1.492 & 0.03442 & 0.8536 & $<$0.07070  & $<$2.621 & $>$29.25 & $>$0.8231 & $<$3.050 & $<$111.1 & $>$25.10 & $>$11.03 & 0.01422  \\
HD 154857 b & 408.6 & 2.248 & 0.02508 & 0.9309 & $<$0.05492  & $<$1.499 & $>$27.03 & $>$1.014 & $<$4.918 & $<$134.9 & $>$24.71 & $>$13.10 & 0.009493  \\
HD 154857 c & 3452 & 2.579 & 0.1193 & 0.9309 & $<$0.2729  & $<$8.817 & $>$25.98 & $>$0.9194 & $<$5.905 & $<$175.3 & $>$27.78 & $>$10.97 & 0.009493  \\
HD 164595 b & 40.00 & 0.05078 & 0.0003920 & 0.9341 & $<$0.0008560  & $<$0.1093 & $>$27.67 & $>$0.2476 & $<$0.11103 & $<$12.86 & $>$24.46 & $>$22.70 & 0.01065  \\
HD 177830 b & 410.1 & 1.320 & 0.01953 & 0.8723 & $<$0.04035  & $<$1.278 & $>$28.82 & $>$1.019 & $<$2.738 & $<$78.92 & $>$24.47 & $>$13.43 & 0.01443  \\
HD 185269 b & 6.838 & 0.9542 & 0.001040 & 0.9694 & $<$0.002002  & $<$0.01755 & $>$31.18 & $>$4.624 & $<$1.820 & $<$12.34 & $>$19.00 & $>$18.87 & 0.005830  \\
HD 190228 b & 1136 & 5.942 & 0.1278 & 0.8628 & $<$0.5136  & $<$2.375 & $>$14.25 & $>$3.188 & $<$24.418 & $<$111.4 & $>$27.35 & $>$14.34 & 0.02329  \\
HD 197037 b & 1036 & 0.8073 & 0.04322 & 0.9947 & $<$0.12112  & $<$3.087 & $>$21.09 & $>$0.8682 & $<$2.2696 & $<$55.87 & $>$27.13 & $>$22.28 & 0.008977  \\
HD 4203 b & 431.9 & 2.082 & 0.02595 & 0.8539 & $<$0.05646  & $<$1.390 & $>$27.46 & $>$1.116 & $<$4.533 & $<$110.2 & $>$24.02 & $>$11.82 & 0.02863  \\
HD 7449 b & 1275 & 1.313 & 0.07578 & 0.9430 & $<$0.16248  & $<$6.525 & $>$30.30 & $>$0.8955 & $<$2.845 & $<$104.2 & $>$27.17 & $>$11.48 & 0.01279  \\
\tablefootmark{$\dagger$}HD 95127 b & 482.0 & 5.036 & 0.006734 & 1.220 & $<$0.017628  & $<$0.3977 & $>$26.57 & $>$2.027 & $<$11.863 & $<$170.2 & $>$23.04 & $>$6.939 & 0.001910  \\
\tablefootmark{$\dagger$}K2-34 b & 2.996 & 1.683 & 0.0001870 & 0.9982 & $<$0.0004070  & $<$0.002572 & $>$28.51 & $>$3.946 & $<$3.525 & $<$23.91 & $>$13.63 & $>$12.40 & 0.002133  \\
WASP-11 b & 3.722 & 0.5398 & 0.0002100 & 1.064 & $<$0.0004260  & $<$0.004463 & $>$29.21 & $>$3.117 & $<$1.1059 & $<$9.719 & $>$15.62 & $>$15.27 & 0.003389  \\
WASP-131 b & 5.322 & 0.2724 & 0.00006800 & 0.9476 & $<$0.00014300  & $<$0.001557 & $>$28.30 & $>$4.120 & $<$0.5756 & $<$3.710 & $>$15.31 & $>$14.62 & 0.01006  \\
WASP-157 b & 3.952 & 0.5592 & 0.00008400 & 0.8699 & $<$0.0001670  & $<$0.001673 & $>$30.85 & $>$3.765 & $<$1.1127 & $<$8.444 & $>$14.31 & $>$13.06 & 0.01517  \\
WASP-17 b & 3.735 & 0.5077 & 0.00005200 & 0.8509 & $<$0.00010000  & $<$0.0007792 & $>$32.14 & $>$5.985 & $<$0.9619 & $<$4.768 & $>$13.48 & $>$12.70 & 0.006831  \\
\multicolumn{13}{l|}{} \\
\multicolumn{13}{l|}{Secondary dataset:} \\
\tablefootmark{$\dagger$}K2-110 b & 13.86 & 0.05293 & 0.00006000 & 1.278 & $<$0.00069800  & $<$0.01073 & $>$4.96 & $>$0.3551 & $<$0.62324 & $<$6.732 & $>$18.03 & $>$17.49 & 0.0007182  \\
\hline
\end{tabular}
\tablefoot{\\
\tablefoottext{$\dagger$}{After 1,000,000 iterations MCMC did not reach convergence, with a final maximum autocorrelation length larger than $N_\text{step}/50$.}
}
\end{sidewaystable*}
\twocolumn

\begin{table*}\tiny\centering
\caption{\label{tab:results_nondet} GASTON results for 27 exoplanet candidates from the non-detection sample. Their 3-$\sigma$ upper-limit on mass is 
smaller than 13.5\,M$_\text{J}$ and the convergence criterion $N_\text{step}/\text{max}(\tau_\lambda$$\geq $$50$. We expand this table in the Appendix, 
Table~\ref{tab:results_nondet_comp}, with the rest of the exoplanet candidates among the non-detection sample and for which 
$M_{\text{max,} 3\sigma}$>13.5\,M$_\text{J}$.}
\begin{tabular}{@{}lcccccccccc@{}}
Planet name & Period & $m\sin i$ & $a\sin i$ & $\epsilon$ & \multicolumn{1}{c}{$a_\text{phot}$} & \multicolumn{1}{c}{$I_c$} & \multicolumn{1}{c}{$M_\text{c,true}$}  & \multicolumn{1}{c}{$\Delta V$} & MCMC  \\
			& (days) & (M$_\text{J}$) & (mas)  & (mas)  	& (mas)		 & ($^\circ$) &   \multicolumn{1}{c}{(M$_\text{J}$)}  &   Acceptance   \\
	 		&		&			&		&		 & 3-$\sigma$ &		3-$\sigma$		&	3-$\sigma$  & 3-$\sigma$& rate\\
\hline
\multicolumn{10}{c}{3-$\sigma$ limits, with $M_\text{c,true}$$<$$13.5$\,M$_\text{J}$ at 3-$\sigma$} \\
\multicolumn{10}{l}{Primary dataset: 24 confirmed exoplanets} \\
BD -06 1339 b & 3.873 & 0.02680 & 0.00007800 & 0.5190 & $<$0.03085 & $>$0.3243 & $<$4.792 & $>$19.77 & 0.1357  \\
BD -08 2823 b & 5.600 & 0.04594 & 0.00008000 & 0.4011 & $<$0.02921 & $>$0.2746 & $<$9.277 & $>$18.80 & 0.1914  \\
HD 10180 c & 5.760 & 0.04151 & 0.00006200 & 0.4662 & $<$0.01925 & $>$0.2841 & $<$8.626 & $>$19.26 & 0.1585  \\
HD 10180 d & 16.36 & 0.03766 & 0.0001120 & 0.4662 & $<$0.06159 & $>$0.2005 & $<$10.37 & $>$20.77 & 0.1592  \\
HD 10180 g & 602.0 & 0.06738 & 0.002221 & 0.4662 & $<$0.4390 & $>$0.3663 & $<$10.62 & $>$25.92 & 0.1496  \\
HD 125595 b & 9.674 & 0.04168 & 0.0001510 & 0.4106 & $<$0.08048 & $>$0.2243 & $<$11.11 & $>$20.48 & 0.2058  \\
HD 154345 b & 3342 & 0.9569 & 0.2360 & 0.3491 & $<$3.454 & $>$4.652 & $<$11.94 & $>$26.02 & 0.1737  \\
HD 175607 b & 29.03 & 0.02626 & 0.0001440 & 0.4171 & $<$0.07016 & $>$0.1865 & $<$7.728 & $>$21.26 & 0.1692  \\
HD 176986 b & 6.490 & 0.02002 & 0.00005500 & 0.2559 & $<$0.02388 & $>$0.2581 & $<$4.681 & $>$19.98 & 0.1911  \\
HD 176986 c & 16.82 & 0.02814 & 0.0001450 & 0.2559 & $<$0.05738 & $>$0.2419 & $<$6.601 & $>$21.35 & 0.1939  \\
HD 179079 b & 14.48 & 0.08378 & 0.0001240 & 0.3794 & $<$0.03729 & $>$0.3766 & $<$13.20 & $>$19.21 & 0.1853  \\
HD 181433 b & 9.374 & 0.02373 & 0.00008600 & 0.2972 & $<$0.04037 & $>$0.2542 & $<$5.376 & $>$20.49 & 0.1210  \\
HD 181433 c & 962.0 & 0.6404 & 0.05114 & 0.2972 & $<$0.6246 & $>$5.196 & $<$6.944 & $>$27.15 & 0.1185  \\
HD 181433 d & 2172 & 0.5355 & 0.07359 & 0.2972 & $<$1.747 & $>$2.665 & $<$11.28 & $>$25.41 & 0.1186  \\
HD 215152 b & 5.760 & 0.005720 & 0.00001900 & 0.3057 & $<$0.009679 & $>$0.1871 & $<$1.779 & $>$20.36 & 0.1786  \\
HD 215152 c & 7.282 & 0.005408 & 0.00002100 & 0.3057 & $<$0.01000 & $>$0.2035 & $<$1.475 & $>$20.70 & 0.1799  \\
HD 215152 d & 10.86 & 0.008816 & 0.00004400 & 0.3057 & $<$0.02144 & $>$0.1998 & $<$2.424 & $>$21.27 & 0.1783  \\
HD 215152 e & 25.20 & 0.009052 & 0.00008000 & 0.3057 & $<$0.04964 & $>$0.1869 & $<$3.069 & $>$22.48 & 0.1794  \\
HD 215497 b & 3.934 & 0.02085 & 0.00002600 & 0.4831 & $<$0.01194 & $>$0.2301 & $<$4.999 & $>$18.50 & 0.1164  \\
HD 7199 b & 615.0 & 0.2950 & 0.01192 & 0.3742 & $<$0.5161 & $>$1.496 & $<$11.45 & $>$25.73 & 0.1817  \\
HD 7924 b & 5.398 & 0.02737 & 0.0001050 & 0.5727 & $<$0.04275 & $>$0.2472 & $<$6.476 & $>$20.81 & 0.1021  \\
HD 7924 c & 15.30 & 0.02484 & 0.0001900 & 0.5727 & $<$0.08133 & $>$0.2575 & $<$5.546 & $>$22.31 & 0.1015  \\
HD 7924 d & 24.45 & 0.02038 & 0.0002130 & 0.5727 & $<$0.1022 & $>$0.2188 & $<$4.934 & $>$22.99 & 0.1091  \\
HIP 57274 b & 8.135 & 0.03657 & 0.0001300 & 0.4507 & $<$0.06026 & $>$0.1980 & $<$10.57 & $>$20.32 & 0.1738  \\
\\
\multicolumn{10}{l}{Secondary dataset: 3 confirmed exoplanets} \\
HD 40307 b & 4.311 & 0.01291 & 0.00006000 & 0.3337 & $<$0.03068 & $>$0.2022 & $<$3.741 & $>$20.72 & 0.1891  \\
HD 40307 c & 9.620 & 0.02115 & 0.0001690 & 0.3337 & $<$0.07895 & $>$0.2109 & $<$5.879 & $>$21.69 & 0.1899  \\
HD 40307 d & 20.46 & 0.02808 & 0.0003710 & 0.3337 & $<$0.1125 & $>$0.2855 & $<$5.586 & $>$22.81 & 0.1927  \\
\hline
\end{tabular}
\end{table*}

Among the 200 other candidate planets, as summarized in Table~\ref{tab:results_nondet_comp}, 103 companions can be confirmed substellar but may be as 
massive as brown dwarfs with a mass strictly smaller than 85\,M$_\text{J}$ at 3-$\sigma$, and 59 others have a mass upper-limit within the M-dwarf domain. For 
the remaining 48 companions, GASTON could not converge within the 50,000 steps, with an autocorrelation length larger than 1000. At the end of the GASTON 
run, the posterior distributions for all of them led to an upper-limit on the mass larger than 13.5\,M$_\text{J}$. This non-convergence is due to a large 
astrometric excess noise but smaller than the detection limit. Simulations are less often compatible with $\varepsilon_\text{DR1}$, GASTON thus needs more time 
to converge. Their nature is undetermined between planet, BD and M-dwarf. We do not publish GASTON results for those 48 candidates.

While most of companions with a mass possibly greater than 13.5\,M$_\text{J}$ have large orbital periods, 30 of them have an orbital period smaller than 
100\,days. Those are possible BD located within the driest region of the brown-dwarfs detection desert (Kiefer et al. 2019). They are particularly interesting 
objects that need to be further characterized in order to better constrain the shores of the BD mass-period phase space. 

\section{Discussion}
\label{sec:discussion}

\subsection{A revised mass for 9 companions}

\begin{figure}
\includegraphics[width=89.3mm]{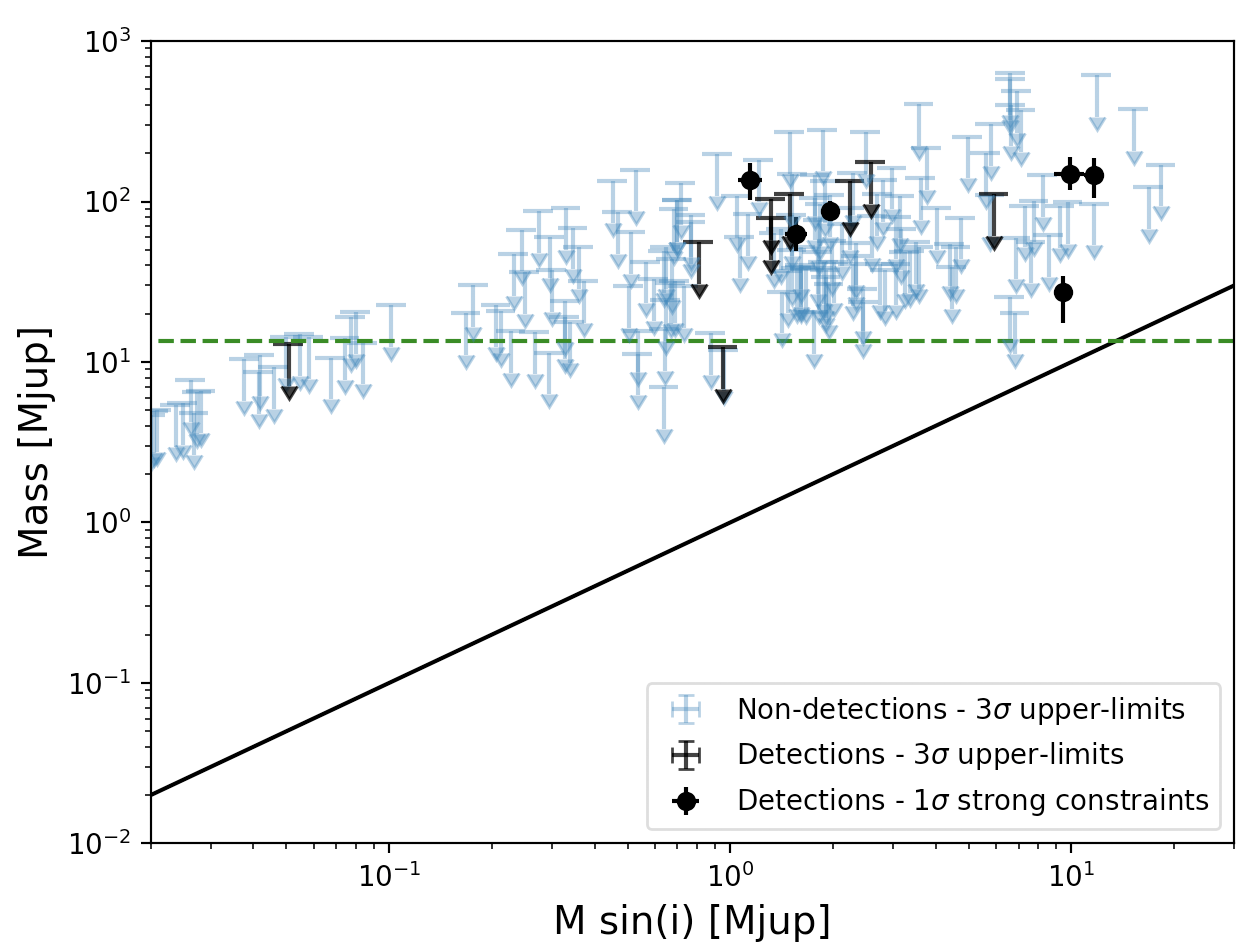}
\caption{\label{fig:mass_period_init} Masses and upper-limits derived with GASTON for 191 companions directly compared to the $m\sin i$ derived from RVs. 
The black solid line represents the relation $M_\text{true}$=$m\sin i$ obtained for an edge-on inclination of the orbits. The green dashed line marks the 
13.5-M$_\text{J}$ deuterium-burning limit.}
\end{figure}

Figure~\ref{fig:mass_period_init} summarizes the corrected mass derived with GASTON compared to the initial $m\sin i$ as given in the Exoplanets.org database. 
The firm measurements for the 9 companions identified in Section~\ref{sec:case1} lead to true masses significantly different from the $m\sin i$ with an 
non edge-on inclination. Their revised mass is generally comprised between 10 and 500\,M$_\text{J}$, as are the 3-$\sigma$ upper-limits reported for companions 
from situation \#2 and in the non-detection sample.

This shows that Gaia will be best at detecting astrometric motions due to companions beyond $\sim$10\,M$_\text{J}$. But with improved precision in the future 
releases and the use of time series, it will certainly allow the detection of Jupiter mass planets. 

\subsection{Small inclinations $<$4$^\circ$}
\label{sec:small_inc}
To our knowledge, no exoplanet RV candidate from the \verb+exoplanets.org+ database were yet found with an inclination strictly lower than 4$^\circ$. 
The exoplanet with the smallest known orbital inclination is Kepler-419\,c with $I_c$=2.5$\pm$3$^\circ$, thanks to transit timing variations (Dawson et al. 2014). In 
Table~\ref{tab:results_det}, among the 9 non-transiting systems with a firmly detected inclination, and accounting for their 3-$\sigma$ bounds, we find
zero companion with $I_c$ strictly smaller than $1^\circ$, one companion, HD\,148427\,b, with $I_c$ strictly smaller than $2^\circ$, and four others with 
$I_c$ strictly smaller than 4$^\circ$. Many other companions from the detection and non-detection samples could have 
such small inclinations, but also posibly larger than 1, 2 or $4^\circ$. Assuming isotropy of orbits within the $\sim$600 known 
non-transiting RV exoplanets in the \verb+exoplanets.org+ database leads to less than 0.4 orbits with $I_c$$\leq $$2^\circ$ and less than 1.5 orbits with $I_c$$\leq $$4^\circ$. 
Finding more than 1 system with an orbit 
less inclined than 2$^\circ$ and more than 4 with $I_c$$<$$4^\circ$ suggests that the distribution of inclinations within exoplanet candidates deviates from a uniform 
distribution at least below $4^\circ$. This questions the isotropy of orbits within the population of discovered RV exoplanets advocated in e.g. Zucker \& Mazeh (2001b), 
Jorissen et al. (2001) and Tabachnik \& Tremaine (2002). It was indeed already noticed in Halbwachs et al. (2000) and Han et al. (2001) that RV-planet systems 
are possibly biased towards small $\sin i$. 

The uniform distribution of inclinations certainly applies on a larger sample of systems than just RV exoplanets hosting systems. SB1 binary companions 
($M_c$$<$$0.6$\,$M_\star$) on orbits with inclinations $<$2$^\circ$ would most likely fall within exoplanetary domain. A $\sin i$$<$$0.035$ 
would indeed lead to $M \sin i$$<$$20 (M_\star/M_\odot)$\,M$_\text{J}$. Several thousands of binary 
systems and stars were, and are still being, followed-up for RV variations for many years. About 1300 SB1 binaries are collected in the SB9 database 
(Pourbaix et al. 2004) and about 600 RV systems (excluding all transiting planets that are biased to 90$^\circ$). Probably twice as many are still being 
followed-up, neither characterized nor published yet. We thus estimate the full population of RV-monitored systems possibly, or actually, harboring planets or 
hidden binary component to reach at least 4000 individuals, among which FGK stars are the dominating class of stellar primaries. Assuming isotropy of orbits in 
this larger sample, we expect to observe at least $\sim$0.6 systems with an inclination $\leq$1$^\circ$, $\sim$2.4 systems with $I_c$$\leq $$2^\circ$ and 
$\sim$9.7 systems with $I_c$$\leq $$4^\circ$. This is in better agreement with our findings, validating the GASTON determination of orbital inclinations for RV 
companions with the Gaia DR1.

\subsection{Systems with edge-on transiting orbits}
\label{sec:edgeon}

In Section~\ref{sec:noise_distribution}, we found 9 over 246 systems (i.e. 3.7\% of them) with transiting-only planets to have an astrometric excess noise larger 
than the defined thresholds for a significant astrometric motion. Six of them are systems around sources from the primary dataset. 
Given their edge-on orbit, with an expected astrometric semi-major axis of the photocenter on the order of a few $\mu$as, it is technically impossible to detect 
their host star's reflex motion with Gaia. But the measurement of $\varepsilon_\text{DR1}$ beyond the threshold suggests on the contrary that a significant 
astrometric motion was actually detected. 

Running GASTON brings a solution to this inconsistency. For 7 among 9 companions, an edge-on astrometric orbit is compatible with the value of $\varepsilon_\text{DR1}$, 
because instrumental and measurement noises can pile up to produce an astrometric excess noise above the threshold. However, this compatibility is far from being
frequent in the MCMC runs, as shown by the low acceptance ratio in both cases ($<$1.5\%; see  Table~\ref{tab:results_det}). Could it be that 
$\varepsilon_\text{DR1}$ is actually not due to instrumental and measurement noise for some of them, and leads to truly incompatible astrometric 
motion?

To test this possibility, we simulated many Gaia observations of non-accelerating stars, with $N_\text{FoV}$ and $N_\text{AL}$ drawn from the sample of 133 
systems with transiting-only planets from the primary dataset, and the 113 systems with transiting-only planets from the secondary dataset. In the primary dataset, 
this results into 6 over 133 astrometric excess noise values beyond 0.85\,mas, produced only from noise, with a probability of 0.0014, i.e. below 3\,$\sigma$. 
Producing 5 over 133 astrometric excess noise values beyond 0.85\,mas has a probability of 0.0072, i.e. above 3\,$\sigma$. In the secondary dataset, 
the probability to produce 2 or 3 over 113 astrometric excess noise values beyond 1.2\,mas is smaller than 0.0001, while producing 1 over 113 is significantly 
more likely with a probability of 1.2\%. 

Thus for at least one system from the primary dataset, possibly K2-34 with the lowest MCMC acceptance rate of 0.002, Gaia detected a signal that 
cannot be fully explained by the combination of the published orbit and noise. This could be the sign of an unseen and unknown outer companion in the 
system of K2-34, or a measurable effect of outliers (see the discussion in Section~\ref{sec:case5} for WASP-43\,b and WASP-156\,b). 

In the secondary dataset, no more than one system among 113 could be simulated with an astrometric excess noise as large as 1.2\,mas. Consistently, we recall
obtaining a marginal compatibility of GASTON simulations with $\varepsilon_\text{DR1}$ only for K2-110\,b (Section~\ref{sec:case3}), and no compatibility at all 
for two other planets, WASP-43\,b and WASP-156\,b (Section~\ref{sec:case5}). We conclude that noise is the likely explanation for the astrometric excess noise of 
K2-110\,b. As already discussed in Section~\ref{sec:case5}, the most reasonable explanations for the large inconsistent $\varepsilon_\text{DR1}$ of WASP-43 
and WASP-156 are either unmodeled outliers, or the presence of an outer companion in both systems. 

We conclude that follow-ups of K2-34, WASP-43 and WASP-156 should be conducted to search for outer companions in these edge-on systems.

\subsection{Outer companions}
\label{sec:outer}

In the cases with the lowest acceptance rates in Table~\ref{tab:results_det}, it could be that the astrometric signal is better 
explained by the influence of another outer companion to the system, especially if as for HD\,4203, a long-period RV-drift is detected (Kane et al. 2014). With a 
minimum mass of at least 2\,M$_\text{J}$, this outer companion with an orbital period of several thousands of days (tens of year) might also be at the origin of the 
astrometric signal. 

As mentioned in Section~\ref{sec:edgeon}, the astrometric excess noise of few compact edge-on systems (K2-34, WASP-43 and WASP-156) is 
difficult, and even impossible, to produce with our simulations. Thus, while outliers could explain the discrepancy of the RV orbit and Gaia astrometry, a more simple 
explanation could be the presence of an outer companion. There are no clues of long RV drifts in neither of these systems, but this does not invalidate the hypothesis 
of an outer companion because the orbital inclination can be adjusted to make the RV signal vanish.

More generally, for any system in situation \#1 leading to a mass measurement of an RV companion, the presence of an unknown outer companion cannot be 
totally excluded. Nevertheless, as far as we know these systems, the solution with the less complexity is preferred, i.e. the 
known RV-orbit with a realistic inclination is responsible for the Gaia measurement of the star motion. Fine-tuned mass, semi-major axis 
and inclination of an unknown unseen companion's orbit would be necessary to explain the astrometric excess noise, while using the existing known companion 
on a known orbit only require to fit a single parameter, the inclination. 

We have shown in Section~\ref{sec:noise_distribution} that the astrometric excess noise is not correlated to the presence of an RV-drift, and thus to the 
presence of an outer companion on an undetermined orbit. Thus, a large astrometric excess noise does not imply the presence of an outer companion and 
conversely the presence of an outer companion does not imply a large astrometric excess noise. It was already shown for HD\,114762\,b (Kiefer 2019) that the 
binary companion HD\,114762\,B, a low-mass star at several hundred au separation and with orbital period much larger than the 
Gaia DR1 campaign duration of 416\,days, only has a minor impact on the motion of HD\,114762\,A. It was better explained by a small orbital inclination 
and larger mass of HD\,114762\,A\,b. 

Solving this issue is not the scope of the present paper, but learning from the specific case of HD\,114762, we expect that outer companions, moreover not 
observed in the RV variations, with period much greater than 416\,days could be neglected. 

\subsection{Comparison with already published mass}

\begin{figure}
\includegraphics[width=89.3mm]{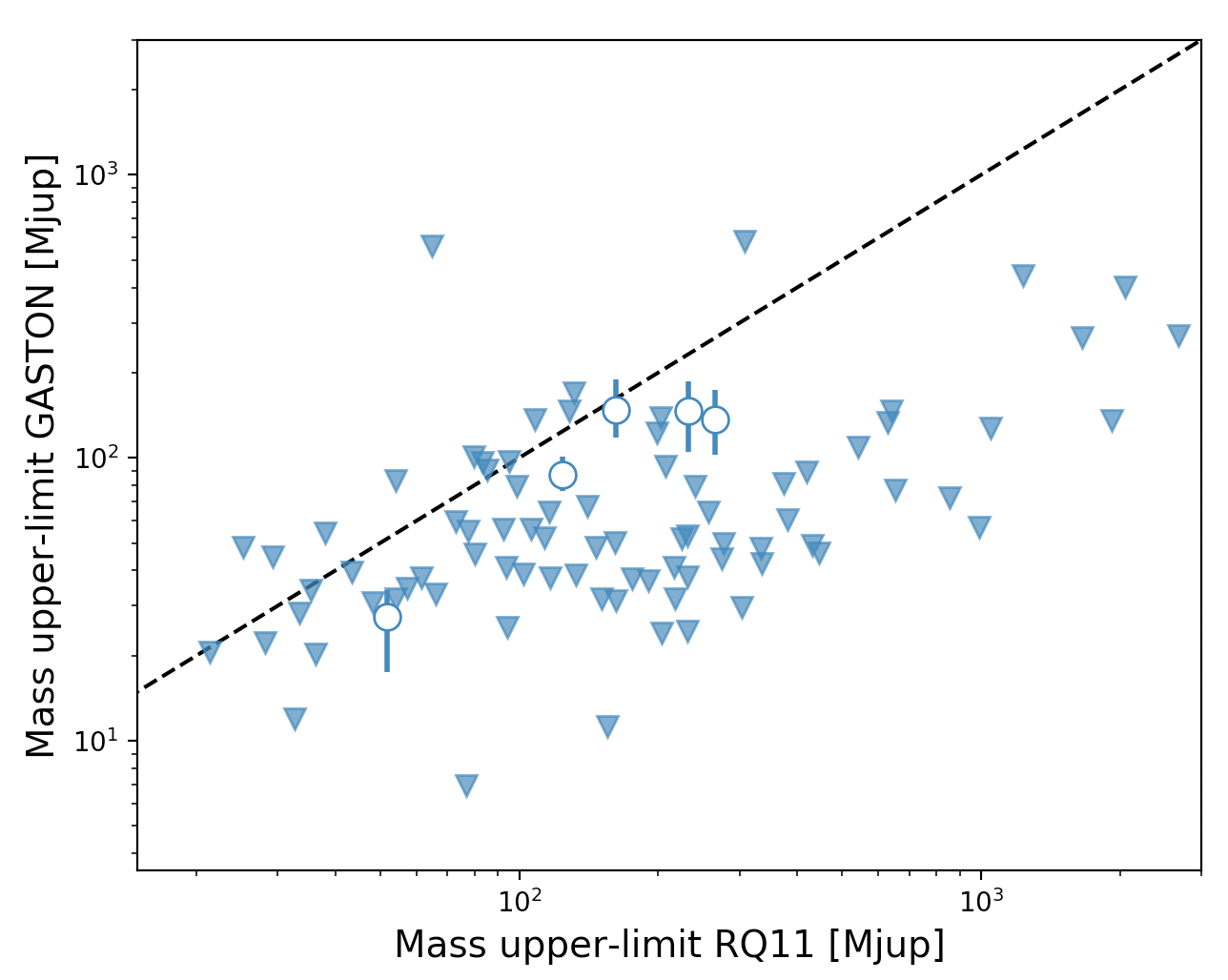}
\caption{\label{fig:comp_reffert} A comparison plot of GASTON mass measurements (open blue circles) and other 3-$\sigma$ upper-limits (plain blue triangles) 
with Reffert \& Quirrenbach (2011), or RQ11, 3-$\sigma$ mass upper-limits for the same companions. The black dashed line represents  the relation
$M_{3\sigma,\text{GASTON}}$=$M_{3\sigma,\text{RQ11}}$.}
\end{figure}

The true mass for 86 exoplanet candidates of our samples were also constrained in Reffert \& Quirrenbach (2011) using Hipparcos-2 data.
The results of the two studies are compared together in Fig.~\ref{fig:comp_reffert}. This comparison shows that GASTON leads to better 
contraints with generally lower upper-limits on the mass for 69 over the 86 companions. The Gaia DR1 astrometric excess noise is thus compatible the Hipparcos-2
astrometry even revealing smaller scatter and better astrometric precision. Among these 86 exoplanet candidates, Table~\ref{tab:comp_reffert} lists 5 companions 
with a well-constrained mass in the present study -- 30\,Ari\,B\,b, HD\,114762\,b, HD\,141937\,b, 
HD\,148427\,b, and HD\,5388\,b -- for which Reffert \& Quirrenbach (2011) analysis only led to upper-limits on mass. The true mass and inclination that we obtain 
with GASTON all stand within the bounds they derived. We also list 5 companions for which GASTON could only derive limits -- HD\,190228\,b, HD\,87883\,b,
HD\,142022\,b, HD\,181720\,b, and HD\,131664\,b -- but Reffert \& Quirrenbach (2011) published upper and lower bounds for both inclination and mass. 
We reduce the interval of possible mass for HD\,142022 b and HD\,181720\,b, now respectively within 4.6--39\,M$_\text{J}$ and 6--32\,M$_\text{J}$, i.e. 
in the giant planet or BD domain. We also confirm the upper-bound on HD\,87883\,b mass, which ranges within 3--21\,M$_\text{J}$ at 3-$\sigma$. HD\,87883\,b is thus 
most likely a giant planet on a long-period 7.5-years orbit. 

\begin{table*}\centering
\caption{\label{tab:comp_reffert} Comparison of results in the present studies to 1 or 3-$\sigma$ upper-limits published in other articles. Inclinations
should be compared modulo $180^\circ$ since in the present study the prograde or retrograde orientation of the orbital motion cannot be determined. We present 
the 1-$\sigma$ confidence interval for the inclinations and masses when they are well-constrained in the present study and obtained with a $\sin I_c$ prior
distribution on the inclination, and the 3-$\sigma$ limits obtained with a flat prior distribution (Table~\ref{tab:results_nondet_comp}) otherwise.}
\begin{tabular}{l|cc|cc|cc}
Companion & \multicolumn{2}{c|}{Reffert \& Quirrenbach (2011)} & \multicolumn{2}{c|}{Other publications} &  \multicolumn{2}{c}{Present study} \\
	           & $I_c$  & true mass &  $I_c$  & true mass & $I_c$ & true mass\\
	           & (3-$\sigma;\,^\circ)$ & (3-$\sigma$;\,M$_\text{J}$) & (1-$\sigma;\,^\circ)$ & (1-$\sigma$;\,M$_\text{J}$) & (1 or 3$\sigma;\,^\circ)$ & (1 or 3$\sigma$;\,M$_\text{J}$) \\
\hline
\multicolumn{7}{c}{Well-constrained mass with GASTON at 1-$\sigma$} \\
\\
30\,Ari\,B\,b & (3.8,174.8) & $<$162.2 & & & $4.2^{+1.0}_{-0.9}$  & $147^{+41}_{-29}$\\
HD\,114762\,b & (4.1,176.7) & $<$233.0 &  \tablefootmark{a}$4.87^{+1.11}_{-0.91}$ & \tablefootmark{a}$141^{+35}_{-28}$ & $4.9^{+1.8}_{-0.9}$ & $147^{+39}_{-42}$) \\
HD\,141937\,b & (11.0,166.2) & $<$51.7 & & & $20.5^{+12.5}_{-4.2}$ & $27.4^{+6.8}_{-9.9}$ \\
HD\,148427\,b & (4.8,176.6)\tablefootmark{$\dagger$} & $<$265 & & & $0.51^{+0.16}_{-0.11}$  & $137^{+37}_{-34}$ \\
HD\,5388\,b & (1.0,179.0) & $<$124.3 & \tablefootmark{b} 178.3$^{+0.4}_{-0.7}$  & \tablefootmark{b} 62.2$\pm$19.9 & $1.4\pm0.2$ & $87^{+14}_{-11}$ \\
\hline
\multicolumn{7}{c}{3-$\sigma$ limits with GASTON} \\
\\
HD 33636 b & (2.9,176.8) & $<$207.3	& \tablefootmark{c}4.0$\pm$0.1	& \tablefootmark{c}142$\pm$11	& $>$6.02 	& $<$93.5 \\ 
HD 87883 b & (4.9,34.9)  & (3.1,21.4)   & &  & $>$4.9 & $<$21 \\
HD 92788 b & (2.5,178.0) & $<$113.6  & \tablefootmark{d}6.3 ; \tablefootmark{e}(0,22)  & \tablefootmark{d}45 ; \tablefootmark{e}(9,28) & $>$4.0  & $<$52 \\
HD 102195 b & (0.0	,180.0) & --- &  \tablefootmark{f}(72.5,84.79) & \tablefootmark{f}0.46$\pm$0.03  & $>$0.21 & $<$133 \\
HD 128311 c & (7.5,171.1) & $<$25.2 & \tablefootmark{g}55.950$\pm$14.553  & \tablefootmark{g}3.789$^{+0.924}_{ -0.432}$ & $>$4.0 & $<$49 \\
HD 130322 b & (0.0	,180.0) & --- & \tablefootmark{e}76$^{+14}_{-42}$ & \tablefootmark{e}1.1$^{+1.0}_{-0.1}$  & $>$0.59 & $<$108 \\
HD 131664 b & (153.9,171.5)	& (42.3,131.6)	& \tablefootmark{h}55$\pm$33 & \tablefootmark{h}23$^{+26}_{-5}$ & $>$6.6 & $<$170 \\
HD 136118 b & (138,172.7)	& $<	$95.3 & \tablefootmark{i}163.1$\pm$3.0	&  \tablefootmark{i}42$^{+11}_{-18}$ & $>$7.25	& $<$97 \\
HD 142022 b & (4.2,49.1) & (4.6,102.2) &	&	& $>$4.348 & $<$38.84 \\
HD 154345 b & (1.7	,178.1) &$<$32.6 & \tablefootmark{e}50$^{+40}_{-26}$ & \tablefootmark{e}1.2$^{+1.3}_{-0.4}$  & $>$4.7 & $<$12 \\
HD 177830 b & (0.9,179.6)  & $<$225.2 & \tablefootmark{d}1.3 & \tablefootmark{d}55 & $>$1 & $<$79 \\
HD 181720 b & (0.1	,3.4) & (6.1,217.9)  & & & $>$0.68 & $<$32 \\
HD 190228 b & (2.5,40.8) & (9.1,142.9) & \tablefootmark{j}4.3$^{+1.8}_{-1.0}$ & \tablefootmark{j}49$\pm$15 & $>$1.56 & $<$111.4 \\
\hline
\end{tabular}
\tablefoot{\\
\tablefoottext{$\dagger$}{We believe there is an issue in Reffert \& Quirrenbach (2011) for this companion, since its $m\sin i$ is $\sim$1\,M$_\text{J}$ and the 
inclination and true mass 3-$\sigma$ constraints are not compatible with each other.} \\
\tablefoottext{*}{Adopting the $\sin I_c$ prior distribution on inclinations. With a flat prior, the 3-$\sigma$ limit on the mass of HD\,185269\,b increases to 
722\,M$_\text{J}$.} \\
\tablefoottext{a}{From Kiefer (2019) ; this values does not account for the astrometric motion of HD\,114762\,A due to the wide-binary component HD\,114762\,B.} \\ 
\tablefoottext{b}{From Sahlmann et al. 2011b.} \\
\tablefoottext{c}{From Bean et al. (2007) with astrometry from the HST/FGS.} \\
\tablefoottext{d}{From Han et al. (2001). No confidence intervals are given.} \\
\tablefoottext{e}{From Simpson et al. (2010). Derived from parent star rotation axis inclination, assuming coplanarity of stellar equator and planet orbit.} \\
\tablefoottext{f}{From Guilluy et al. (2019) by extracting the spectra emitted by this non-transiting planet.} \\
\tablefoottext{g}{From McArthur et al. (2014) with astrometry from the HST/FGS.} \\
\tablefoottext{h}{From Sozzetti \& Desidera (2010) with Hipparcos astrometry.}\\
\tablefoottext{i}{From Martioli et al. (2010) with astrometry from the HST/FGS.} \\
\tablefoottext{j}{From Sahlmann et al. (2011a)  with Hipparcos astrometry.} \\
}
\end{table*}

Besides Reffert \& Quirrenbach (2011), we found several other publications of true mass and inclinations for 12 companions. They are summarized on Table~\ref{tab:comp_reffert} and discussed individually below. 

\paragraph{\bf HD\,5388\,b} The mass of this exoplanet candidate was already measured to be 62.2$\pm$19.9\,M$_\text{J}$ with Hipparcos measurements 
(Sahlmann et al. 2011b). Our new mass estimation of 87.02$^{+13.99}_{-10.80}$\,M$_\text{J}$ is 2-$\sigma$ compatible with Sahlmann et al. (2011b) estimation. 
This companion is thus indeed a likely massive brown dwarf. 

\paragraph{\bf HD\,33636\,b} Bean et al. (2007) rejected the planetary nature of this candidate, with a mass determined in the M-dwarf domain 
$M$=140$\pm$11\,M$_\text{J}$ with an orbital inclination of $\sim$4$^\circ$. Interestingly, the small astrometric excess noise measured by Gaia 
$\varepsilon_\text{DR1}$=0.53\,mas leads to a probability for the mass of this companion to be higher than 93.5\,M$_\text{J}$ is
0.27\%. The mass measurements from Gaia and FGS astrometry are thus incompatible at 3$\sigma$. However, with an inclination of 4$^\circ$, our simulations 
could produce $\varepsilon_\text{DR1}$ smaller than 0.53\,mas with a probability of 0.7\%. Thus, the Gaia DR1 astrometric excess noise is compatible with 
Bean et al. (2007) results at 3$\sigma$. The disagreement between the parallax of this measured by FGS and Hipparcos ($\sim$35-36\,mas) with the parallax 
measured by Gaia ($\sim$34) may also explain the small $\varepsilon_\text{DR1}$ if part of the orbital motion was wrongly fitted as parallax motion.

\paragraph{\bf HD\,92788\,b} Han et al. (2001) proposed a true mass of 45\,M$_\text{J}$, with an inclination of 6.3$^\circ$ for this Jupiter-mass candidate
($m\sin i$=3.6\,M$_\text{J}$) on an Earth-like orbit ($P$=325\,days). 
Simpson et al. (2010) later proposed a derivation of the orbit inclination based on the assumption of coplanarity of the stellar equator and the companion orbit and 
by measuring the rotation speed of the star compared to its $v\sin i$. This led to a lower mass of 9-28\,M$_\text{J}$. This method is however not fully reliable as 
coplanarity of the stellar equator and the companion orbit is never a robust assumption. Both results are compatible with the 3-$\sigma$ limit that we derived 
here ($I_c$$>$$3.9^\circ$, $M_b$$<$$54$\,M$_\text{J}$) with $\varepsilon_\text{DR1}$=0.32\,mas. It confirms that this companion is most likely a 
brown-dwarf if not a massive planet.

\paragraph{\bf HD\,102195\,b} This companion on a 4-days orbit was determined as planetary by Guilluy et al. (2019) by extracting the emitted 
spectrum of its atmosphere. They determined a Jovian mass of 0.46\,M$_\text{J}$ with an orbit inclination of 72-85$^\circ$. GASTON cannot confirm the 
planetary nature of HD\,102195\,b, with a 3-$\sigma$ limit on its mass of 187\,M$_\text{J}$ and a minimum inclination of 0.15$^\circ$. 

\paragraph{\bf HD\,114762\,b} Its true mass was already published in Kiefer (2019). The results that we find here for this companion are compatible with those reported in 
Kiefer (2019) when HD\,114762\,B is not taken into account, validating that the modifications brought to GASTON (see Section~\ref{sec:gaston}) kept the results of such 
well-behaved case unchanged. The tentative estimation of the mass of this companion obtained by Han et al. (2001) from few 
Hipparcos points led to an inclination of 4.3$^\circ$ and a companion mass of 145\,M$_\text{J}$, in full agreement with our result.

\paragraph{\bf HD\,128311\,b \& c} This system has two known companions $b$ and $c$, with periods of respectively 454 and 924 days, in an almost 2:1 
resonance, with minimum masses of 1.5 and 3.2\,M$_\text{J}$. The outermost companion $c$ was shown to be most likely planetary by McArthur et al. (2014) using 
HST/FGS precise astrometry, with an orbit inclination of 55.95$\pm$14.55$^\circ$ and a mass of 3.789$^{+0.92}_{-0.43}$\,M$_\text{J}$. There are no constraint 
on the inclination or true mass of companion $b$, but an assumed coplanarity with planet $c$ orbit would imply a planetary nature as well with a mass close to 
2\,M$_\text{J}$. Coplanarity is not generic, and it remains thus possible that planet $c$ is actually circumbinary, possibly leading to an interesting configuration 
in 2:1 resonance. The non-detection of the astrometric motion of the host star by the Gaia DR1 astrometric excess noise with $\varepsilon_\text{DR1}$=0.6\,mas 
puts a 3-$\sigma$ upper-limit on the mass at 46 and 48\,M$_\text{J}$ for companions $b$ and $c$. We can therefore exclude a stellar nature for planet $c$ in 
agreement with McArthur et al. (2014), as well as for object $b$, but it could still be a brown dwarf.

\paragraph{\bf HD\,130322\,b} As for HD\,154345\,b, assuming the coplanarity of HD\,130322's equator and companion 
$b$ orbit, Simpson et al. (2010) proposes a mass of 1.1\,M$_\text{J}$ for this companion. The low astrometric excess 
noise of 0.3\,mas for this source allows deriving with GASTON a 3-$\sigma$ upper-limit on the mass of HD\,130322\,b of 136\,M$_\text{J}$. The planetary 
nature of this object cannot be confirmed here.

\paragraph{\bf HD\,131664\,b} This 5-years period candidate brown dwarf ($m\sin i$=18\,M$_\text{J}$) was characterized using Hipparcos astrometry by Sozzetti 
\& Desidera (2010) and Hipparcos-2 data by Reffert \& Quirrenbach (2011). Both found a possible small orbital inclinations for this companion down to $\sim$10-20$^\circ$. 
Sozzetti \& Desidera (2010) could not reject an edge-on inclination while Reffert \& Quirrenbach (2011) obtain inclinations smaller than 30$^\circ$ at 3$\sigma$
with a mass of at least 42\,M$_\text{J}$. GASTON cannot help settling the true mass of HD\,131664\,b but constrains its mass to less than 170\,M$_\text{J}$
at 3$\sigma$. \\

\paragraph{\bf HD\,136118\,b} Using the Hubble Space Telescope Fine Guidance Sensor (HST/FGS), Martioli et al. (2010) could measure the astrometric motion of 
the F9V star HD\,136118. They obtained an inclination of 163$\pm$3$^\circ$ and a true mass for the exoplanet candidate of 42$^{11}_{-18}$\,M$_\text{J}$ 
instead of the 12\,M$_\text{J}$ deduced from RV assuming an orbit seen edge-on. The 3-$\sigma$ upper-limit that we derived with GASTON for the true mass 
of HD\,136118\,b is close to the 3-$\sigma$ upper-limit derived from Hipparcos-2 measurements by Reffert \& Quirrenbach (2011) about 95-97\,M$_\text{J}$.
The Gaia DR1 astrometric excess noise of HD\,136118 (0.51\,mas) is thus compatible with the true astrometric motion $\sim$1.45\,mas of the host star due to 
companion 'b'.

\paragraph{\bf HD\,154345\,b} This long-period (3325\,days) companion is a planet confirmed by Simpson et al. (2010) by measuring the rotation axis angle of 
the host star. But this conclusion relies on the hypothesis of coplanar orbital and stellar equator planes, which is never guaranteed. Our 3-limit based on 
$\varepsilon_\text{DR1}$=0.35\,mas shows that HD\,154345\,b is indeed a planet with a mass smaller than 11.6\,M$_\text{J}$.

\paragraph{\bf HD\,177830\,b} Its true mass was tentatively determined at 55\,M$_\text{J}$, with an inclination of 1.3$^\circ$ by Han et al. (2001) using 
Hipparcos data. This result is within the 3-$\sigma$ limit that we derived here ($I_c$$>$$1.0^\circ$, $M_b$$<$$79$\,M$_\text{J}$). It confirms that 
the $\varepsilon_\text{DR1}$=0.87\,mas for this source incorporates a consequent fraction of real astrometric orbital motion.

\paragraph{\bf HD\,190228\,b} Using Hipparcos astrometry, this companion was previously identified as a brown dwarf with a mass of 67$\pm$29\,M$_\text{J}$ 
(Zucker \& Mazeh 2001a). Its mass was then reduced to 49$\pm$18\,M$_\text{J}$ and an inclination of 4.3$^\circ$\,\!$^{+1.8}_{-1.0}$ (Sahlmann et al. 2011a). 
The orbit significance they obtained for this star was 2-3$\sigma$. Using GASTON and a $\sin i$-prior on the inclination, we measure a 1-$\sigma$ upper-limit 
for the same mass of $<$24\,M$_\text{J}$ and a 3-$\sigma$ upper-limit of 111\,M$_\text{J}$. The inclination is $>$14$^\circ$ at 1-$\sigma$ and 
$>$3.2$^\circ$ at 3-$\sigma$. Our result agrees with the most precise measurements of the mass of HD\,190228\,b, but cannot bring significant 
improvements. Given the astrometric orbit semi-major axis as large as 2\,mas, Gaia will certainly provide the best measurement for this brown-dwarf once the 
astrometric series will be available.

The global compatibility of the true masses derived with GASTON with the true masses already published for these systems validates the GASTON method and 
confirms it can lead to better characterize candidate planetary systems.

\subsection{An updated mass-period diagram}
\label{sec:mass-period}
\begin{figure}
\includegraphics[width=89.3mm]{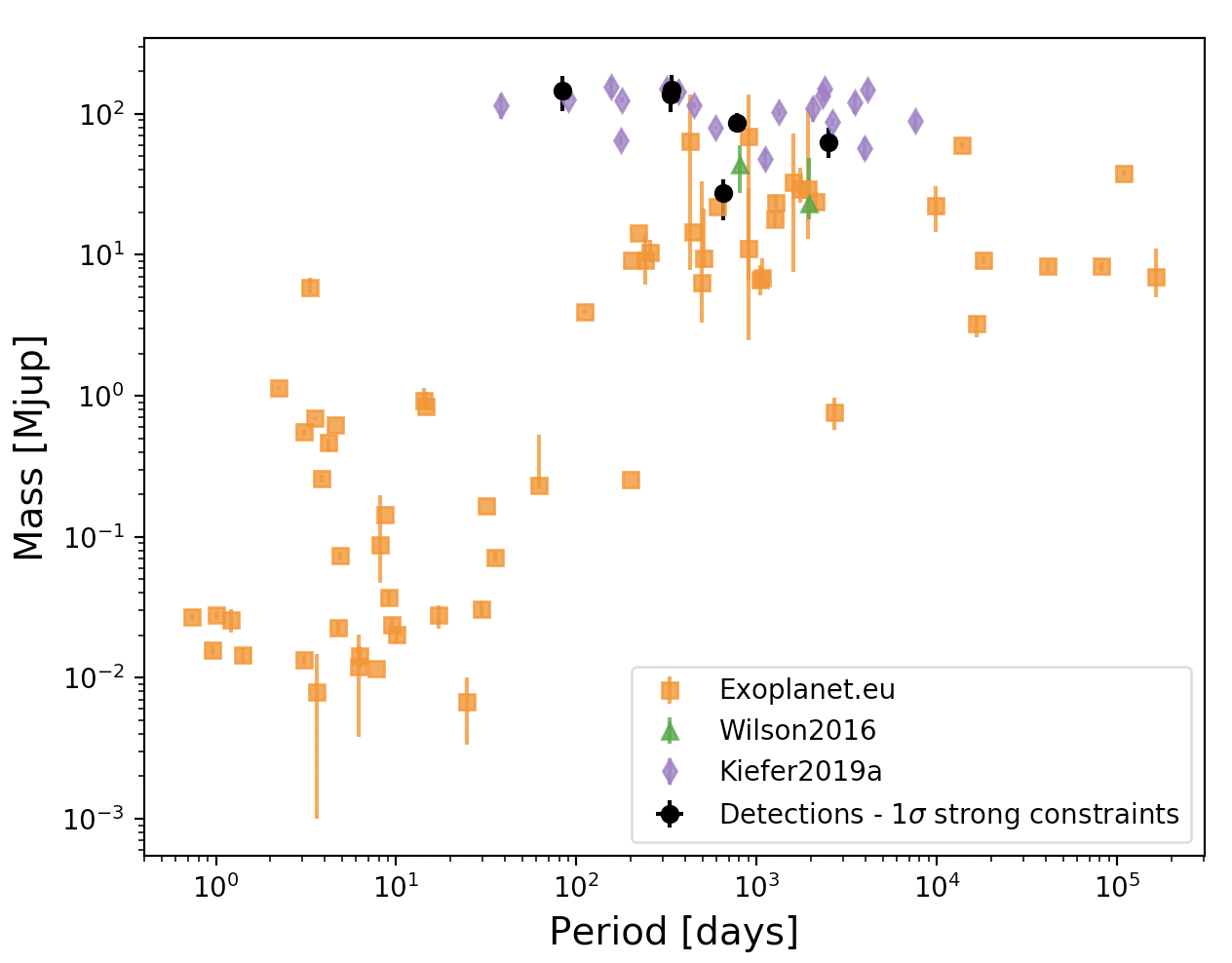}
\caption{\label{fig:mass_period_revised} Mass-period diagram from planets up to low-mass M-dwarfs 
($\sim$150\,M$_\text{J}$). The true masses derived with GASTON for non-transiting planets and systems without RV-drift are represented as black circles. 
They are compared to the true mass and period of companions in the exoplanet.eu database (orange squares), and of supplementary BD and low-mass M-dwarfs 
published in Wilson et al. (2016) in green, and Kiefer et al. (2019) in purple. Only systems within 60\,pc from the Sun and surrounding FGK host stars 
(0.52$<$$M_\star$$<$$1.7$\,M$_\odot$) are represented.}
\end{figure}

The masses derived with GASTON allow us to update the mass-period diagram of planet and brown dwarf companions. It is represented in 
Fig.~\ref{fig:mass_period_revised}, compared to companions with true mass from the \verb+Exoplanet.eu+ database and massive companions reported in 
Wilson et al. (2016) and Kiefer et al. (2019). We selected only systems within 60\,pc from the Sun, surrounding FGK host stars with masses within 
0.52--1.7\,M$_\odot$, with a published inclination measurement. Such systems are objects of extensive surveys (e.g. Sahlmann et al. 2011a, H\'ebrard et al. 
2016, Kiefer et al. 2019) with better observational completion and detection of planets and BD with mass larger than 1\,M$_\text{J}$. 
We include any mass compatible with as much as 150\,$M_\text{J}$ in order to encompass the surroundings of the substellar domain. We exclude GASTON masses 
of transiting planets, which are better determined in the \verb+Exoplanet.eu+ database. We also exclude the GASTON mass of candidates which host star RVs have 
a long-term drift, in order to remove possible bias due to an outer companion. Upper-limits are not represented.

The mass measurements of the present study add new points to the $M$-$P$ diagram in the BD-to-stellar domain at orbital periods larger than 100\,days. There
still remains blank regions: the BD domain below 100-days period (Kiefer et al. 2019a), the short-period Neptunian desert (Mazeh et al. 2016), and the 
observationally biased triangular area from short-period Earth-mass planets to long-period Jupiter mass planets.

\begin{figure}
\includegraphics[width=89.3mm]{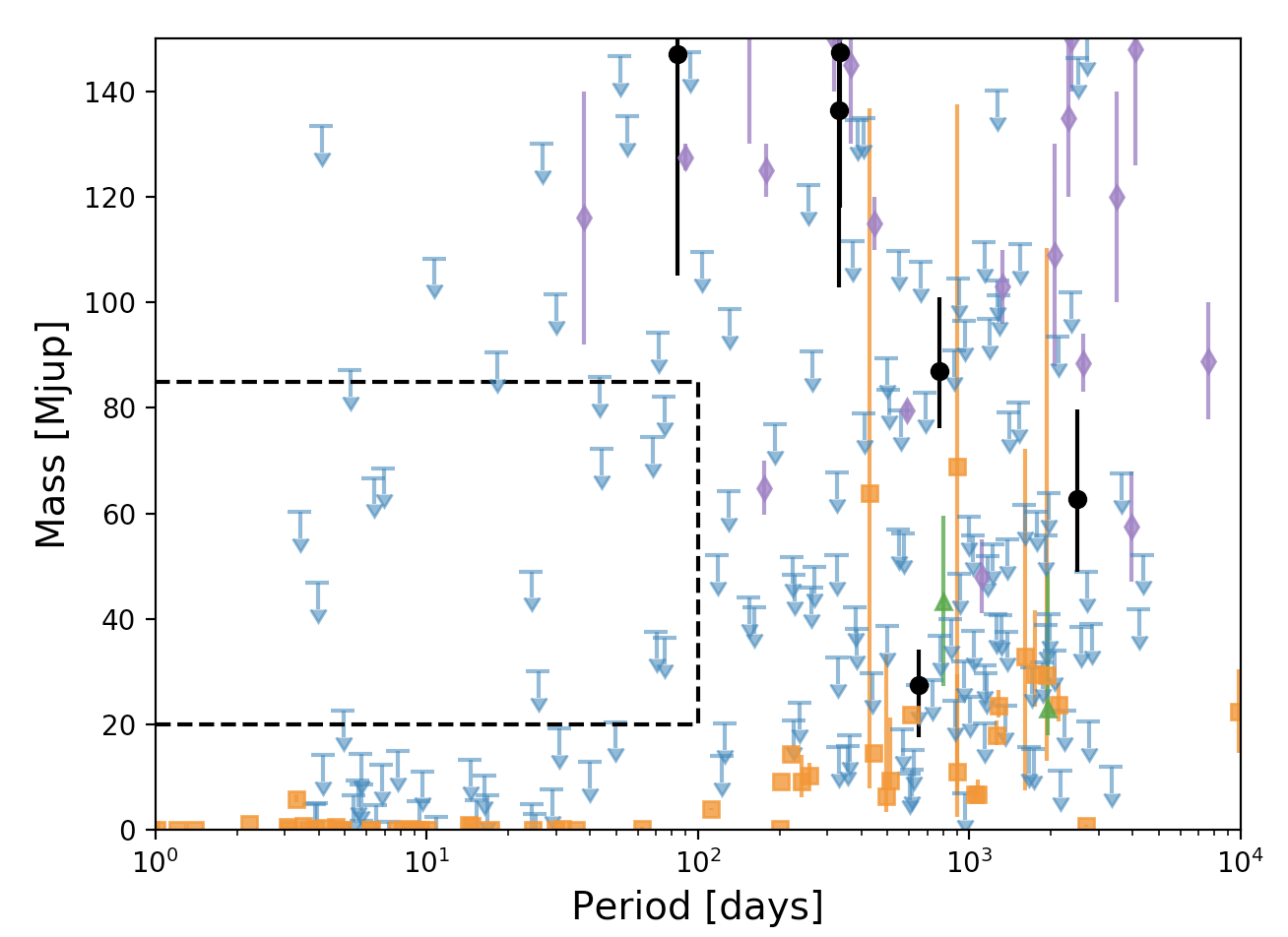}
\caption{\label{fig:mass_period_revised_zoom} Same diagram as Fig.~\ref{fig:mass_period_revised}, but focused on the 0-90\,M$_\text{J}$ region with a linear 
scaling of the y-axis and centered on the BD domain. The 3-$\sigma$ upper-limits on mass derived by GASTON are shown as blue downward arrows.}
\end{figure}

In the BD domain, the $M$-$P$ distribution presents a strong cut in the region of brown dwarf companions at $\sim$100\,days. But below 
100\,days, several tens of other companions which mass cannot be well-constrained may reside in the BD mass regime. 
In Fig.~\ref{fig:mass_period_revised_zoom}, we included the 3-$\sigma$ upper-limits derived with GASTON around and in the BD domain. Even then, the 
region bounded by masses 20-85\,M\,$_\text{J}$ and periods 0-100\,days remains significantly emptier than its surrounding. This tends to confirm the 
most recent estimation of the brown dwarfs desert boundaries (Ma \& Ge 2014, Ranc et al. 2015, Kiefer et al. 2019).

\subsection{Star-host metallicity and orbital eccentricity distributions in the BD domain}

Brown dwarfs companion stand at the boundary between stellar binaries and giant planets. It remains unknown if some BD companions belong to one or the other 
population, or if BDs have a main formation channel. 

Core-accretion scenarios (Pollack et al. 1996) predict that giant planets are more difficult to form around 
metal-deficient stars (Ida \& Lin 2005) and strong observational evidences show that giant planets indeed occur less frequently about stars with sub-solar metallicity 
(Fischer \& Valenti 2005, Mayor et al. 2011). On the other hand, the formation of stellar binaries by gravitational instability does not depend on the metallicity of 
the host star and it is expected that binary companions have the same metallicity distribution whatever their mass (Maldonado et al. 2017). 

The study of eccentricity distributions in the BD domain by Ma \& Ge (2014) revealed the existence of a sharp transition at $\sim$42.5\,M$_\text{J}$.
Below 42.5\,M$_\text{J}$ the eccentricity distribution is consistent with mass-limited eccentricity pumping by planet-planet scattering (Rasio \& Ford 1996), and
beyond 42.5\,M$_\text{J}$ the eccentricity distribution of BD companions is similar to that of binary stars (Halbwachs et al. 2003). 
A consistent transition at similar mass was found by Maldonado et al. (2017) in the distribution of host star metallicities. The host-star metallicity of BD with a 
mass $>$42.5\,M$_\text{J}$ spans a large range of values from sub-solar to super-solar, while those with mass $<$42.5\,M$_\text{J}$ have host-stars 
metallicities more similar to those of giant exoplanets with a prevalence for metal-rich hosts. 

This limiting mass of $\sim$42.5\,M$_\text{J}$ could thus be separating low-mass BDs formed like planets by core-accretion from high-mass BDs formed like 
stars by gravitational instability in molecular clouds. 

\begin{figure}
\includegraphics[width=89.3mm]{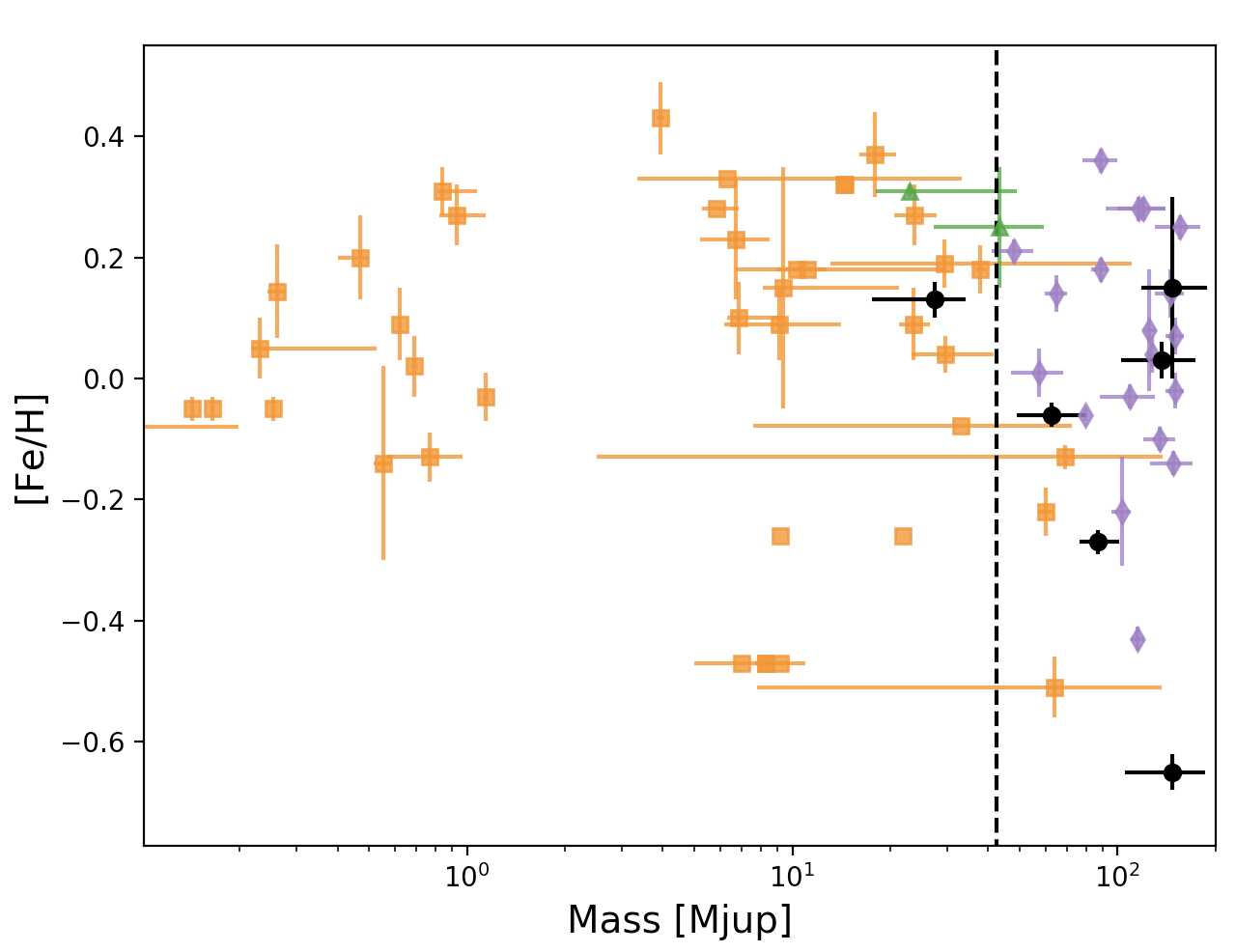}
\caption{\label{fig:mass_metallicity} Mass-metallicity diagram from planets up to low-mass M-dwarfs ($\sim$150\,M$_\text{J}$). The symbols are the same as 
in Figure~\ref{fig:mass_period_revised}. The dashed line indicates the 42.5\,M$_\text{J}$ mass-limit derived by Ma \& Ge (2014).}
\end{figure}

\begin{figure}
\includegraphics[width=89.3mm]{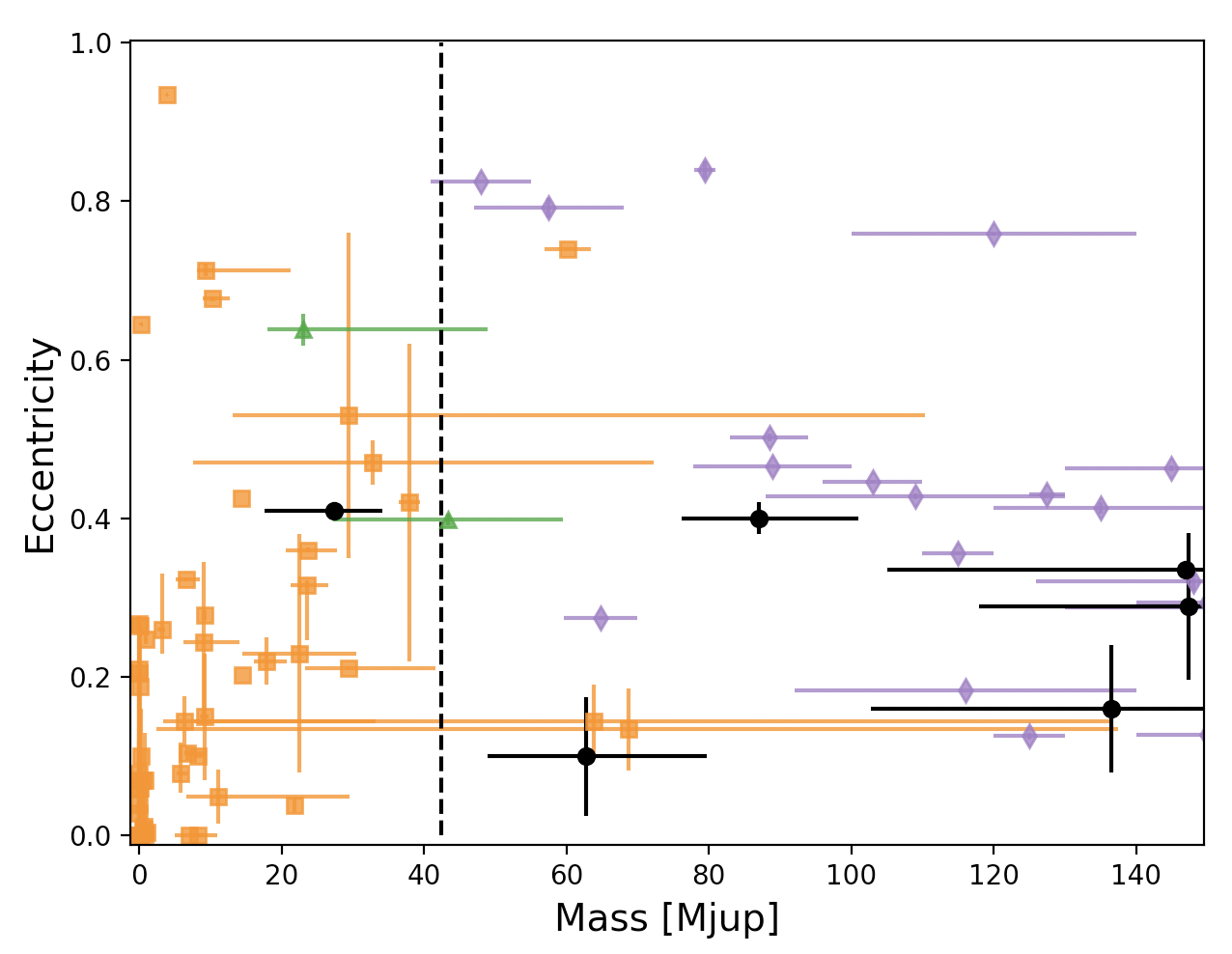}
\caption{\label{fig:mass_eccentricity} Mass-eccentricity diagram from planets up to low-mass M-dwarfs ($\sim$150\,M$_\text{J}$), with a linear scale as in 
Ma \& Ge (2014). The symbols are chosen the same as in Figure~\ref{fig:mass_metallicity}.}
\end{figure}

Here, we add the new GASTON measurements to \verb+exoplanet.eu+ companions, and the published companions in Wilson et al. (2016) and Kiefer et al. (2019a)
to obtain metallicity and eccentricity distributions with respect to true masses in Figs.~\ref{fig:mass_metallicity} and~\ref{fig:mass_eccentricity}. 
We select, as in Section~\ref{sec:mass-period}, systems within 60\,pc from the Sun, surrounding FGK host stars with masses within 
0.52--1.7\,M$_\odot$, and with a published inclination measurement. Metallicity, or [Fe/H], measurements are taken 
from the \verb+exoplanets.org+ database for the sample studied in the present paper, from \verb+exoplanet.eu+ for the corresponding sample, and from the 
Wilson et al. (2016) and Kiefer et al. (2019a) for the rest of the considered companions.

Beyond the brown dwarf domain, metallicity reaches sub-solar values, while giant planets are indeed found preferably around stars with super-solar 
metallicity. No clear boundary can be derived from the still sparse distribution of measured companion mass, although a 
transition could be occurring about 50\,M$_\text{J}$ in agreement with the 42.5\,M$_\text{J}$ found by Maldonado et al. (2017). 
The distribution of eccentricity with companion mass in Fig.~\ref{fig:mass_eccentricity} does not exhibit a well-defined transition at 42.5\,M$_\text{J}$, as 
reported in Ma \& Ge (2014). Nevertheless, four BD with $M$$>$$45$\,M$_\text{J}$ stand above $e$$=$$0.7$ while all BD companions have $e$$<$$0.7$ below 
$45$\,M$_\text{J}$. The eccentricity distribution within our sample thus seems to match with that of Ma \& Ge (2014). 

Our current sample of exoplanets, BD and low-mass M-dwarf companions around FGK stars at less than 60\,pc of the Sun with a true mass measured still 
need to be populated, but it agrees with previous non-volume limited studies on a transition in the brown dwarf domain at a mass of 
$\sim$42.5\,M$_\text{J}$. This critical mass possibly separates two populations of BD, those formed like stars from those formed like planet and mainly following 
predictions of the core-accretion scenario.

\section{Conclusions}
\label{sec:conclusion} 

We use the GASTON method developed in Kiefer et al. (2019) \& Kiefer (2019) with Gaia DR1 data to determine the true mass of the 911 RV-detected exoplanet candidates published in the 
\verb+exoplanets.org+ database. Reliable DR1 data were found for the host stars of 755 companions. Among those, a total of 29 companions induce an 
orbital motion of their host star significant enough to be detected as large astrometric excess noise, constituting a 'detection sample'. With GASTON, an 
inclination, and thus a true mass could be determined for 8 of them. For the remaining 21 companions we could only constrain an upper-limit on the true mass, 
with an astrometric motion compatible with the edge-on inclination within measurement noise. For other 227 candidates, the astrometric excess noise is not large 
enough to imply a firm detection of the astrometric motion of their host star, but it allows deriving an upper-limit on their true mass. They constitute a 'non-detection
sample'. 

We found that among the detection sample, 30 Ari B b, HD 114762 b, HD 148427 b, HD 5388 b, HD 6718 b, HD 16760 b, and HIP 65891 b are not planets, but 
brown dwarfs or M-dwarfs. Moreover, we measured a true mass of HD 141937 b, within 9-50\,M$_\text{J}$ compatible at 3-$\sigma$ with a planetary nature, 
altthough more likely a brown-dwarf.

Among the 227 candidates of the non-detection sample, GASTON applied on the small astrometric excess noise measured by Gaia DR1 confirms that 27 exoplanet 
candidates are indeed planets. The lower-limit on their inclination deduced from the small value of their astrometric excess noise led to an upper-limit on mass 
below 13.5\,M$_\text{J}$ at 3-$\sigma$. 

These new measurements populate the mass-period diagram in the BD-to-M-dwarf domain constraining the driest region of the desert of brown-dwarf 
companions detection, also-known-as the brown-dwarf desert, to orbital periods smaller than 100-days and mass larger than 20\,M$_\text{J}$. We 
thus confirm previous estimates of the period threshold of the brown-dwarf desert, $\sim$100\,days, obtained by Ma \& Ge (2014) accounting for 
the true mass of whole set of detected BD companions as of 2014, and Kiefer et al. (2019) analysing the $m\sin i$ distribution of companions to FGK-stars at less 
than 60\,pc from Earth. Moreover, the distributions of eccentricities and metallicities among brown dwarf companions are consistent with a transition from 
planet-like formation to star-like formation at about 40-50\,M$_\text{J}$ (Ma \& Ge 2014, Maldonado et al. 2017).

Since GASTON allowed determining companion masses of few tens of M$_\text{J}$ using only preliminary Gaia data products at a precision of $\sim$1\,mas, we 
can rejoice that future orbital solutions from Gaia astrometric time series at a precision of a few 10-100\,$\mu$as, will allow measuring the orbital 
inclination and masses of many RV Jupiter-mass exoplanets and brown dwarfs, as well as new detections among the several billions of sources monitored with 
Gaia.

\begin{acknowledgements}
We thank the anonymous referee for her/his fruitful comments. This work has been funded by the Centre National d'Etudes Spatiales (CNES), and by Paris Science et Lettre (PSL) University. 
\end{acknowledgements}

\begin{appendix}
\section{The magnitude of the companion}
\label{app:magnitude}

To calculate the impact of the light emitted from the companion on the apparent primary semi-major axis, as observed by Gaia, also known as the photocenter 
semi-major axis, we have to take into account two effects: the emission from the companion (c) itself if not planetary and the star ($\star$) light reflected by the 
companion towards the observer. We recall the equation of the photocenter semi-major axis given in Kiefer et al. (2019) (see also van de Kamp 1975 for the original calculation):

\begin{equation}
a_\text{phot}=(B-\beta) a_\text{tot} \qquad \text{with } a_\text{tot}=a_c+a_\star
\end{equation}

We introduced the luminosity fraction $\beta$=$L_c/(L_c+L_\star)$ and the mass fraction $B$=$q/(1+q)$ with $q$=$M_c/M_\star$ the mass ratio. The key-parameter
here is the luminosity fraction. 

\begin{figure}
\includegraphics[width=89.3mm]{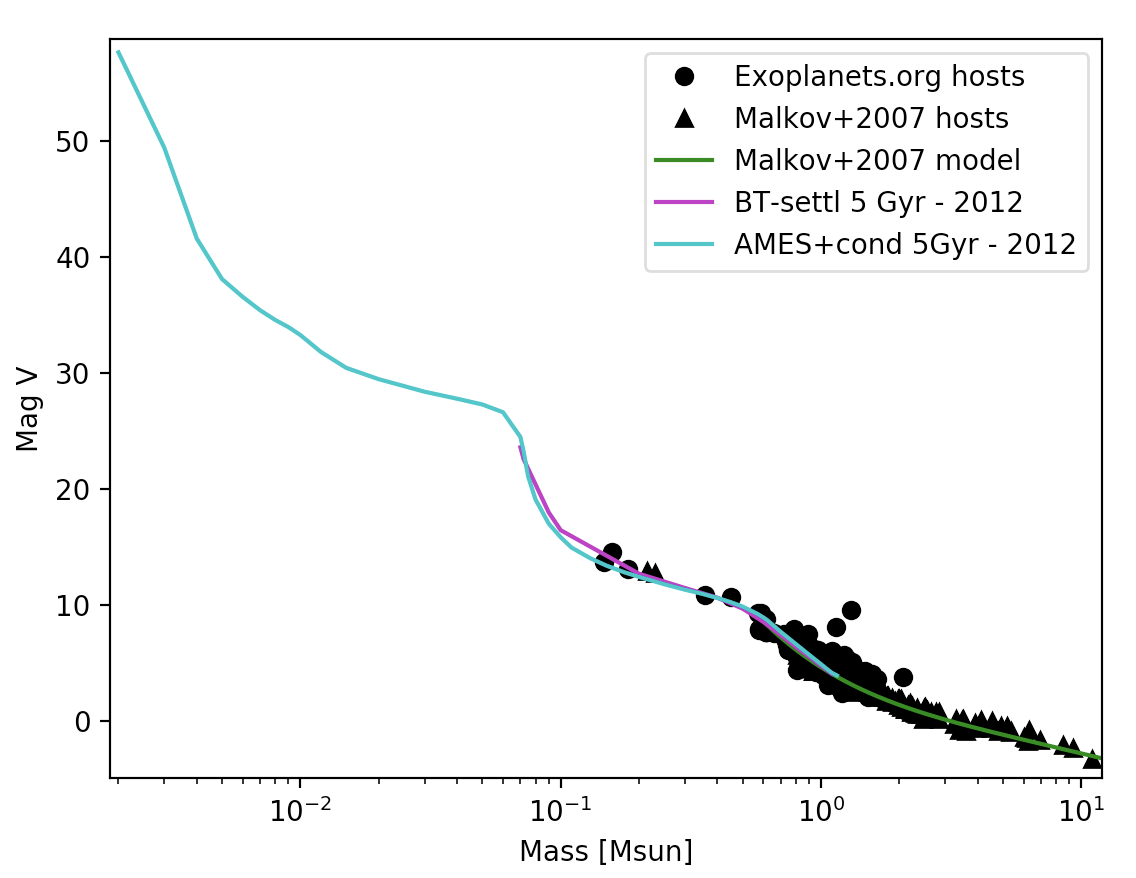}
\caption{\label{fig:mass_mag}Absolute visual magnitude of emitted light against mass from low-mass brown-dwarf up to massive stars. Data are taken from the Exoplanets.org catalog and 
Malkov et al. (2007). Solid lines are models, as presented in the legend.}
\end{figure}

In order to calculate the emitted $V$-magnitude of the components, we use different empirical models. We use the given $V$ magnitude as an approximation of 
the $G$ magnitude. Since only the $\delta G$ is important for our study, we do not expect strong deviations due to this approximation. They are listed below and 
presented on Fig.~\ref{fig:mass_mag}, compared to published data from \verb+exoplanets.org+ and Malkov (2007) in the Johnson $V$-band:
\begin{itemize}
\item Allard et al. (2012) AMES-cond isochrone at 5\,Gyr, for the BD-to-dM regime, 
\item BT-Settl (Allard et al. 2012, and references therein) isochrone at 5\,Gyr, for the stellar regime up to 1.1\,M$_\odot$,
\item Malkov (2007) empirical model above 1.1\,M$_\odot$ up to 10\,M$_\odot$.
\end{itemize} 

Only visual magnitudes for objects with mass larger than 0.1\,M$_\odot$ could be found in the literature, due to the difficulty for observing faint objects 
such as brown dwarfs and exoplanets in the optical band. Therefore, we can only assess the validity of the $V$-mag models in the stellar domain and we will assume 
their validity in the $G$-band down to the planetary domain owing moreover to the relative compatibility of the AMES-cond models with observations in the 
infrared (Chabrier et al. 2000, Baraffe et al. 2003, Allard et al. 2012). The BT-settl model seems more accurate than AMES-cond in the IR for massive brown-dwarfs 
(Allard et al. 2012), but it  does not give magnitudes for objects with mass below 0.75\,M$_\odot$. In order to insure continuity of the mass-magnitude relation 
we prefer using the AMES-cond model from masses of 0.1\,M$_\odot$ down to 10\,M$_\text{J}$. 

\begin{figure}
\includegraphics[width=89.3mm]{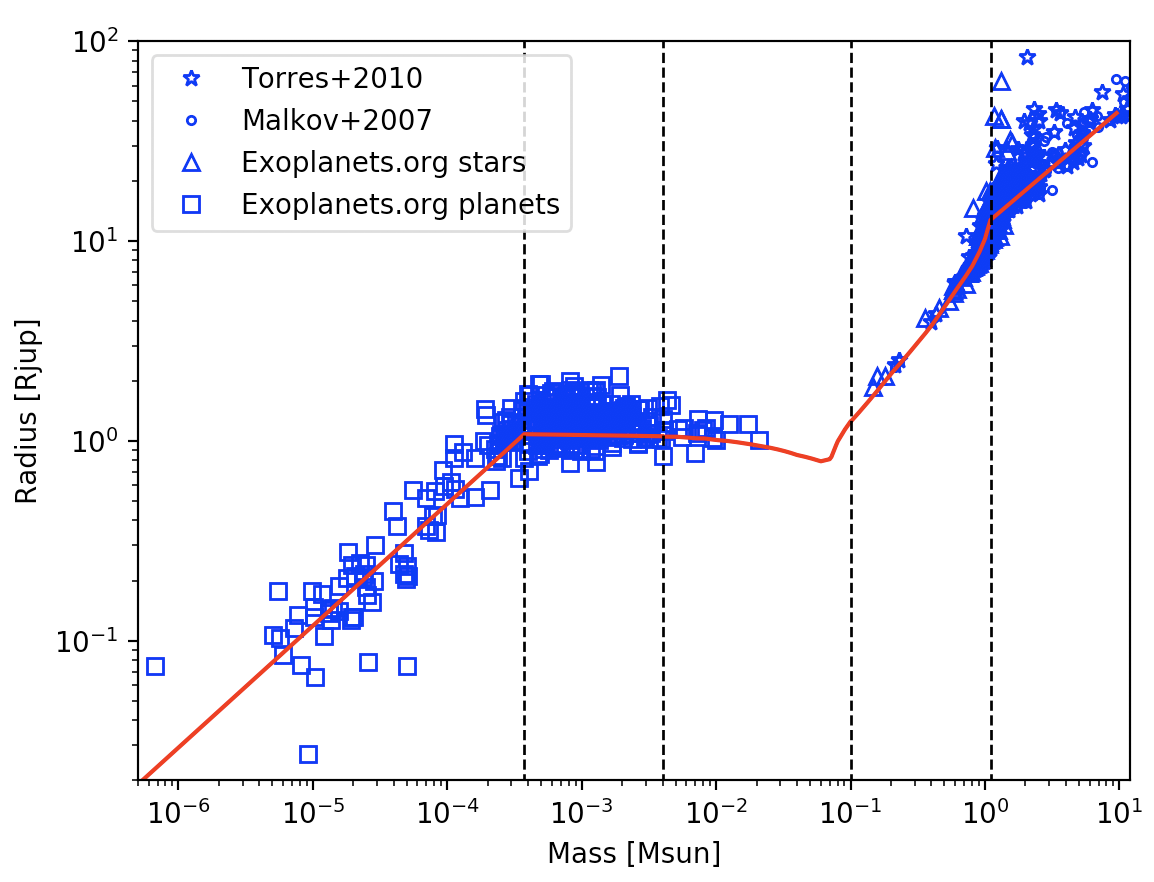}
\caption{\label{fig:mass_radius}Radius against mass from terrestrial mass planets up to massive stars. Data are taken from the Exoplanets.org catalog, Malkov et al. (2007) 
and Torres et al. (2010). The red solid line represents the continuous model, comprised of Bashi et al. (2017) from 10$^{-6}$ to 0.004\,M$_\odot$, Baraffe et al. (2003) from 
0.004 to 0.1\,M$_\odot$, BT-settl from 0.1 to 1.1\,M$_\odot$ and the empirical relation $R$$\propto $$M^{0.57}$ above 1.1\,M$_\odot$ (e.g. Demircan 1997).}
\end{figure}

For the reflected light, we calculate the mean reflection along an entire orbit. For the whole domain from planets to stars, we assume the body is a Lambertian sphere with a 
typical Bond albedo of 0.3. The radius of the sphere is related to the mass of the body. There is no continuous exact law $R$=$f(M)$ on the whole 
planet-to-star domain, with a wide diversity of densities for planets and also for stars if we account for (sub-)giants stars. However, for the sake of continuousness 
and simplicity, we will assume a common continuous law relating the radius of a body to its mass. This will avoid issues with gaps in the MCMC and allow us to 
derive well-behaved posterior distributions. This law is established using several segments and is presented in Fig.~\ref{fig:mass_radius}: 
\begin{itemize}
\item Low-mass planets, up to 0.5\,M$_J$ with the empirical relations of Bashi et al. (2017),
\item Jovian-mass planets and brown-dwarfs up to 0.1\,M$_\odot$ stars along an AMES-cond isochrone at 5\,Gyr (Allard et al. 2012),
\item Low-mass stars from 0.1 to 1.1\,M$_\odot$ with the BT-settl model (Allard et al. 2012, and references therein) along an isochrone at 5\,Gyr,
\item More massive stars up to 10\,M$_\odot$ with the empirical $R\propto M^{0.57}$ relationship (e.g. Demircan \& Kahraman 1991, Torres et al. 2010).
\end{itemize}

\begin{figure}
\includegraphics[width=89.3mm]{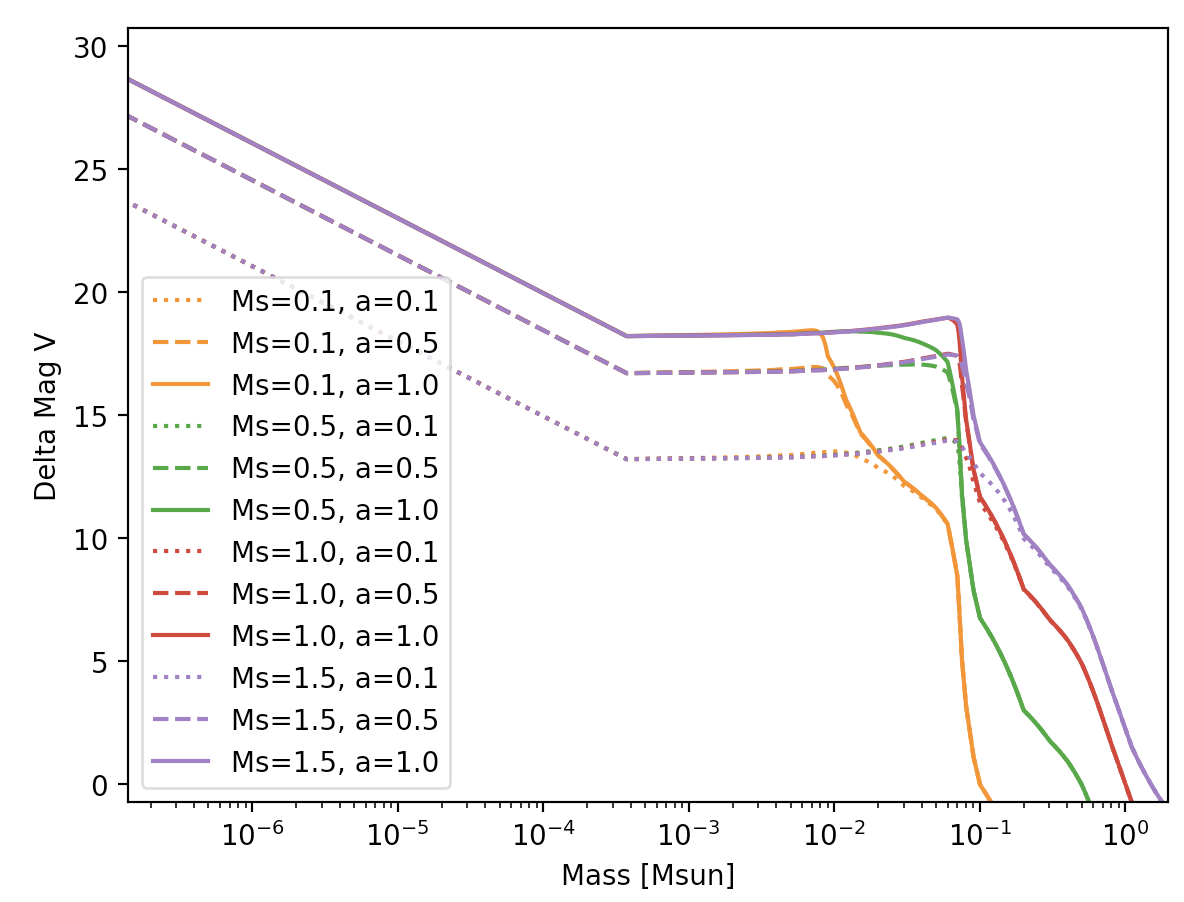}
\caption{\label{fig:delta_mag_mass}The adopted continuous model of $\Delta V$ with respect to the mass ratio. On this plot, the primary is assumed with 
a mass ranging from 0.1 to 1.5\,M$_\odot$, and the secondary is a Lambertian sphere with average Bond albedo of 0.3 on an orbit with semi-major axis between 0.1 and 1\,au.}
\end{figure}

\begin{figure}
\includegraphics[width=89.3mm]{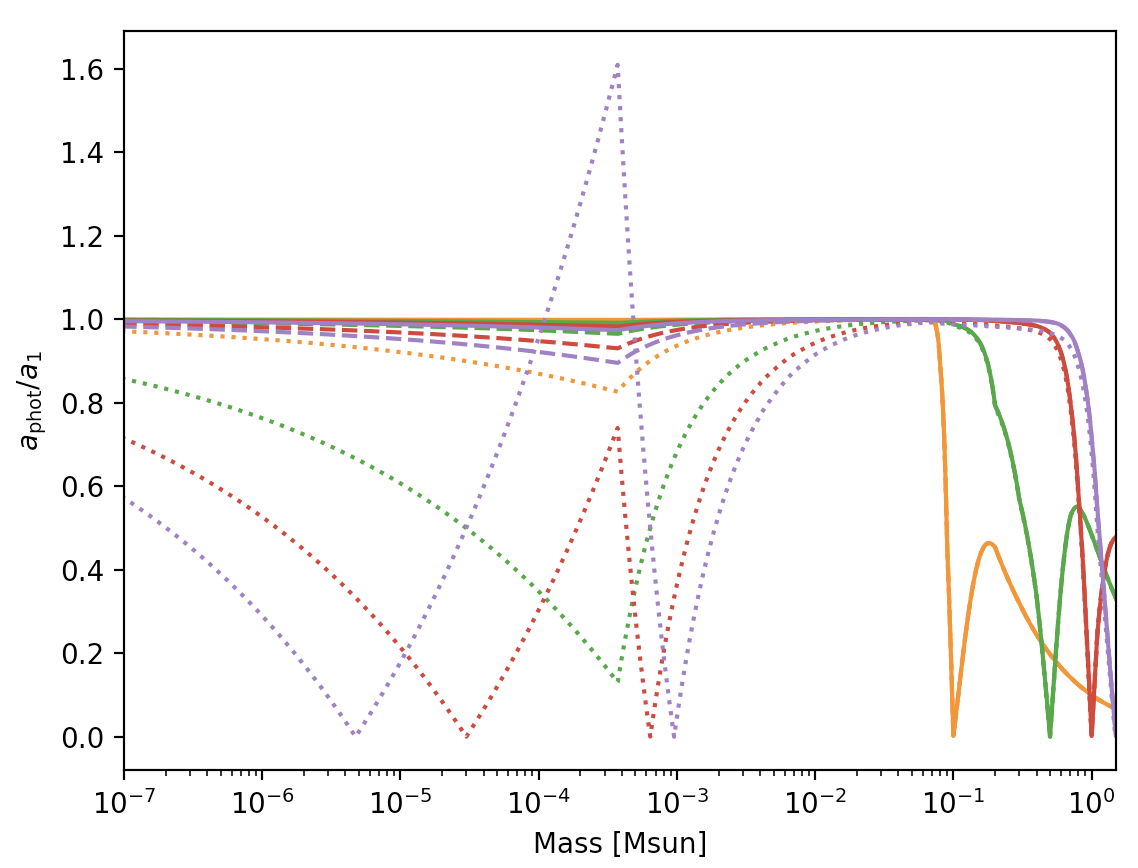}
\includegraphics[width=89.3mm]{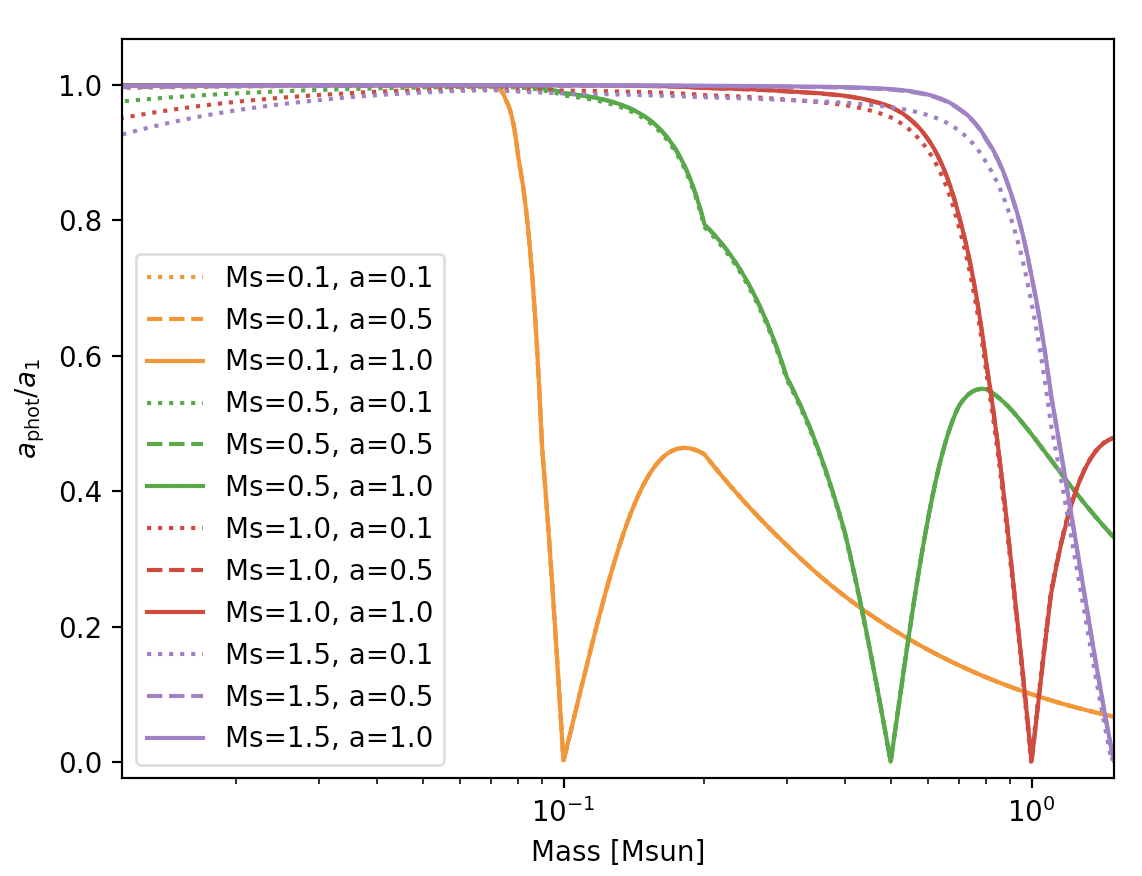}
\caption{\label{fig:aphot_a1_mass}The effect of emitted and reflected light by the secondary companion on the photocenter semi-major axis. Top: on the full mass ratio range.
Bottom: beyond $M$=0.1\,M$_\odot$. Captions are identifical in the two panels.}
\end{figure}

Taking into account emission and reflection from the companion, we can calculate magnitude differences between the primary and the companion for different 
values of their mass and for different companion's orbit semi-major axis. This is plotted on Fig.~\ref{fig:delta_mag_mass}. The net impact on the apparent primary 
semi-major axis can be measured by comparing the photocenter semi-major axis to the primary semi-major axis. This is presented on 
Fig.~\ref{fig:aphot_a1_mass}. In general the semi-major axis of the photocenter orbit decreases with increasing contribution of the companion in the total luminosity
of the system.

In the stellar regime, the impact of the companion starts to be significant on the apparent primary semi-major axis for companion masses larger than about 20\% 
of the primary star mass. For a primary of mass 0.5\,M$_\odot$ the magnitude of the secondary has a measurable effect for a mass larger than about 
0.1\,M$_\odot$. If the secondary is too luminous compared to the primary, the semi-major axis of the photocenter can even reach 0. However this effect remains 
hidden in the present study, because we impose that the companion be dark with ($\Delta V$$>$$2.5$). 

In the planetary mass regime on the other hand, the impact of the companion on the photocenter can be strong if the semi-major axis of the companion orbit is smaller 
than 0.5\,au and the mass ratio $q$=$M_c/M_\star$$\sim $$10^{-5}-10^{-3}$. The impact is stronger for earlier primaries. This is due to the shortening of the 
primary star orbit with decreasing companion mass, while the companion radius is relatively constant about 1\,R$_\text{J}$ down to a few fractions of Jupiter mass 
(Bashi et al. 2017). In this regime, the astrometric motion of the system observed from Earth, although of small extent, can actually follow the motion of the companion rather than that of the stellar host. 
This happens precisely when the luminosity ratio $L_c/L_\star$$>$$q$. 

Nevertheless, this effect will not have a strong role in the present study, because $q$$\sim $$10^{-5}-10^{-3}$ with $a_c$$<$$0.5$\,a.u. implies 
$a_\star$$\ll $$a_c$ and $a_\text{ph}$$<$$10^{-3}$\,a.u. In the worst case scenario, it could only lead at most to an astrometric motion of $\sim$0.5\,mas if
the parallax is $\sim$500\,mas. This is well below the detection thresholds ($\varepsilon_\text{thresh,prim}$=0.85\,mas and $\varepsilon_\text{thresh,second}$=1.2\,mas) defined in 
Section~\ref{sec:threshold} and this situation is thus undetectable within the diverse noises accounted for in Gaia measurements. However, it will be an important 
effect to account for in future analysis of Gaia time series when this precision will be reached. Neglecting it might lead to strongly underestimate the star 
semi-major axis, and thus the mass of the companion.


\section{Complementary tables of GASTON results for the  non-detection sample}
\label{app:non-detection}

\onecolumn
\thispagestyle{empty}
\newpage
\newgeometry{margin=1mm,top=1mm}
\begin{landscape}
\begin{flushleft}
\begin{tiny}
\begin{longtable}{@{}l@{}|c@{~}c@{~}c@{~}c@{~}c@{~}c@{~}c@{~}c@{~}c@{~}c@{}c@{~~}c@{~}|c@{~~}c@{~~}c@{~~}c@{~~}c@{}} 
\caption{\label{tab:list_planets_nondet} List of the 227 selected planets in the non-detection sample. Where uncertainties are missing we assume 10\% errors 
on the corresponding parameter. The parallax with uncertainties are all taken from the DR1, while those without uncertainties are taken from SIMBAD. }  \\
\multicolumn{13}{c|}{RV data} & \multicolumn{5}{c}{Gaia DR1 data} \\
\hline
Name & $P$          	& $m\sin i$  		& $K$ 			& $e$ 	& $\omega$ 	& $T_p$ 	& $M_\star$ 	& $a_p$ 	& $\pi$    	& $a_\star\sin i$  & Transit  & Drift  	& $\varepsilon_\text{DR1}$ 	& $N_\text{pts}$ 	& $N_\text{FoV}$ 	& Gaia   & Duplicate \\
	  & (days) 		& (M$_\text{J}$) 	&  (m\,s$^{-1}$) 	& 		& ($^\circ$) 	& -2,450,000	& ($M_\odot$)	&   (au)      & (mas)	&   (mas) 			& flag & flag &  (mas)          	&                      		&				&  dataset & source \\
\hline
\endfirsthead
\caption{Continued...}  \\
\hline
Name & $P$          	& $m\sin i$  		& $K$ 			& $e$ 	& $\omega$ 	& $T_p$ 	& $M_\star$ 	& $a_p$ 	& $\pi$    	& $a_\star\sin i$  & Transit  & Drift  	& $\varepsilon_\text{DR1}$ 	& $N_\text{pts}$ 	& $N_\text{FoV}$ 	& Gaia   & Duplicate \\
	  & (days) 		& (M$_\text{J}$) 	&  (m\,s$^{-1}$) 	& 		& ($^\circ$) 	& -2,450,000	& ($M_\odot$)	&   (au)      & (mas)	&   (mas) 			& flag & flag &  (mas)          	&                      		&				&  dataset & source \\
\hline
\endhead
8 UMi b	& 93.4$\pm$4.5 	& 1.52$\pm$0.15 	& 46.1$\pm$4.0 	& 0.060$\pm$0.090 	& 91$\pm$85 	& 4108$\pm$23 	& 1.80$\pm$0.10 	& 0.490$\pm$0.018 	& 6.14$\pm$0.21 	& 0.00243$\pm$0.00030 	& n 	& n 	& 0.26 	& 68 	& 16 	& 1 	& n\\
BD -06 1339 b	& 3.87280$\pm$0.00040 	& 0.0268$\pm$0.0038 	& 4.40$\pm$0.60 	& 0 	& 90 	& 5220.50$\pm$0.10 	& 0.70 	& 0.043 	& 49.78$\pm$0.28 	& 0.000078 	& n 	& n 	& 0.52 	& 70 	& 9 	& 1 	& n\\
BD -06 1339 c	& 125.94$\pm$0.44 	& 0.168$\pm$0.054 	& 9.1$\pm$2.9 	& 0.31$\pm$0.11 	& 41$\pm$55 	& 5265$\pm$18 	& 0.70 	& 0.44 	& 49.78$\pm$0.28 	& 0.0050 	& n 	& n 	& 0.52 	& 70 	& 9 	& 1 	& n\\
BD -08 2823 b	& 5.600$\pm$0.020 	& 0.0459$\pm$0.0032 	& 6.50$\pm$0.10 	& 0.15$\pm$0.15 	& 30$\pm$100 	& 4637.7$\pm$1.6 	& 0.740$\pm$0.070 	& 0.0558$\pm$0.0018 	& 24.28$\pm$0.24 	& 0.0000803$\pm$0.0000098 	& n 	& n 	& 0.40 	& 59 	& 9 	& 1 	& n\\
BD -08 2823 c	& 237.6$\pm$1.5 	& 0.328$\pm$0.033 	& 13.4$\pm$1.0 	& 0.190$\pm$0.090 	& 127$\pm$21 	& 4193$\pm$13 	& 0.740$\pm$0.070 	& 0.679$\pm$0.022 	& 24.28$\pm$0.24 	& 0.00698$\pm$0.00099 	& n 	& n 	& 0.40 	& 59 	& 9 	& 1 	& n\\
BD -10 3166 b	& 3.48777$\pm$0.00011 	& 0.430$\pm$0.017 	& 60.9$\pm$1.4 	& 0.019$\pm$0.021 	& 334 	& 1171.22$\pm$0.66 	& 0.920$\pm$0.035 	& 0.04378$\pm$0.00056 	& 12.17$\pm$0.45 	& 0.000238$\pm$0.000016 	& n 	& n 	& 0.28 	& 77 	& 11 	& 1 	& n\\
BD -11 4672 b	& 1667$\pm$32 	& 0.538$\pm$0.044 	& 13.4$\pm$1.0 	& 0.050$\pm$0.050 	& 231$\pm$110 	& 5899$\pm$42 	& 0.571$\pm$0.014 	& 2.283$\pm$0.035 	& 36.72$\pm$0.26 	& 0.0754$\pm$0.0066 	& n 	& n 	& 0.40 	& 53 	& 6 	& 1 	& n\\
BD +14 4559 b	& 268.94$\pm$0.99 	& 1.52$\pm$0.19 	& 55.2$\pm$2.3 	& 0.290$\pm$0.030 	& 87.6$\pm$7.9 	& 3293.7$\pm$6.3 	& 0.86$\pm$0.15 	& 0.775$\pm$0.045 	& 20.10$\pm$0.24 	& 0.0263$\pm$0.0058 	& n 	& n 	& 0.28 	& 35 	& 5 	& 1 	& n\\
\hline
\end{longtable}
\end{tiny}
\end{flushleft}
\end{landscape}
\newgeometry{margin=1.3cm,top=2cm,bottom=2.5cm}
\twocolumn

\onecolumn
\begin{longtable}{lcccccccc|c}
\caption{\label{tab:results_nondet_comp} Continuation of Table~\ref{tab:results_nondet} with the GASTON results for 162 exoplanet candidates from the 
non-detection sample. For those 162, the 3-$\sigma$ upper-limit on mass is larger than 13.5\,M$_\text{J}$ and the convergence criterion 
$N_\text{step}/\text{max}(\tau_\lambda)$$>$$50$ (Section~\ref{sec:general}). A flat-prior distribution on inclination is assumed and the 
MCMC runs on only 50,000 steps (Sections~\ref{sec:gaston} and~\ref{sec:case3}).} \\
Planet name & Period & $m\sin i$ & $a\sin i$ & $\epsilon$ & \multicolumn{1}{c}{$a_\text{phot}$} & \multicolumn{1}{c}{$I_c$} & \multicolumn{1}{c}{$M_\text{c,true}$}  & \multicolumn{1}{c|}{$\Delta V$} & MCMC \\
			& (days) & (M$_\text{J}$) & (mas)  & (mas)  	& (mas)		 & ($^\circ$) &   \multicolumn{1}{c}{(M$_\text{J}$)}  &  & Acceptance  \\
	 		&		&			&		&		 & 3-$\sigma$ &		3-$\sigma$		&	3-$\sigma$  & 3-$\sigma$& rate  \\
\hline
\endfirsthead
\caption{Continued from previous page...} \\
Planet name & Period & $m\sin i$ & $a\sin i$ & $\epsilon$ & \multicolumn{1}{c}{$a_\text{phot}$} & \multicolumn{1}{c}{$I_c$} & \multicolumn{1}{c}{$M_\text{c,true}$}  & \multicolumn{1}{c|}{$\Delta V$} & MCMC  \\
			& (days) & (M$_\text{J}$) & (mas)  & (mas)  	& (mas)		 & ($^\circ$) &   \multicolumn{1}{c}{(M$_\text{J}$)}  &  & Acceptance  \\
	 		&		&			&		&		 & 3-$\sigma$ &		3-$\sigma$		&	3-$\sigma$  & 3-$\sigma$& rate \\
\hline
\endhead
\multicolumn{10}{c}{Non-detection sample} \\
\multicolumn{10}{c}{3-$\sigma$ limits, with 13.5$<$$M_\text{max,3$\sigma$}$$<$85\,M$_\text{J}$} \\
\multicolumn{9}{l|}{Primary dataset: 92 possible brown dwarfs} \\
BD +14 4559 b & 268.9 & 1.519 & 0.02629 & 0.2845 & $<$0.9991 & $>$1.769 & $<$49.83 & $>$20.72 & 0.2238  \\
BD -06 1339 c & 125.9 & 0.1682 & 0.004987 & 0.5190 & $<$0.6764 & $>$0.4732 & $<$20.14 & $>$22.23 & 0.1410  \\
BD -08 2823 c & 237.6 & 0.3281 & 0.006979 & 0.4011 & $<$0.6344 & $>$0.7914 & $<$24.12 & $>$22.05 & 0.2054  \\
BD -11 4672 b & 1667 & 0.5377 & 0.07535 & 0.3952 & $<$2.540 & $>$1.977 & $<$15.65 & $>$21.61 & 0.2018  \\
HD 10180 e & 49.75 & 0.07979 & 0.0004990 & 0.4662 & $<$0.2196 & $>$0.2366 & $<$20.44 & $>$22.37 & 0.1572  \\
\end{longtable}

\end{appendix}

\end{document}